\documentclass[a4paper,11pt]{article}   
\newcount\driver
\newcount\bozza

\font\msytw=msbm10 scaled\magstep1

\font\msytwww=msbm7 scaled\magstep1
\font\indbf=cmbx10 scaled\magstep2

{\count255=\time\divide\count255 by 60 \xdef\hourmin{\number\count255}
        \multiply\count255 by-60\advance\count255 by\time
   \xdef\hourmin{\hourmin:\ifnum\count255<10 0\fi\the\count255}}

\let\a=\alpha \let\b=\beta  \let\g=\gamma     \let\d=\delta  \let\e=\varepsilon
\let\z=\zeta  \let\h=\eta   \let\th=\vartheta \let\k=\kappa   \let\l=\lambda
\let\m=\mu    \let\n=\nu    \let\x=\xi        \let\p=\pi      \let\r=\rho
\let\s=\sigma \let\t=\tau   \let\f=\varphi     \let\c=\chi
\let\ps=\psi   \let\o=\omega 
 \let\D=\Delta     \let\L=\Lambda

\def\PP{{\cal P}}\def\MM{{\cal M}}\def\VV{{\cal V}}
\def\CC{{\cal C}}\def\WW{{\cal W}}
\def\TT{{\cal T}}\def\NN{{\cal N}}\def\BB{{\cal B}}
\def\RR{{\cal R}}\def\LL{{\cal L}}\def\QQ{{\cal Q}}
\def\DD{{\cal D}}\def\SS{{\cal S}}

\def\pp{{\bf p}}\def\xx{{\bf x}}
\def\yy{{\bf y}}\def\kk{{\bf k}}\def\nn{{\bf n}}
\def\zz{{\bf z}}
 \def\bP{{\bf P}}

       \def\oo{{\underline \omega}}
\def\ee{{\underline \varepsilon}}  
          
          \def\ux{{\underline\xx}}
\def\ue{{\underline \e}}           \def\uy{{\underline\yy}}
            \def\uo{{\underline \o}}
\def\us{{\underline \s}}           \def\xxx{{\underline \xx}}

\def\RRR{\hbox{\msytw R}}          
\def\rrr{\hbox{\msytwww R}}        
        
\def\NNN{\hbox{\msytw N}}

\let\dpr=\partial

\let\==\equiv

\let\io=\infty
\let\0=\noindent

\def\ie{\hbox{\it i.e.\ }}

\def\lft{\left}
\def\rgt{\right}
\def\der{\hbox{\rm d}}
\def\la{{\langle}}
\def\ra{{\rangle}}

\def\tende#1{\,\vtop{\ialign{##\crcr\rightarrowfill\crcr
             \noalign{\kern-1pt\nointerlineskip}
             \hskip3.pt${\scriptstyle #1}$\hskip3.pt\crcr}}\,}
\def\otto{\,{\kern-1.truept\leftarrow\kern-5.truept\to\kern-1.truept}\,}
\def\fra#1#2{{#1\over#2}}

\def\defi{{\buildrel def \over =}}

\def\wt#1{\widetilde{#1}}
\def\wh#1{\widehat{#1}}
\def\hat#1{\wh{#1}}
\def\sqt[#1]#2{\root #1\of {#2}}

\def\hf{{\widehat \f}}
\def\hW{{\widehat W}}

\def\hW{{\widehat W}}

\def\hp{{\widehat \ps}}

\def\hj{{\widehat J}}
\def\hg{{\widehat g}}

\def\hD{{\widehat \D}}       
                        
\def\hG{{\widehat G}}

\def\hR{{\widehat R}}

\def\PP{{\cal P}}\def\MM{{\cal M}}\def\VV{{\cal V}}
\def\CC{{\cal C}}\def\WW{{\cal W}}
\def\TT{{\cal T}}\def\NN{{\cal N}}\def\BB{{\cal B}}
\def\RR{{\cal R}}\def\LL{{\cal L}}\def\QQ{{\cal Q}}
\def\DD{{\cal D}}\def\SS{{\cal S}}

\def\T#1{{#1_{\kern-3pt\lower7pt\hbox{$\widetilde{}$}}\kern3pt}}
\def\VVV#1{{\underline #1}_{\kern-3pt
\lower7pt\hbox{$\widetilde{}$}}\kern3pt\,}
\def\W#1{#1_{\kern-3pt\lower7.5pt\hbox{$\widetilde{}$}}\kern2pt\,}

\def\indica{\leaders \hbox to 0.5cm{\hss.\hss}\hfill}
\def\guida{\leaders\hbox to 1em{\hss.\hss}\hfill}
\mathchardef\oo= "0521

\def\pp{{\bf p}}\def\xx{{\bf x}}
\def\yy{{\bf y}}\def\kk{{\bf k}}\def\nn{{\bf n}}
\def\zz{{\bf z}}
 \def\bP{{\bf P}}
 
\def\oo{{\underline \omega}}

\def\xxx{{\underline\xx}}

\def\qed{\raise1pt\hbox{\vrule height5pt width5pt depth0pt}}

\def\indic{\hbox{\raise-2pt \hbox{\indbf 1}}}
\def\bk#1#2{\bar\kk_{#1#2}}

\def\RRR{\hbox{\msytw R}} 
\def\rrr{\hbox{\msytwww R}} 
 
\def\NNN{\hbox{\msytw N}}

\def\ins#1#2#3{\vbox to0pt{\kern-#2 \hbox{\kern#1 #3}\vss}\nointerlineskip}

\newdimen\xshift \newdimen\xwidth \newdimen\yshift
\newcount\griglia

\def\insertplot#1#2#3#4#5#6{%
\xwidth=#1pt \xshift=\hsize \advance\xshift by-\xwidth \divide\xshift by 2%
\begin{figure}[ht]
\vspace{#2pt}
\hspace{\xshift}
\begin{minipage}{#1pt}
#3
\ifnum\driver=1 \griglia=#6
\ifnum\griglia=1
\openout13=griglia.ps
\write13{gsave .2 setlinewidth}
\write13{0 10 #1 {dup 0 moveto #2 lineto } for}
\write13{0 10 #2 {dup 0 exch moveto #1 exch lineto } for}
\write13{stroke}
\write13{.5 setlinewidth}
\write13{0 50 #1 {dup 0 moveto #2 lineto } for}
\write13{0 50 #2 {dup 0 exch moveto #1 exch lineto } for}
\write13{stroke grestore}
\closeout13
\includegraphics{griglia.ps}
\fi
\includegraphics{#4.ps}\fi%
\ifnum\driver=2 \fi
\end{minipage}
\caption{#5}
\end{figure}
}

\newdimen\shift \shift=0.1truecm
\def\lb#1{%
\ifnum\bozza=1
\label{#1}\hbox{\hskip\shift$\scriptstyle#1$}
\else\label{#1}
\fi}

\def\be{\begin{equation}}
\def\ee{\end{equation}}
\def\bea{\begin{eqnarray}}\def\eea{\end{eqnarray}}
\def\bean{\begin{eqnarray*}}\def\eean{\end{eqnarray*}}
\def\bfr{\begin{flushright}}\def\efr{\end{flushright}}
\def\bc{\begin{center}}\def\ec{\end{center}}
\def\ba#1{\begin{array}{#1}} \def\ea{\end{array}}
\def\bd{\begin{description}}\def\ed{\end{description}}
\def\bv{\begin{verbatim}}\def\ev{\end{verbatim}}
\def\nn{\nonumber}
\def\Halmos{\hfill\vrule height10pt width4pt depth2pt \par\hbox to \hsize{}}
\def\pref#1{(\ref{#1})}
 
\def\virg{\quad,\quad}

\newtheorem{theorem}{Theorem}
\newtheorem{corollary}{Corollary}
\newtheorem{lemma}{Lemma}

\begin{document}

\title{{Functional Integral Construction of the
Thirring model: axioms verification and massless limit}}

\author{ G. Benfatto \and  P. Falco \and   V. Mastropietro}

\date{}
\maketitle

\begin{center}
Dipartimento di Matematica, Universit\`a di Roma ``Tor Vergata''\\
via della Ricerca Scientifica, I-00133, Roma
\end{center}

\begin{abstract} We construct a QFT for the
Thirring model for any value of the mass in a functional integral
approach, by proving that a set of Grassmann integrals converges,
as the cutoffs are removed and for a proper choice of the bare
parameters, to a set of Schwinger functions verifying the
Osterwalder-Schrader axioms. The corresponding Ward Identities
have anomalies which are not linear in the coupling and which
violate the anomaly non-renormalization property. Additional
anomalies are present in the closed equation for the interacting
propagator, obtained by combining a Schwinger-Dyson equation with
Ward Identities.
\end{abstract}

\section{ Introduction and Main result}\lb{ss1}

\subsection{Historical Introduction}\lb{ss1.1}

Proposed by Thirring \cite{T} half a century ago, the {\it
Thirring model} is a Quantum Field Theory (QFT) of a spinor field
in a two dimensional space-time, with a self interaction of the
form $(\l/4) \int\! d\xx (\bar\psi_\xx\g^\m\psi_\xx)^2$. The
interest of such a model, witnessed by the enormous number of
papers devoted to it, is mainly due to the fact that it has a non
trivial behavior, similar to the one of more realistic models, but
at the same time it is simple enough to be in principle accessible
to an analytic investigation. Hence the validity of several
properties of QFT models, which in general can be verified at most
by perturbative expansions, can be checked in principle in the
Thirring model at a non-perturbative level. The Thirring model has
been studied along the years following different approaches and we
will recall here briefly the main achievements.

\*

\noindent {\it Exact approach.} After a certain number of
"solutions" of the model fell into disrepute after inconsistences
were encountered, Johnson \cite{J} was able to derive, {\it in the
massless case}, an exact expression for the two point function; if
$\la T(\psi_\xx\bar\psi_{\bf 0})\ra$ is the two-point function in
the Minkowski space, he found that $\la T(\psi_\xx\bar\psi_{\bf
0})\ra= i(\bar\g^\m\partial_\m)^{-1} (|\xx|/x_0)^{-\h_z}$, where
$\h_z=2 (\l/4\pi)^2 [1-(\l/4\pi)^2]^{-1}$, $\bar\g_\m$ are the
Minkowski gamma matrices and $x_0$ is an arbitrary constant with
the dimension of a length. This result was followed shortly
\cite{K} by the general $n$-point function at non-coinciding
points. The Johnson solution, based on operator techniques, is
essentially a {\it self-consistency} argument: a number of
reasonable requirements on the correlations is assumed from which
their explicit expression can be determined. The first assumpion
is the validity of {\it Ward-Takahashi Identities} (WTi) of the
form
\bea
&&i\partial_\m \la T (j^{\m}_\zz \psi_\xx\bar\psi_\yy)\ra
=a[\d(\zz-\xx)-\d(\zz-\yy)] i\la T(\psi_\xx\bar\psi_\yy)\ra\;, \lb{1.1a}\\
&&i\partial_\m \la T (j^{\m,5}_\zz \psi_\xx\bar\psi_\yy)\ra =\bar
a[\d(\zz-\xx)-\d(\zz-\yy)]\g^5 i\la T(\psi_\xx\bar\psi_\yy)\ra\;,\nn
\eea
where the current $j^{\m}_\xx$ and pseudocurrent $j^{\m,5}_\xx$
are operators, formally defined respectively as
$\bar\psi_\xx\bar\g^\m\psi_\xx$ and
$\bar\psi_\xx\bar\g^\m\g^5\psi_\xx$, and the coefficients
$a^{-1}-1$ and $\bar a^{-1}-1$ are called {\it anomalies}; they
would vanish in the naive WTi which one would expect from the
classical conservation laws, see for instance \cite{A1}. The
second assumption was the validity of {\it Schwinger-Dyson}
equations (the analogue of the equations of motion), and,
combining them with the WTi, closed equations for the $n$-point
functions were found; from them an explicit expression for the
$n$-point function at distinct points was derived and, by a
self-consistency argument, the following explicit values for the
anomalies:
\be
a^{-1}=1-{\l\over 4\pi}\quad\quad\quad \bar a^{-1}=1+{\l\over
4\pi}\;.\lb{1.1b}
\ee
The anomalies are then {\it linear in the coupling}, that is no
higher orders contributions are present; this property is called
{\it anomaly non-renormalization} or {\it Adler-Bardeen theorem}, and it holds, as a statement
valid {\it at all orders in perturbation theory} and with suitable
regularizations \cite{AB}, in realistic models like $QED$ or the
Electroweak model in $d=4$ (in the last model it plays a crucial
role in the proof of its perturbative renormalizability). The
validity of \pref{1.1b} in the Thirring model is particularly
significant, as it has been considered \cite{GR} as a {\it
non-perturbative verification} of the perturbative analysis of
\cite{AB} adapted to this case; however the applicability of the
\cite{AB} analysis to the Thirring model has been also questioned
\cite{AF}. Another remarkable relation found in the exact analysis
in \cite{J} is
\be
\h_z={\l\over 2\pi}(a-\bar a)\;, \lb{1.1c}
\ee
relating the anomalous exponent of the two point function with the
anomalies; it is an immediate consequence of the the {\it closed
equation} for the two point functions obtained by inserting the
WTi \pref{1.1a} in the Schwinger-Dyson equation. The outcome of
this exact analysis is an explicit expression for the $n$-point
function at non-coinciding points and for the WTi. However, as
stressed by Wigthmann \cite{W}, the procedure is {\it not
satisfactory} from a mathematical point of view, as it involves
several formal manipulations of diverging quantities; even the
meaning of the basic equation \pref{1.1a} is unclear as the
averages in the l.h.s. and r.h.s. has to be (formally) divided by
a vanishing constant to be not identically vanishing.

\*

\noindent {\it Axiomatic approach.} The Johnson analysis still
left as an open problem the {\it rigorous} construction of a QFT
corresponding to the massless Thirring model. Wigthmann \cite{W}
proposed to construct the massless Thirring model following an
axiomatic approach; one can start directly from the explicit
expressions of the $n$-point functions at non-coinciding points
derived in \cite{J}, \cite{K} (forgetting how they were derived)
and try to verify the axioms necessary for the reconstruction
theorem. Indeed all axioms can be easily verified except {\it
positive definiteness}, which was proved later on in \cite{DFZ}
and \cite{CRW}; the idea was to define certain field operators,
depending on a certain number of parameters, whose expectations
verify the positivity property {\it by construction} and such that
their $n$-point functions coincide, for a suitable choice of the
parameters, with the expression found in \cite{J}, \cite{K}. As
the axioms are verified by the $n$-point functions of \cite{J},
\cite{K}, a rigorous construction of a QFT corresponding to
massless Thirring model is then obtained. Note however that the
fermionic mass cannot be included in this approach; moreover
quadratic fermionic operators at coinciding points, like
$j^\m_\xx$ or $j^{\m,5}$ cannot be considered, hence the WTi
\pref{1.1a} cannot be rigorously derived.

\*

\noindent {\it Perturbative approach.} The massive case is much
less understood; Coleman \cite{C} considered a perturbation
expansion in the mass showing that it was order by order
coinciding with the expansion of the Sine-Gordon model, if
suitable identification of parameters is done; however an explicit
expression for the n-point functions was not obtained, hence a QFT
corresponding to the massive Thirring model has never been
constructed.

\*

\noindent {\it Bosonic functional integral approach.} If the
coupling $\l$ is positive, the partition function and the
generating functional of the massless (Euclidean) Thirring model
can be written as bosonic functional integrals \cite{FGS} by a
{\it Hubbard-Strato\-novich} transformation; one can then
integrate the fermion variables and it turns out that the
partition function of the Thirring model can be written as
\be
\int P(d A) {{\rm det}(\g_\m[\partial_\m+A_\m])\over {\rm
det}(\g_\m\partial_\m)}\;,\lb{1.1e}
\ee
where $A_{\m,\xx}$ is a two-dimensional  Gaussian field with
covariance $\la A_{\m,\xx} A_{\n,\yy}\ra=\l \d_{\m,\n}\d(\xx-\yy)$
and $\g_\m$ are the Euclidean gamma matrices. A similar expression
holds for the generating functional. It is well known \cite{S}
that, {\it under suitably regularity conditions over $A$}, $\log
{\rm det}(\g_\m\partial_\m +\g_\m A_\m) - \log {\rm
det}(\g_\m\partial_\m )$ is quadratic in $A$; by replacing the
determinant with a quadratic exponential, one then gets an
explicitly solvable integral, from which the n-point functions can
be derived. As stressed in \cite{FGS}, in this way one gets in a
very simple way the results of the exact approach found in
\cite{J} and \cite{K}. In particular the relation \pref{1.1c} for
the two point critical index $\h_z$ is verified and the anomalies
\pref{1.1b} can be easily computed. If a dimensional
regularization is adopted, one finds $a=1$ and $\bar
a^{-1}-1=\l/(2\pi)$, while with a momentum regularization
\pref{1.1b} holds; in both cases the anomaly non-renormalization
holds. Of course in the above derivation an approximation is
implicit; the logarithm of the fermionic determinant in
\pref{1.1e} is given by a quadratic expression {\it only if} $A$
is sufficiently regular, but {\it the integral is over all
possible fields $A$}, hence one is neglecting the contributions of
the irregular fields and there is {\it no guarantee at all} that
such contribution is negligible. This approximation is usually
supported by the fact that one gets in this way the same results
found in \cite{J} and \cite{K}.

\*

\noindent {\it Fermionic functional integral approach.} This is
the approach we will follow in this paper. The generating
functional for the Euclidean Thirring model is the following {\it
Grassmann integral} (see below for a more precise definition)
\be
e^{\WW(\phi,J)}={1\over \NN}\int P_N(d\psi) e^{ \int d\xx
[-{\l\over 4} (\bar\psi_\xx\g^\m\psi_\xx)^2+
J_{\m,\xx}\bar\psi_{\xx}\g_\m
\psi_{\xx}+{\phi_\xx\bar\psi_\xx\over\sqrt{Z_N}}+
{\bar\phi_\xx\psi_\xx \over \sqrt{Z_N}}]} \;,\lb{1.1d}
\ee
where $\NN$ is a normalization constant, $\phi,J$ are external
fields, $Z_N$ is {\it the wave function renormalization},
$\psi_\xx,\bar\psi_\xx$ are Grassmann variables, $P_N(d\psi)$ is
the fermionic integration corresponding to a fermionic propagator
with mass $\m_N$ and a (smooth) momentum ultraviolet cut-off
$\g^N$, with $\g>1$.  Note that the averages of
$\bar\psi_{\xx}\g^\m\g^5 \psi_{\xx}$ can be obtained by the
derivatives with respect to $J_\m$, using the relation
$\bar\psi_{\xx}\g^\m\g^5
\psi_{\xx}=-i\e_{\m,\n}\bar\psi_{\xx}\g^\m\psi_{\xx}$ with
$\e_{\m,\n}=-\e_{\n,\m}$ and $\e_{1,0}=1$. When $J=\phi=0$ and
$\m_N=0$ the r.h.s. of \pref{1.1d} coincides with \pref{1.1e} (if
$\l$ is positive) in the limit $N\to \io$.

We will show that, by properly choosing the bare wave function
renormalization $Z_N$ and the bare mass $\m_N$, the Schwinger
functions at non-coinciding points obtained from \pref{1.1d}
converge, for $N\to\io$, to a set of functions verifying the {\it
Osterwalder-Schrader axioms} \cite{OS2} for an Euclidean QFT.
These functions depend on three parameters, the physical mass, the
physical wave function renormalization and the physical coupling,
but they are independent on the way the ultraviolet
cutoff is explicitly realized. On the contrary, the relation
between the physical and the bare parameters depends on the
details of the ultraviolet cutoff.

In this way we have obtained for the first time a construction of
a QFT for the Thirring model for any value of the (physical) mass.
Moreover, even if in the massless case other constructions were
known, we find in any case interesting to reach a complete
construction of the Thirring model relying only on a functional
integral approach, which could be the only possible one at higher
dimensions or for more realistic models.

The analysis of the functional integral \pref{1.1d} is performed
by a multiscale analysis using a (Wilsonian) Renormalization Group
approach as in \cite{G}. After each iteration step an effective
theory with new couplings, mass, wave function and current
renormalizations is obtained. The effective parameters obey to a
recursive equation called {\it Beta function}, and a major
technical problem is that this iterative procedure can be
controlled only by proving non trivial cancellations in the Beta
function. Such cancellations are established by suitable WTi valid
at each scale and reflecting the symmetries of the formal action;
contrary to the WTi formally valid when all cutoffs are removed,
they have corrections due to the cutoffs introduced for performing
the multiscale integration. The crucial role of WTi in the
construction of the theory is a feature that the functional
integral \pref{1.1d} shares with realistic models like $QED$ or
the Electroweak theory in $d=4$, requiring WTi even to prove the
perturbative renormalizability, which is absent in the models
previously rigorously constructed by functional integral methods,
like the massive {\it Yukawa} model \cite{Le} or the massive {\it
Gross-Neveu} model \cite{GK,FMRS}. From the functional integral
\pref{1.1d} we obtain, for $N\to\io$ and in the massless limit,
WTi of the same form as the one postulated in \cite{J}:
\bea
&&\partial_\m \la \bar\psi_\zz\g^\m\psi_\zz;\psi_\xx\bar\psi_\yy\ra
=a[\d(\zz-\xx)-\d(\zz-\yy)]\la\psi_\xx\bar\psi_\yy\ra\;,\lb{1.1aa}\\
&&\partial_\m \la \bar\psi_\zz\g^\m\g^5\psi_\zz;
\psi_\xx\bar\psi_\yy\ra =\bar
a[\d(\zz-\xx)-\d(\zz-\yy)]\g^5\la\psi_\xx\bar\psi_\yy\ra\;,\nn
\eea
where $\la\psi_\xx\bar\psi_\yy\ra=\lim_{N\to\io}
{\partial^2\over\partial\bar\phi_\xx
\partial\phi_\yy}\WW|_{0}$ (similar definitions
hold for the other averages); however the anomaly coefficients in
\pref{1.1aa} are given by the following expression
\be
a^{-1}=1-{\l\over 4\pi}+c_+\l^2+O(\l^3)\virg \bar
a^{-1}=1+{\l\over 4\pi}+c_+\l^2+O(\l^3)\;,\lb{1.1h}
\ee
where $c_+$ is a {\it non-vanishing} constant, (its explicit value
is calculated in Appendix B). The anomaly coefficients are not
linear in the bare coupling ({\it the anomaly non-renormalization
is violated }), contrary to what happens in the values
\pref{1.1b}, found in the exact approach. Indeed the
regularizations used in the exact solution are different with
respect to the ones used in the functional integral approach, and
it is not too surprising to get different properties (despite
often is guessed that the same results should be obtained by the
two approaches). In particular, the constant $c_+$, not only is
different from $0$, but even depends on the way the ultraviolet
cutoff is realized. The difference of \pref{1.1h} with respect to
\pref{1.1b} also implies that the approximation in \pref{1.1e} of
the determinant with a quadratic exponential does not lead to
correct results, at least if a momentum regularization is used.

In \pref{1.1d} a {\it bare wave function} $Z_N$ for the fermionic
fields has been introduced, to be fixed so that the "physical"
renormalization has a fixed value at the "laboratory scale";
analogously we can introduce a (finite) {\it bare charge} also for
the current, defining it as $\x\bar\psi\g^\m\psi$. A physically
meaningful choice for $\x$ could be $\x=a^{-1}$, implying that the
current has no anomalies; this choice fix the renormalization even
of the pseudocurrent (remember that $\bar\psi\g^\m\g^5\psi=
i\e_{\m,\n}\bar\psi\g^\m\psi$), which has then still anomalies.

Note that \pref{1.1h} is not in contrast to the Adler-Bardeen \cite{AB} analysis,
as they consider a boson-fermion interaction with a
massive boson, which corresponds to require a non local
current-current interaction. If we replace in \pref{1.1d} the
local current-current interaction with a {\it non local short
ranged} one, still a WTi like
\pref{1.1aa} is found for $N\to\io$, but the anomalies are linear
in $\l$ and identical to the ones found in the exact approach,
that is they are given by \pref{1.1b} instead of \pref{1.1h}, see
\cite{M}.

Finally we will show that a closed equation for the 2-point
function is indeed valid starting from the functional integral
\pref{1.1d}; it is however {\it different} with respect to the one
postulated in \cite{J} (which was the natural one obtained
inserting the WTi in the Schwinger-Dyson equation) for the
presence of {\it additional anomalies}. As a consequence, we get a
relation between the critical index of the two point function and
the anomalies different with respect to \pref{1.1c}, namely
\be
\h_z={\l\over 4\pi}(a-\bar a)[1-c_0\l+O(\l^2)]\;, \lb{1.1cc}
\ee
with $c_0>0$ nonvanishing. This additional anomalies says that the
closed equation for the 2-point function is not simply obtained
inserting the WTi in the Schwinger-Dyson equation.

In the rest of this section we will define more precisely our
regularized functional integral and we state our main results. We
will find more convenient, from the point of view of the notation,
to introduce the {\it Weyl spinors} $\psi^\pm_{\xx,\o}$, with
$\o=\pm$, such that $\psi_\xx=(\psi^-_{\xx,+},\psi^-_{\xx,-})$,
$\bar\psi\= \ps^\dagger \g^0$ and
$\psi_\xx^+=(\psi^+_{\xx,+},\psi^+_{\xx,-})$; the $\g$'s matrices
are explicitly given by
$$\g^0=
 \pmatrix{0&1\cr
          1&0\cr}\;,
 \qquad
 \g^1=
 \pmatrix{0&-i\cr
          i&0\cr}\;,
 \qquad
 \g^5=-i\g^0\g^1
 =
 \pmatrix{1&0\cr
          0&-1\cr}\;.$$

\subsection{Thirring model with cutoff}\lb{ss1.3}

We introduce in $\L=[-L/2, L/2]\times [-L/2, L/2]$ a lattice
$\L_a$ whose sites are given by the space-time points
$\xx=(x,x_0)=(na,n_0 a)$, with ${L/2 a}$ integer and $n,n_0=-L/2
a,\ldots,L/2 a-1$. We also consider the set $\DD_a$ of space-time
momenta $\kk=(k,k_0)$, with $k=(m+{1\over 2}){2\p\over L}$ and
$k_0=(m_0+{1\over 2}){2\pi\over L}$ and $m,m_0=-L/2 a,\ldots,L/2
a-1$. In order to introduce an ultraviolet and an infrared cutoff,
we fix a number $\g>1$, a positive integer $N$ and a negative
integer $h$; then we define the function $C_{h,N}^{-1}(\kk)$ in
the following way; let $\chi_0\in C^\io(\RRR_+)$ be a
non-negative, non-increasing smooth function such that
\be \chi_0(t)\defi
\lft\{\matrix{ 1\hfill&\hfill{\rm if\ }0\le t\le 1\cr
0\hfill&\hfill{\rm if\ } t\ge \g_0\;,}\rgt.\lb{1.5bis}\ee
for a fixed choice of  $\g_0:1<\g_0\le\g$; then we define, for any
$h\le j\le N$,
\be f_j(\kk)\defi
\chi_0\lft(\g^{-j}|\kk|\rgt)-\chi_0\lft(\g^{-j+1}|\kk|\rgt)\lb{1.5}\ee
and $C_{h,N}^{-1}(\kk)=\sum_{j=h}^N f_j(\kk)$; hence
$C_{h,N}^{-1}(\kk)$ acts as a smooth cutoff for momenta $|\kk|\ge
\g^{N+1}$ (ultraviolet region) and $|\kk|\le \g^{h-1}$ (infrared
region). It is useful for technical reasons to choose for
$\chi_0(t)$ a Gevrey function, for example one of class $2$, that
is a function such that, for any integer $n$,
\be
|d^n \chi_0(t)/ dt^n| \le C^n (n!)^2\;.,\lb{1.5a}\ee
where $C$ is a symbol we shall use regularly in the following to
denote a generic constant. With each $\kk\in\DD_a$ we associate
four {\it Grassmann variables} $\big\{ \hp_{\kk,\o}^{[h,N]\s},
\s,\o=\pm \big\}$, to be called {\it field variables}; we define
$\DD^{[h,N]}\defi \lft\{\kk\in \DD_a:C_{h,N}^{-1}(\kk)\neq
0\rgt\}$. On the {\it finite Grassmannian algebra} generated from
these variables we define a linear functional $\der\hp^{[h,N]}$
(the {\it Lebesgue measure}), so that, given a monomial $\QQ(\ps)$
in the field variables, $\int \der\hp^{[h,N]} \QQ(\ps)=0$ except
in the case $\QQ(\ps)$ is equal to $\QQ_0(\ps)=
\prod_{\kk\in\DD^{[h,N]}}\prod_{\o=\pm} \hp_{\kk,\o}^{[h,N]-}
\hp_{\kk,\o}^{[h,N]+}$ or to one of the monomials obtained from
$\QQ_0(\ps)$ by a permutation of the field variables; in these
cases the value of $\int \der\hp^{[h,N]} \QQ(\ps)$ is determined
by the condition $\int \der\hp^{[h,N]} \QQ_0(\ps)=1$ and the
anticommuting properties of the field variables.

We also define a Grassmann field on the lattice $\L_a$ by Fourier
transform, according to the following convention:
\be
 \ps^{[h,N]\s}_{\xx,\o}\defi
 {1\over L^2} \sum_{\kk\in\DD_a} e^{i\s\kk\xx}
 \hp^{[h,N]\s}_{\kk,\o}\;,
 \qquad \xx\in \L_a\;.
\ee
By the definition of $\DD_a$, $\ps^{[h,N]\s}_{\xx,\o}$ is
antiperiodic both in time and in space coordinate.

The {\it Generating Functional} of the {\it Thirring model with
cutoff} is
\bea
&&\qquad \WW(\f,J) = \log \int\!\!P_{Z_N}(d\ps)
\exp \Big\{-\l V(\sqrt{Z_N}\ps) +\lb{1.6}\\
&&+ Z_N \sum_\o \int\!d\xx\
J_{\xx,\o}\ps^{[h,N]+}_{\xx,\o}\ps^{[h,N]-}_{\xx,\o}
+\sum_\o\int\!d\xx\
\lft[\f^{+}_{\xx,\o}\ps^{[h,N]-}_{\xx,\o}+\ps^{[h,N]+}_{\xx,\o}\f^{-}_{\xx,\o}\rgt]
\Big\}\;,\nn\eea
where $\int\!d\xx$ is a short hand notation for
$a^2\sum_{\xx\in\L_a }$,
\bea
&& P_{Z_N}(d\ps)\defi \der\hp^{[h,N]} \cdot
\prod_{\kk\in\DD^{[h,N]}} \left[ L^{-4} Z^2_N|(-|\kk|^2-\m_N^2)
C^2_{h,N}(\kk) \right]^{-1} \cdot\nn\\
&& \exp\left\{-Z_N {1\over L^2}\sum_{\o,\o'=\pm} \sum_{\kk\in
\DD^{[h,N]}} {T_{\o,\o'}(\kk)\over C^{-1}_{h,N}(\kk)}
\hp^{[h,N]+}_{\kk,\o}\hp^{[h,N]-}_{\kk,\o'}\right\}\;,
\lb{1.7}\eea
\be  T_{\o,\o'}(k) \defi
\pmatrix{ D_+(\kk) & \m_N \cr \m_N & D_-(\kk)\cr}_{\o,\o'}\;;
\qquad D_\o(\kk)\defi-ik_0+\o k_1\;,\lb{1.8}\ee
\be V(\ps)\defi{1\over 2}
 \sum_{\o=\pm}\int\!d\xx\
 \hp^{[h,N]+}_{\xx,\o}
\hp^{[h,N]-}_{\xx,\o}\hp^{[h,N]+}_{\xx,-\o}\hp^{[h,N]-}_{\xx,-\o}\lb{1.9}\ee
and $\{J_{\xx,\o}\}_{\xx,\o}$ are commuting variables, while
$\{\f^\s_{\xx,\o}\}_{\xx,\o,\s}$ are anticommuting.
$\{J_{\xx,\o}\}_{\xx,\o}$ and $\{\f^\s_{\xx,\o}\}_{\xx,\o,\s}$ are
the {\it external field variables}.

\vskip.5cm

{\it Remark.} It is immediate to check that \pref{1.6} coincides
with \pref{1.1d}, if the notational conventions adopted at the end
of \S 1.1. are used and up to the trivial rescaling $\psi\to
\sqrt{Z}\psi$ of the Grassmann variables. Note also that the {\it
continuum regularization} we have introduced is very suitable to
derive WTi and SDe ; its main disadvantage is that the
positive definiteness property is not automatically ensured; such
a property will be recovered indirectly later by introducing a
different regularization preserving positive definiteness and such
that, by a proper choice of the bare parameters, the Schwinger
functions in the limit of removed cutoffs are coinciding.
\vskip.5cm Setting ${\underline\xx}\defi \xx_1,\ldots,\xx_n$, and
$\uy\defi \yy_1,\ldots,\yy_m$, for any given choice of the labels
$\us\defi(\s_1\ldots,\s_m)$, $\uo\defi(\o_1\ldots,\o_n)$ and
$\ue\defi(\e_1\ldots,\e_n)$, the Schwinger functions are defined
as
\be
 S^{N,h,a;(m;n)}_{\us;\uo,\ue}(\uy;\ux)\defi
\lim_{L\to\io} {\partial^{n+m}\WW\over\partial
 J_{\yy_1,\s_1}\cdots\partial J_{\yy_m,\s_m}
 \partial\f^{\e_1}_{\xx_1,\o_1}\cdots\partial\f^{\e_n}_{\xx_n,\o_n}}
 (0,0)\;.\lb{1.10}\ee
We will follow the convention that a missing label means that the
corresponding limit has been performed, for instance
$S^{N,h;(m;n)}_{\us;\uo,\ue}=\lim_{a\to 0}
S^{N,h,a;(m;n)}_{\us;\uo,\ue}$ In particular, in order to shorten
the notation of the most used Schwinger functions, let:
\bea
G^{2,N,h,a}_{\o}(\xx,\yy) &\defi& S^{N,h,a,(0;2)}_{\o,\o,(+,-)}(\xx,\yy)\;, \\
G^{2,1,N,h,a}_{\o',\o}(\zz;\xx,\yy) &\defi&
S^{N,h,a,(1;2)}_{\o';\o,\o,+,-}(\zz;\xx,\yy)\;.\lb{1.11}
\eea
We define the Fourier transforms so that, for example,
\bea
G^2_\o(\xx,\yy) &\defi& \int {d\kk\over (2\p)^2}
e^{-i\kk(\xx-\yy)} \hG^2_\o(\kk)\;,\\
G^{2,1}_{\o',\o}(\zz;\xx,\yy) &\defi& \int {d\kk d\pp\over
(2\p)^4} e^{i\pp(\zz-\yy)}e^{-i\kk(\xx-\yy)}
\hG^{2,1}_{\o',\o}(\pp,\kk)\;.\lb{1.12}\eea
The presence of the cutoffs makes the Schwinger functions
$S^{N,h,a;(m;n)}_{\us;\uo,(\ue)}(\uy;\ux)$ well defined, since the
generating functional is simply a polynomial in the external field
variables, for any finite $L$, and the limit $L\to\io$ gives no
problem, if $h$ is finite. Note that the lattice is introduced
just to give a meaning to the Grassmann integral and it can be
removed safely if $h,N$ are fixed. In \S\ref{ss2} we will prove
the following result.
\\

\begin{theorem} \lb{th1} Given $\l$ small enough and $\m>0$,
there exist functions $Z_N\equiv Z_N(\l)$ $\m_N\equiv \m g_N(\l)$,
such that
\be
Z_N=\g^{-N\h_z}\big(1+{\rm O}(\l^2)\big) \virg
\m_N=\m\g^{-N\h_\m}\big(1+{\rm O}(\l))\;,\lb{1.13}\ee
with $\h_z=a_z\l^2 +{\rm O}(\l^4)$, $\h_\m=-a_\m\l+{\rm O}(\l^2)$,
$a_z, a_\m>0$, and the following is true.

\bd

\item[1.] The limit
\be \lim_{N,-h,a^{-1}\to\io} S^{N,h,a;(m;n)}_{\us;\uo,\ue}(\uy;\ux)=
S^{(m;n)}_{\us;\uo,\ue}(\uy;\ux)\;,\lb{1.14}\ee
exist at non coinciding points.

\item[2.] The family of functions $S_{2n,\uo}(\ux)$, defined as equal to
$S^{(0;2n)}_{\uo,\ue}(\ux)$, with $\e_i=+1$ for $i=1,\ldots,n$ and
$\e_i=-1$ for $i=n+1,\ldots,2n$, fulfills the OSa.

\item[3.] The two point Schwinger function verifies the following
bound
\be \left|G^2_{\o}(\xx,\yy)\right|\leq
{C\over |\xx-\yy|^{1+\h_z}}
 e^{-c\sqrt{\k\m^{1+ \h'_\m}|\xx-\yy|}}\;,\lb{1.15}\ee
with $\h'_\m=a_\m\l+O(\l^2)$. Moreover $G^2_{\o}(\xx,\yy)$ is
singular for $\xx\to \yy$ and it diverges as
$|\xx-\yy|^{-1-\h_z}$.

\item[4.] In the massless limit $\mu\to 0$ two point Schwinger
function can be written as
\be \hG^{2}_{\o}(\kk)= (1+f(\l))
{|\kk|^{\h_z}\over -i k_0+\o k}\;,\lb{1.16}\ee
with $f(\l)=O(\l)$ and independent from $\kk$.

\ed
\end{theorem}


{\it Remark.} It is an easy consequence of our proof that the
Schwinger functions {\it do not} depend on the parameter $\g$, but
are only functions functions of $\l$ and $\m$.

\subsection{WTi and chiral anomalies}\lb{ss1.3a}

Once that the model is constructed and the OSa are verified, we
can compute the WTi in the massless limit. We will show that
\bea
&&\qquad D_\o(\pp) \hG_{\o,\o'}^{2,1,N,h}(\pp;\kk) =\nn\\
&& =\d_{\o,\o'}[\hG^{2,N,h}_{\o}(\kk-\pp)-\hG^{2,N,h}_{\o}(\kk)]+
\hD^{2,1,N,h}_{\o,\o'}(\pp;\kk)\lb{1.17}\eea
where $\hD^{2,1,N,h}_{\o,\o'}(\pp;\kk)$ is a {\it correction term}
which is formally vanishing if we replace $C_{h,N}^{-1}(\kk)$ by
$1$.

\insertplot{330}{80}%
{\footnotesize
\ins{0pt}{55pt}{$D_{\o}(\pp)
\hskip 70pt
=\d_{\o,\o'}\Bigg[
\hskip 38pt
-
\hskip 42pt
\Bigg]
\hskip 5pt  +
$}
\ins{40pt}{60pt}{$\o$}
\ins{40pt}{25pt}{$\o$}
\ins{70pt}{70pt}{$\kk,\o'$}
\ins{60pt}{10pt}{$\kk-\pp,\o'$}
\ins{130pt}{70pt}{$\kk,\o'$}
\ins{138pt}{10pt}{$\o'$}
\ins{170pt}{70pt}{$\kk-\pp,\o'$}
\ins{193pt}{10pt}{$\o'$}

\ins{260pt}{60pt}{$\o$}
\ins{260pt}{25pt}{$\o$}
\ins{290pt}{70pt}{$\kk,\o'$}
\ins{280pt}{10pt}{$\kk-\pp,\o'$}
}%
{f1}{\lb{f1}: Graphical representation of \pref{1.17}; the small
circle in the last term represents the function in the r.h.s. of
\pref{3.3}.}{0}

The anomaly manifests itself in the fact that
$\hD^{2,1,N,h}_{\o,\o'}$ is nonvanishing in the limit $N,-h\to
\io$; we will prove in fact the following Theorem.

\begin{theorem} \lb{th2} Under the same conditions of Theorem \ref{th1},
in the massless limit, i.e. $\m=0$, it holds that for finite
nonvanishing $\kk,\kk-\pp,\pp$
\bea
&&\qquad \hD^{2,1,N,h}_{\o,\o'}(\pp;\kk) = D_\o(\pp)
\hR^{2,1,N,h}_{\o,\o'}(\pp;\kk) + \lb{1.18}\\
&& +\n^+_{h,N} D_\o(\pp) \hG^{2,1,N,h}_{\o,\o'}(\pp;\kk)
+\n_{-,h,N} D_{-\o}(\pp)
\hG^{2,1,N,h}_{-\o,\o'}(\pp;\kk)\;,\nn\eea
where all the quantities appearing in this identity admit a
$N,-h\to\io$-limit, such that
\be \n_{-}={\l\over 4\pi}+O(\l^2)\;,
\qquad \n_{+}=c_+\l^2+O(\l^2)\;,\lb{1.19}\ee
with $c_+<0$, $|\hG^{2,1}_{\o,\o'}(\pp;\kk)|$ satisfies the bound
\pref{2.64} below, and
\be \hR^{2,1}_{\o,\o'}(\pp;\kk)=0\;.\lb{1.20}\ee
\end{theorem}

It is immediate to check that the above result implies the WTi
\pref{1.1aa}, with $a^{-1}=1-\n^--\n^+$ and $\bar
a^{-1}=1+\n^--\n^+$.

\insertplot{300}{90}%
{\footnotesize

\ins{100pt}{75pt}
{$=\hskip 10pt D_{\o}(\pp)$}

\ins{28pt}{25pt}
{$+\n^+_{h,N}D_{\o}(\pp)
\hskip 85pt
+\n^-_{h,N}D_{-\o}(\pp)$}

\ins{30pt}{90pt}{$\o$}
\ins{30pt}{55pt}{$\o$}
\ins{55pt}{95pt}{$\kk,\o'$}
\ins{45pt}{50pt}{$\kk-\pp,\o'$}

\ins{165pt}{90pt}{$\o$}
\ins{165pt}{55pt}{$\o$}
\ins{190pt}{95pt}{$\kk,\o'$}
\ins{180pt}{50pt}{$\kk-\pp,\o'$}

\ins{95pt}{40pt}{$\o$}
\ins{95pt}{5pt}{$\o$}
\ins{120pt}{45pt}{$\kk,\o'$}
\ins{110pt}{0pt}{$\kk-\pp,\o'$}

\ins{240pt}{40pt}{$\o$}
\ins{240pt}{5pt}{$\o$}
\ins{265pt}{45pt}{$\kk,\o'$}
\ins{255pt}{0pt}{$\kk-\pp,\o'$}
}%
{f2}{\lb{f2}: Graphical representation of \pref{1.18}; the filled
circle in the second term is the operator implicitly defined in
\S\ref{ss4.1}}{0}

\subsection{Closed equation and additional anomaly}\lb{ss1.5}


It is easy to see (see for instance \cite{BM4}) that the Schwinger
functions of \pref{1.6} in the massless limit verify the following
SDe
\be \hG^{2,N,h}_\o(\kk)={\hg^{N,h}_\o(\kk)\over Z_N}-
\l \hg^{N,h}_\o(\kk)\int {d\pp\over (2\p)^2}
 \bar\chi_N(\pp)\hG^{2,1,N,h}_{-\o,\o}
(\pp;\kk)\;,\lb{1.21}\ee
where $\hg^{N,h}_\o(\kk) = C_{h,N}^{-1}(\kk) D_\o(\kk)^{-1}$ and
$\bar\chi_N(\pp)$ is a smooth function with support in $|\pp|\le
3\g^{N+1}$, equal to $1$ if $|\pp|\le 2\g^{N+1}$ (we can insert it
freely in the SDe, thanks to the support properties of the
propagator).

\insertplot{330}{70}%
{\footnotesize
\ins{90pt}{43pt}
{$=\hskip 60pt + $}

\ins{10pt}{50pt} {$\kk,\o$}
\ins{60pt}{50pt} {$\kk,\o$}
\ins{120pt}{50pt} {$\kk,\o$}
\ins{190pt}{50pt} {$\kk,\o$}
\ins{190pt}{50pt} {$\kk,\o$}
\ins{290pt}{50pt} {$\kk,\o$}
\ins{220pt}{70pt} {$\kk-\pp,\o$}
\ins{230pt}{47pt} {$-\o$}
\ins{240pt}{23pt} {$-\o$}
}%
{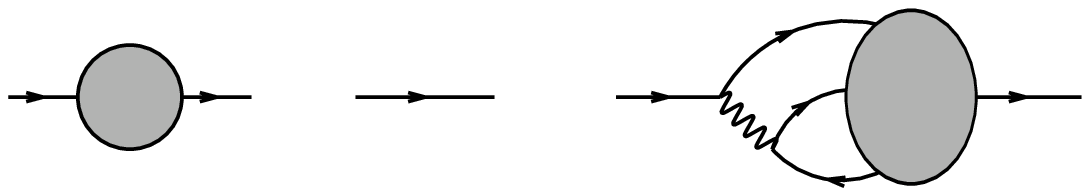}{\lb{f3}: Graphical representation of \pref{1.21}.}{0}

Inserting the WTi \pref{1.17} in SDe \pref{1.21} and using
\pref{1.18}, we get
\bea
\hG^{2,N,h}_{\o}(\kk) &=&{\hg^{N,h}_\o(\kk)\over Z_N}- \l
\hg^{N,h}_\o(\kk) \int {d\pp\over (2\p)^2} \bar\c_h(\pp)
\hG^{2,1,h,N}_{-\o,\o}(\pp;\kk) - \nn\\
&-& \l A_{+,h,N} \hg^{N,h}_\o(\kk) \int {d\pp\over (2\p)^2}
\wt\c(\pp) {\hG_{\o}^{2,N,h}(\kk-\pp)\over D_{-\o}(\pp)}+\lb{1.22}\\
&+& \hg^{N,h}_\o(\kk) \sum_{\e}\l A_{\e,h,N} \int {d\pp\over
(2\p)^2} \wt\c(\pp) {{D_{\e\o}(\pp)} \over D_{-\o}(\pp)}
\hR^{2,1,N,h}_{\e\o,\o} (\pp;\kk)\;,\nn\eea
where
\bea
A_{\e,h,N} &\defi& {a_{h,N}-\e \bar a_{h,N}\over 2}\;, \lb{1.23}\\
a_{h,N}={1\over 1-\n_{h,N}^- -\n_{h,N}^+} \quad&,& \bar
a_{h,N}={1\over 1+\n_{h,N}^- -\n_{h,N}^+}\;, \nn\eea
$\bar\c_h(\pp)$ is defined as $\bar\c_N(\pp)$, with $h$ in place
of $N$, and $\wt\c(\pp)= \bar\chi_N(\pp) - \bar\c_h(\pp)$ (so that
the support of $\wt\c(\pp)$ is only for $2\g^{h+1}\le|\pp|\le
3\g^{N+1}$). The bound \pref{2.64} below implies that, if $\kk$ is
fixed to a non vanishing value, $\hG^{2,1,h,N}_{-\o,\o}(\pp;\kk)$
diverges more slowly that $|\pp|^{-1/8}$, as $\pp\to 0$; hence the
second addend in the r.h.s. of \pref{1.22} is vanishing in the
limit $h=-\io$.


{\it If} the last term in \pref{1.22} were vanishing for
$-h,N\to\io$ (as the second addend), one would get a closed
equation for $\hG_{\o}^{2}$ which is identical to the one
postulated in \cite{J}; it is just the formal Schwinger-Dyson
equation combined with the WTi in the limit of removed cutoffs.

{\it However this is not what happens}; despite both WTi and
Schwinger-Dyson equation are true in the limit, one cannot simply
insert one in the other to obtain a closed equation. The last term
is {\it non vanishing} and this is a {\it additional anomaly
effect which seems to be unnoticed in the literature}.

Despite the presence of the additional anomaly, a closed equation
({\it different} with respect the one in \cite{J}) holds, as shown
from the following theorem.

\begin{theorem} \lb{th3} Under the same conditions of Theorem \ref{th1},
in the massless limit there exist functions $\a_{\e,h,N}$,
$\r_{\e,h,N}$ such that, for non vanishing $\kk$,
\bea
&&\qquad \hg^{N,h}_\o(\kk) \int {d\pp\over (2\p)^2}
\wt\c(\pp){{D_{\e\o}(\pp)} \over D_{-\o}(\pp)}
\hR^{2,1,N,h}_{\e\o,\o}(\pp;\kk)= \nn\\
&& =-\a_{\e,h,N} {\hg^{N,h}_\o(\kk)\over Z_N}+
\big(\a_{\e,h,N}+\r_{\e,h,N}) \hG^{2,N,h}_\o(\kk)
+\hR^{4,N,h}_{\e}(\kk)\;,\lb{1.24}\eea
with
\be \lim_{N,-h\to\io} \hR^{4,N,h}_{\e}(\kk)=0\;,\lb{1.25}\ee
and, in the limit of removed cutoff,
\bea
&&\a_+= c_1\l+O(\l^2)\;,\qquad \r_{+}=c_2\l+O(\l^2)\;,\nn\\
&& \phantom{**} \a_-= c_3+O(\l)\;, \qquad
\r_{-}=c_4+O(\l)\;.\lb{1.26}\eea
\end{theorem}


The above result says that, up to a vanishing term, the last
addend in the r.h.s. of \pref{1.22} can be written in terms of $g$
and $G^2$, so that a closed equation still holds in the limit, but
different with respect to the one postulated in \cite{J}; in
particular one gets a relation between the critical index $\h_z$
and the anomalies $a,\bar a$, which is different with respect to
the \pref{1.1c}, that found in \cite{J}.

\begin{corollary} \lb {cor1} The critical index of the massless two
point Schwinger function \pref{1.16} verifies
\be
\h_z = {\l \over 2\p}{a-\bar a \over 1 -\l \sum_{\e}
A_\e(a_\e+\r_{\e})}\;.\lb{1.27}\ee
with $\sum_{\e} A_\e(a_\e+\r_{\e})=c_0+O(\l^2)$ with $c_0>0$.
\end{corollary}

\subsection{Lattice fermions and positive definiteness}\lb{ss1.3b}

There are of course several ways to introduce a functional
integral formulation of the Thirring model corresponding to
different ways of regularizing the theory. The choice
corresponding to \pref{1.6} is the closest to the formal continuum
limit (the regularized propagator is  linear in $\kk$) and this is
convenient under many respects, for instance in the derivation of
WTi and closed equation for the two point Schwinger function which
we will discuss below. However such a choice has the big
disadvantage that the crucial property of {\it positive
definiteness} is quite difficult to prove; such a property is
however automatically fulfilled with a {\it lattice
regularization}. There is an extensive literature on the lattice
fermions \cite{MM}; if one simply replaces $k,k_0$ in the
propagator with $a^{-1}\sin k a$ and $a^{-1}\sin k_0 a$ the well
known {\it fermion doubling} problem is encountered, namely that
the massless fermion propagator has {\it four} poles instead of a
single one. In the continuum limit $a\to 0$ this means that there
are four fermion state per field component and such extra unwanted
fermions influence possibly the physical behavior in a non trivial
way. Several solutions have been proposed; we will follow here the
Wilson formulation of adding a term to the free action, called
{\it Wilson term}, to cancel the unwanted poles \cite{MM}. Then in
the Wilson lattice regularization the fermionic integration is
given by
\bea
&& P_{Z_a} (d\ps) \defi \exp\left\{-{Z_a\over
L^2}\sum_{\o,\o'=\pm} \sum_{\kk\in\DD_a} \Big({\hat
r}^{-1}(\kk)\Big)_{\o',\o}
\hp_{\kk,\o'}^{+}\hp_{\kk,\o}^{-}\right\}
\cdot\nn\\
&&\quad \cdot \prod_{\kk\in\DD_a} \prod_{\o=\pm}
{\der\hp_{\kk,\o}^{+}\der\hp_{\kk,\o}^{-}\over \bar \NN_a(\kk)}\;,
\lb{1.28}\eea
where the covariance ${\hat r}_{\o,\o'}(k)$ is defined as
\be {\hat r}_{\o,\o'}(\kk)\defi
{1\over e_+(\kk)e_-(\kk) -\m^2_a(\kk)} \pmatrix {e_-(\kk) &
-\m_a(\kk) \cr -\m_a(\kk)  & e_+(\kk)\cr}_{\o,\o'}\;,\lb{1.28a}\ee
with $k_0=(m_0+1/2)2\pi/L$, $k=(m+1/2)2\pi/L$, $n,n_0=-L/2
a,1,\ldots,L/2 a-1$,
\bea
e_\o(\kk) &\defi& -i{\sin(k_0 a)\over a}+\o
{\sin(k a)\over a}\;, \nn\\
\m_a(\kk) & \defi& \m + {1-\cos(k_0a) \over a} + {1-\cos(k a)\over
a} \;, \lb{1.29}
\eea
and $\bar \NN_a$ is the normalization. The generating functional
is given by

\bea
&& \int\! P_{Z_a}(d\ps) \exp\Big\{-\l_aZ^2_aV(\ps)+\n_a
Z_aN(\ps)\Big\}\cdot\lb{1.30}\\
&&\cdot \exp\Big\{Z_a^{(2)}\sum_\o\int d\xx\
J_{\xx,\o}\ps^+_{\xx,\o}\ps^-_{\xx,\o} +\sum_\o
 \int\!d\xx\
\lft[\f^+_{\xx,\o}\ps^-_{\xx,\o}+\ps^+_{\xx,\o}\f^-_{\xx,\o}\rgt]
\Big\}\;,\nn\eea
where $N(\ps)=\sum_{\o=\pm} \int\!d\xx\
\ps^{+}_{\xx,\o}\ps^-_{\xx,-\o}$. Note the presence of the term
$(1-\cos(k_0a))/ a+(1-\cos(ka))/ a$ which has the effect that, in
the massless case, only one pole is present. On the other hand it
is not true, contrary to what happened in the previous case, that
the massless case corresponds simply to $\m=0$; the Wilson term
breaks a parity symmetry leading to the generation though the
interaction of a mass; we introduce then a counterterm $\n_a$ to
fix the mass proportional to $\m$.

We call $S^{N,(m;n)}_{{\us;\uo,\ue}}$ the Schwinger functions
\pref{1.10} (in the limit $a=0$ and $h=-\io$) and  $\bar
S^{a,(m;n)}_{{\us;\uo,\ue}}$ the Schwinger functions corresponding
to \pref{1.30}; in \S\ref{ss4b} we shall prove the following
theorem.

\begin{theorem} \lb{th4} Given $N>0$, let  $a_N=
\pi(4 \g^{N+1})^{-1}$; if $\l$ is small enough, there exist
functions $Z_N(\l)$, $\m g_N(\l)$ and $\l_{a_N}(\l)$,
$Z_{a_N}(\l)$, $\n_{a_N}(\l)$, $\m g_{a_N}(\l)$, such that, if all
the points $\underline\zz,\underline\xx$ are different from each
other, then $\bar S^{a_N, (m;n)}_{{\us;\uo,\ue}}
({\underline\zz};\underline{\xx})$ is well defined in the limit
$N\to\io$ and
\be \lim_{N\to\io}
[S^{N,(m;n)}_{{\us;\uo,\ue}}({\underline\zz};\underline{\xx})
-\bar
S^{a_N,(m;n)}_{{\us;\uo,\ue}}({\underline\zz};\underline{\xx})]=0\;.\lb{1.31}\ee
\end{theorem}

The above result says that in the limit of removed cutoffs the two
{\it different regularizations} of the Thirring model give the
{\it same} Schwinger functions, if the ``bare'' parameters are
suitably chosen.

The proof of the above results is based on many technical
arguments, some of which were already proved in
\cite{BM1}-\cite{BM4}; hence, in this paper we shall discuss in
detail only the arguments not discussed in those papers.

\section{ Continuum fermions with cutoff}\lb{ss2}

\subsection{Renormalization Group analysis}\lb{ss2.1}
The integration of the generating functional \pref{1.6} is done
almost exactly (essentially up to a trivial rescaling) as
described in \cite{BM1}-\cite{BM4}; hence we briefly resume here
such procedure to fix notations. It is possible to prove by
induction that, for any $j:h\le j\le N$, there are a constant
$E_j$, two positive functions $\tilde Z_j(\kk)$, $\tilde\m_j(\kk)$
and two functionals $\VV^{(j)}$ and $\BB^{(j)}$, such that, if
$Z_j =\max_{\kk} \tilde Z_j(\kk)$,
\be e^{\WW(\f,J)}=e^{-L^2 E_j}
\int P_{\tilde Z_j,\tilde\m_j,C_{h,j}} (d\psi^{[h,j]})
e^{-\VV^{(j)}(\sqrt{Z_j}\psi^{[h,j]})+\BB^{(j)}
(\sqrt{Z_j}\psi^{[h,j]},\f,J)}\,,\lb{2.1}\ee
where:

\bd
\item[1.] $P_{\wt Z_j,\tilde\m_j,C_{h,j}}(d\psi^{[h,j]})$ is the {\it
effective Grassmannian measure at scale $j$}, equal to
\bea
&& P_{\wt Z_j,\tilde\m_j,C_{h,j}}(d\psi^{[h,j]})
\prod_{\kk:C^{-1}_{h,j}(\kk)>0} \prod_{\o,\o'=\pm1}
{d\hat\psi^{[h,j]+}_{\kk,\o}d\hat\psi^{[h,j]-}_{\kk,\o}\over
\NN_j(\kk)} \cdot \lb{2.2}\\
&&\cdot\; \exp \left\{-{1\over L^2} \sum_{\kk:
C^{-1}_{h,j}(\kk)>0} \, C_{h,j}(\kk) \tilde Z_j(\kk)\sum_{\o\pm1}
\hat\psi^{[h,j]+}_{\kk,\o} T^{(j)}_{\o,\o'} (\kk)
\hat\psi^{[h,j]-}_{\kk,\o'}\right\} \;,\nn\eea
with $T^{(j)}_{\o,\o'}$ given by \pref{1.8} with $\tilde\m_j(\kk)$
replacing $\m_N$, $C_{h,j}(\kk)^{-1}=\sum_{r=h}^j f_r(\kk)$ and
$\NN_j(\kk)$ a suitable normalization constant.
\item[2.]
The {\it effective potential on scale $j$}, $\VV^{(j)}(\psi)$, is
a sum of monomial of Grassmannian variables multiplied by suitable
kernels. \ie it is of the form
\be \VV^{(j)}(\psi) = \sum_{n=1}^\io {1\over L^{4n}}
\sum_{\kk_1,\ldots,\kk_{2n} \atop \o_1,\ldots,\o_{2n}} \left[
\prod_{i=1}^{2n} \hat\psi^{\s_i}_{\kk_i,\o_i} \right] \hat
W_{2n,\uo}^{(j)}(\kk_1,...,\kk_{2n-1})
\d\left(\sum_{i=1}^{2n}\s_i\kk_i\right)\;,\lb{2.3}\ee
where $\s_i=+$ for $i=1,\ldots,n$, $\s_i=-$ for $i=n+1,\ldots,2n$
and $\uo=(\o_1,\ldots,\o_{2n})$;
\item[3.] The {\it effective source term at scale $j$},
$\BB^{(j)}(\sqrt{Z_j}\psi, \f,J)$, is a sum of monomials of
Grassmannian variables and $\f^\pm,J$ field, with at least one
$\f^\pm$ or one $J$ field; we shall write it in the form
\be \BB^{(j)}(\sqrt{Z_j}\psi, \f,J) = \BB_\f^{(j)}(\sqrt{Z_j}\psi)+
\BB_J^{(j)}(\sqrt{Z_j}\psi) +
W_R^{(j)}(\sqrt{Z_j}\psi,\f,J)\;,\lb{2.4}\ee
where $\BB_\f^{(j)}(\psi)$ and $\BB_J^{(j)}(\psi)$ denote the sums
over the terms containing only one $\f$ or $J$ field,
respectively. $\BB^{(j)}(\sqrt{Z_j}\psi, \f,J)$ can be written as
sum over monomials of $\psi,\f,J$ multiplied by kernels $\hat
W_{2n,n_\f,n_J,\uo}^{(j)}$.

Of course \pref{2.1} is true for $j=N$, with $\tilde
Z_N(\kk)=Z_N$, $W_R^{(0)}=0$, and $\VV^{(N)}(\psi), \BB_\f^{(N)},
\BB_J^{(N)}$ given implicitly by \pref{1.6}.
The kernels in $\hat W^{(j)}$, $\VV^{(j)}$ and $\BB^{(j)}$, $j<N$,
are functions of $\m_k$, $Z_k$ and the {\it effective couplings}
$\l_k$ (to be defined later), with $k\ge j$; the iterative
construction below will inductively implies that the dependence on
these variables is well defined.

\ed

We now begin to describe the iterative construction leading to
\pref{2.1}. We introduce two operators $\PP_r$, $r=0,1$, acting on
the kernels $\hat W^{(j)}$ in the following way
\be \lft.\PP_0 \hat W^{(j)}= \hat W^{(j)}\rgt|_{\tilde\m_j,..\m_N=0}\;,
\qquad \lft.\PP_1 \hat W^{(j)}=\sum_{k\ge j,\kk} \tilde\m_k(\kk)
{\partial \hat W^{(j)} \over
\partial \tilde\m_k(\kk)}\rgt|_{\tilde\m_j,..\m_N=0}\;.\lb{2.5}\ee
We introduce also two operators $\LL_r$, $r=0,1$, acting on the
kernels $\hat W^{(j)}$ in the following way:

\bd
\item[1.] If $n=1$,
\bea
\LL_0 \wh W_{2,\uo}^{(j)}(\kk) &\defi& \fra14 \sum_{\h,\h'=\pm
1}\wh W_{2,\uo}^{(j)}\left(\bk\h{\h'}\right)\;, \nn\\
\LL_1\widehat W_{2,\uo}^{(j)}(\kk) &\defi& \fra14 \sum_{\h,\h'=\pm
1}\widehat W_{2,\uo}^{(j)}(\bk\h{\h'}) \big[\h {k_0 L\over \pi} +
\h'{k L\over\pi}\big]\;, \lb{2.6}\eea
where $\bk\h{\h'} = \left(\h{\p\over L},\h'{\p\over L}\right)$.
\item[2.]
If $n=2$, $\LL_1\hW_{4,\uo}\defi 0$ and
\be \LL_0 \hW_{4,\uo}^{(j)}(\kk_1,\kk_2,\kk_3)\defi
\hW_{4,\uo}^{(j)}(\bk++,\bk++,\bk++)\;.\lb{2.7}\ee
\item[3.]
If $n>2$, $\LL_0\hW_{2n,\uo}^{(j)}
\defi\LL_1\hW_{2n,\uo}^{(j)} \defi 0$.

\ed

\0 Given $\LL_j,\PP_j$, $j=0,1$ as above, we define the action of
$\LL$ on the kernels $\hW_{2n,\uo}^{(j)}$ as follows.

\bd
\item[4.] If $n=1$, then
\be \LL \hW_{2,\o,\o'}^{(j)} \defi(\LL_0+\LL_1) \PP_0
\widehat W_{2,\o,\o'}^{(j)}+ \LL_0\PP_1\widehat
W_{2,\o,\o'}^{(j)}\;.\lb{2.8}\ee
\item[5.]
If $n=2$, then $\LL \hW_{4,\uo}^{(j)} \defi
\LL_0\PP_0\hW_{4,\uo}^{(j)}$.
\item[6.]
If $n>2$, then $\LL \hW_{2n,\uo}^{(j)}=0$.

\ed

Note that $\LL_0\PP_0\hW_{2,\o,\o}^{(j)} =0$, because of the
parity properties (in the exchange $\kk\to-\kk$) of the diagonal
propagators, whose number is surely odd in each Feynmann graph
contributing to $W_{2,\o,\o}^{(j)}$;
$\LL_0\PP_1\hW_{2,\o,\o}^{(j)}=0$, because there are no
contributions of first order in $\m_k$; $\PP_0\hW_{2,\o,-\o}^{(j)}
=0$, since the only way to get a contribution to
$\hW_{2,\o,-\o}^{(j)}$ is to use at least one antidiagonal
propagator. Therefore \pref{2.8} reads
\be \LL \widehat W_{2,\o,\o}^{(j)} =\LL_1\PP_0\widehat W_{2,\o,\o}^{(j)}
\;, \qquad \LL \widehat W_{2,\o,-\o}^{(j)} =\LL_0\PP_1\widehat
W_{2,\o,-\o}^{(j)} \;.\lb{2.9}\ee
Note also that $\LL^2\VV^{(j)}=\LL\VV^{(j)}$. The effect of $\LL$
on $\VV^{(j)}$ is, by definition, to replace on the r.h.s. of
\pref{2.3} $\wh W_{2n,\uo}^{(j)}$ with $\LL\wh W_{2n,\uo}^{(j)}$;
we get
\be \LL\VV^{(j)}(\psi^{[h,j]})=z_j F_\z^{[h,j]}+s_j
F_\s^{[h,j]}+l_j F_\l^{[h,j]}\;,\lb{2.10} \ee
where $z_j$, $a_j$ and $l_j$ are real numbers and
\bea
F_\z^{[h,j]} &=& {1\over L^2}\sum_\o
\sum_{\kk:C^{-1}_{h,j}(\kk)>0} D_\o(\kk)
\hat\psi^{[h,j]+}_{\kk,\o} \hat\psi^{[h,j]-}_{\kk,\o}\;,\nn\\
F_\s^{[h,j]} &=& {1\over L^2}\sum_\o
\sum_{\kk:C^{-1}_{h,j}(\kk)>0}
\hat\psi^{[h,j]+}_{\kk,\o} \hat\psi^{[h,j]-}_{\kk,-\o}\;,\lb{2.11}\\
F_\l^{[h,j]} &=& {1\over L^8}
\sum_{\kk_1,...,\kk_4:C^{-1}_{h,j}(\kk_i)>0}
\hat\psi^{[h,j]+}_{\kk_1,+} \hat\psi^{[h,j]-}_{\kk_2,+}
\hat\psi^{[h,j]+}_{\kk_3,-}
\hat\psi^{[h,j]-}_{\kk_4,-}\d(\kk_1-\kk_2+\kk_3-\kk_4)\;. \nn\eea

Analogously, we write $\BB^{(j)}=\LL \BB^{(j)}+\RR \BB^{(j)}$,
$\RR=1-\LL$, according to the following definition. First of all,
we put $\LL W_R^{(j)}=0$. Let us consider now
$\BB_J^{(j)}(\sqrt{Z_j}\psi)$. It is easy to see that the field
$J$ is equivalent, from the point of view of dimensional
considerations, to two $\psi$ fields. Hence, the only terms which
need to be renormalized are those of second order in $\psi$, which
are indeed marginal. We shall use for them the definition
\be
\BB_J^{(j,2)}(\sqrt{Z_j}\psi) = \sum_{\o,\tilde\o} {1\over L^4}
\sum_{\pp, \kk} \hat B_{\o,\tilde\o}^{(j)}(\pp,\kk) \hat
J_{\pp,\o} (\sqrt{Z_j}\hat\psi^{[h,j]+}_{\pp+\kk,\tilde\o})
(\sqrt{Z_j} \hat\psi^{[h,j]-}_{\kk,\tilde\o})\;.\lb{2.12}\ee

We regularize $\BB_J^{(j,2)}(\sqrt{Z_j}\psi)$, in analogy to what
we did for the effective potential, by decomposing it as the sum
of $\LL\BB_J^{(j,2)}(\sqrt{Z_j}\psi)$ and
$\RR\BB_J^{(j,2)}(\sqrt{Z_j}\psi)$, where $\LL$ is defined through
its action on $\hat B_{\o,\tilde\o}^{(j)}(\pp,\kk)$ in the
following way:
\be \LL \hat B_{\o,\tilde\o}^{(j)}(\pp,\kk) = \fra14 \d_{\o,\tilde\o}
\sum_{\h,\h'=\pm 1} \PP_0\hat B_{\o,\tilde\o}^{(j)}(0,
\bar\kk_{\h,\h'}) \;;\lb{2.13}\ee
note that $\LL \hat B_{\o,-\o}^{(j)}=0$ because of the symmetry
property
\be
\hat g^{(j)}_\o(\kk)=-i\o \hat g^{(j)}_\o(\kk^*) \virg
\kk=(k,k_0),\quad \kk^*=(-k_0,k)\;.\lb{2.13a}
\ee
We get
\be \LL\BB_J^{(j,2)}(\sqrt{Z_j}\psi)=
\sum_\o {Z_j^{(2)}\over Z_j} \int\!\der\xx\ J_{\xx,\o}
\left(\sqrt{Z_j}\psi^{+}_{\xx,\o}\right)
\left(\sqrt{Z_j}\psi^{-}_{\xx,\o}\right) \;,\lb{2.14}\ee
which defines the renormalization constant $Z_j^{(2)}$; we shall
extend this definition to $j=N$ by putting, in agreement with
\pref{1.6}, $Z_N^{(2)} = Z_N$.

Finally we have to define $\LL$ for
$\BB_\f^{(j)}(\sqrt{Z_j}\psi)$; we want to show that, by a
suitable choice of the localization procedure, if $j\le N-1$, it
can be written in the form
\bea
&&\BB_\f^{(j)}(\sqrt{Z_j}\psi) = \sum_{\o,\o'}
\sum_{i=j+1}^N \int d\xx d\yy\;\cdot\nn\\
&&\hskip-1cm \cdot\;\left[ \f^+_{\xx,\o}
g^{Q,(i)}_{\o,\o'}(\xx-\yy){\dpr\over \dpr\psi^+_{\yy,\o'}}
\VV^{(j)}(\sqrt{Z_j}\psi) + {\dpr\over \dpr\psi^-_{\yy,\o}}
\VV^{(j)}(\sqrt{Z_{j}}\psi)
g^{Q,(i)}_{\o,\o'}(\yy-\xx)\f^-_{\xx,\o'} \right]+\nn\\
&&+ \sum_{\o,\o'} {1\over L^2} \sum_{\kk:C^{-1}_{h,j}(\kk)>0}
\left[ \hat\psi^{[h,j]+}_{\kk,\o} \hat Q^{(j+1)}_{\o,\o'}(\kk)
\hat\f_{\kk,\o'}^- +\hat\f^+_{\kk,\o} \hat Q^{(j+1)}_{\o,\o'}(\kk)
\hat\psi^{[h,j]-}_{\kk,\o'} \right]\;\lb{2.15}\eea
where $ \wh g^{Q,(i)}_{\o,\o'}(\kk)= \sum_{\o''}
\hg^{(i)}_{\o,\o''}(\kk) \wh Q^{(i)}_{\o'',\o'}(\kk) $,
$g^{(i)}_{\o,\o''}$ is the renormalized propagator of the field on
scale $j$ (see \pref{2.20} below for a precise definition) and
$\wh Q^{(j)}_{\o,\o'}(\kk)$ is defined inductively by the
relations
\bea
\wh Q^{(j)}_{\o,\o'}(\kk) &=& \wh Q^{(j+1)}_{\o,\o'}(\kk) - z_j
Z_j D_\o(\kk) \sum_{i=j+1}^N \hat g^{Q,(i)}_{\o,\o'}(\kk)-s_j Z_j
\sum_{i=j+1}^N \hat g^{Q,(i)}_{\o,-\o'}(\kk) \;,\nn\\
\quad \hat Q^{(0)}_{\o,\o'}(\kk) &=& 1\;.\lb{2.16}\eea
The $\LL$ operation for $\BB^{(j)}_\f$ is defined by decomposing
$\VV^{(j)}$ in the r.h.s. of \pref{2.15} as $\LL \VV^{(j)}+\RR
\VV^{(j)}$, $\LL \VV^{(j)}$ being defined by \pref{2.10}.

After writing $\VV^{(j)}=\LL \VV^{(j)}+\RR\VV^{(j)}$ and
$\BB^{(j)}=\LL \BB^{(j)}+\RR\BB^{(j)}$, the next step is to {\sl
renormalize} the free measure $P_{\tilde Z_j,\tilde\m_j,C_{h,j}}
(d\psi^{[h,j]})$, by adding to it part of the r.h.s. of
\pref{2.10}. We get that \pref{2.1} can be written as
\be
e^{-L^2 t_j}\int P_{\tilde Z_{j-1}, \tilde \m_{j-1},
C_{h,j}}(d\psi^{[h,j]}) \, e^{-\tilde
\VV^{(j)}(\sqrt{Z_j}\psi^{[h,j]}) + \tilde\BB^{(j)}
(\sqrt{Z_j}\psi^{[h,j]})}\;,\lb{2.17}\ee
where, since $\tilde Z_j(\kk)=Z_j \=\max_\kk \tilde Z_j(\kk)$ and
$\tilde\m_j(\kk)=\m_j \= (Z_{j+1}/ Z_j)(\m_{j+1}+ s_{j+1})$, if
$C_{h,j}^{-1}(\kk)\not=0$, then
\be \tilde Z_{j-1}(\kk)=Z_j [1+C^{-1}_{h,j}(\kk) z_j]\;,\quad
\tilde\m_{j-1}(\kk)={Z_j\over \tilde
Z_{j-1}(\kk)}[\m_j+C^{-1}_{h,j}(\kk) s_j]\;, \lb{2.18}\ee
\be \tilde\VV^{(j)}(\psi^{[h,j]})= \VV^{(j)}(\psi^{[h,j]})-
z_j F_\z^{[h,j]} -s_j F_\s^{[h,j]}\;, \lb{2.18a}\ee
and the factor $\exp(-L^2 t_j)$ in \pref{2.17} takes into account
the different normalization of the two measures. Moreover
\be \tilde\BB^{(j)}(\sqrt{Z_j}\psi^{[h,j]})=
\tilde\BB^{(j)}_\f(\sqrt{Z_j}\psi^{[h,j]}) +
\BB^{(j)}_J(\sqrt{Z_j} \psi^{[h,j]})+W^{(j)}_R\;,\lb{2.19}\ee
where $\tilde\BB^{(j)}_\f$ is obtained from $\BB^{(j)}_\f$ by
inserting \pref{2.19} in the second line of \pref{2.15} and by
absorbing the terms proportional to $z_j,s_j$ in the terms in the
third line of \pref{2.15}.

If $j>h$, the r.h.s of \pref{2.17} can be written as
\bea
&& e^{-L^2t_j} \int P_{\tilde Z_{j-1},\m_{j-1},C_{h,j-1}}
(d\psi^{[h,j-1]}) \int P_{Z_{j-1},\m_{j-1},\tilde
f_j^{-1}}(d\psi^{(j)})\;\cdot\nn\\
&& \cdot\; e^{-l_j F_\l(\sqrt{Z_j} \psi^{[h,j]})
-\RR\VV(\sqrt{Z_j}\psi^{[h,j]}) +\tilde\BB^{(j)}
(\sqrt{Z_j}\psi^{[h,j]})}\;, \lb{2.20}\eea
where $\tilde f_j(\kk)=f_j(\kk) Z_{j-1} [\tilde
Z_{j-1}(\kk)]^{-1}$.

The above integration procedure is done till the scale
$h^*=\max\{h,\bar h^*\}$, where $\bar h^*$ is the maximal $j$ such
that $\g^j\le \m_j$. If $\bar h^*< j\le N$, by using the Gevray
property \pref{1.5a} of $\chi_0$, see \cite{DR}, we get
\bea
|g^{(j)}_{\o,\o}(\xx,\yy)| &\le& {C\over Z_{j-1}}\g^j
e^{-c\sqrt{\g^j{|\xx-\yy|}}}\;,\nn\\
|g^{(j)}_{\o,-\o}(\xx,\yy)| &\le& {C\over Z_{j-1}} \lft({\m_j\over
\g^j}\rgt)\g^j e^{-c\sqrt{\g^j{|\xx-\yy|}}}\;,\lb{2.21}\eea
where $C$ and $c$ are suitable constants; moreover,
\bea
|g^{(\leq \bar h^*)}_{\o,\o}(\xx,\yy)| &\le& {C\over Z_{\bar
h^*-1}} \g^{\bar h^*} e^{-c\sqrt{\g^{\bar h^*}{|\xx-\yy|}}}\;, \nn\\
|g^{(\leq \bar h^*)}_{\o,-\o}(\xx,\yy)| &\le& {C\over Z_{\bar
h^*-1}} \lft({\m_{\bar h^*}\over \g^{\bar h^*}}\rgt) \g^{\bar h^*}
e^{-c\sqrt{\g^{\bar h^*}{|\xx-\yy|}}}\;.\lb{2.21a}\eea
Note that the propagator $\hat g^{Q,(i)}_{\o}(\kk)$ is equivalent
to $\hat g^{(i)}_{\o}(\kk)$, as concerns the dimensional bounds,
since the sum in the r.h.s. of \pref{2.16} contains at most two
nonvanishing terms. We now {\it rescale} the field so that
\bea
&& l_j F_\l(\sqrt{Z_j}\psi^{[h,j]})+\RR\VV(\sqrt{Z_j}\psi^{[h,j]})
= \hat\VV^{(j)}(\sqrt{Z_{j-1}} \psi^{[h,j]})\;, \nn\\
&&\tilde\BB^{(j)}(\sqrt{Z_j}\psi^{[h,j]})=
\hat\BB^{(j)}(\sqrt{Z_{j-1}} \psi^{[h,j]})\;;\lb{2.23}\eea
it follows that $\LL\hat\VV^{(j)}(\psi^{[h,j]})=\l_j F_\l^{[h,j]}$
where $\l_j=(Z_j Z_{j-1}^{-1})^2 l_j$; we shall extend this
definition to $j=N$ by putting, in agreement with \pref{1.6},
$\l_N=\l$. If we now define
\bea
&& e^{-\VV^{(j-1)} (\sqrt{Z_{j-1}} \psi^{[h,j-1]})+
\BB^{(j-1)}(\sqrt{Z_{j-1}}\psi^{[h,j-1]})-L^2 E_j}=\lb{2.24}\\
&&=\int P_{Z_{j-1},\m_{j-1},\tilde f_j^{-1}}(d\psi^{(j)}) \,
e^{-\hat \VV^{(j)} \big(\sqrt{Z_{j-1}}[\psi^{[h,j-1]} +
\psi^{(j)}]\Big)+ \hat \BB^{(j)}
\big(\sqrt{Z_{j-1}}[\psi^{[h,j-1]} + \psi^{(j)}] \big)}\;,\nn\eea
it is easy to see that $\VV^{(j-1)}$ and $\BB^{(j-1)}$ are of the
same form of $\VV^{(j)}$ and $\BB^{(j)}$ and that the procedure
can be iterated. Note that the above procedure allows, in
particular, to write $\l_j$, $Z_j$, $\m_j$, for any $j$ such that
$N> j\ge h^*$, in terms of $\l_{j'}$,  $Z_{j'}$, $\m_{j'}$, $j'>
j$.

At the end of the iterative integration procedure, we get
\be \WW(\f,J)=-L^2 E_{L} + \sum_{m^\f+n^J\ge 1}
S_{2m^\f,n^J}^{(h)}(\f,J)\;,\lb{2.25}\ee
where $E_{L}$ is the {\it free energy} and
$S_{2m^\f,n^J}^{(h)}(\f,J)$ are suitable functionals, which can be
expanded, as well as $E_{L}$, the effective potentials and the
various terms in the r.h.s. of \pref{2.4} and \pref{2.3}, in terms
of {\it trees}.
We do not repeat here the analysis leading to the
tree expansion, as it is essentially identical to the one for
instance in \S 3 of \cite{BM1}, and we quote the results; it turns
out the kernels $S_{2m^\f,n^J}^{(h)}(\f,J)$ can be written as in
formula (102) of \cite{BM2}:
\bea
&& S_{2m^\f,n^J}^{(h)}(\f,J) = \sum_{n=0}^\io
\sum_{j_0=h^*-1}^{N-1} \sum_{\uo}
\sum_{\t\in\TT_{j_0,n,2m^\f,n^J}} \sum_{\bP\in \PP \atop
|P_{v_0}|=2m^\f} \nn\\
&&\int d\xxx \prod_{i=1}^{2m^\f} \f^{\s_i}_{\xx_i,\o_i}
\prod_{r=1}^{n^J} J_{\xx_{2m^\f+r},\o_{2m^\f+r}}
S_{2m^\f,n^J,\t,\uo}(\xxx)\;, \lb{2.26}\eea
where we refer to \S 3.4 of \cite{BM2} for the notation. In
particular,

\bd

\item{-} $\TT_{j_0,n,2m^\f,n^J}$ is a family of {\it trees}
(identical to the those defined in \S3.2 of \cite{BM2}, up to the
(trivial) difference that the maximum scale of the vertices is
$N+1$ instead of $+1$), with root at scale $j_0$, $n$ normal
endpoints (\ie endpoints not associated to $\f$ or $J$ fields),
$n^\f=2m^\f$ endpoints of type $\f$ and $n^J$ endpoints of type
$J$.

\item{-} If $v$ is a vertex of the tree $\t$, $P_v$ is a set of
labels which distinguish the {\it external fields of $v$}, that is
the field variables of type $\psi$ which belong to one of the
endpoints following $v$ and either are not yet contracted in the
vertex $v$ (we shall call $P_v^{(n)}$ the set of these variables)
or are contracted with the $\psi$ variable of an endpoint of type
$\f$ through a propagator $g^{Q(h_v)}$; note that $|P_v|=
|P_v^{(n)}|+ n_v^\f$, if $n_v^\f$ is the number of endpoints of
type $\f$ following $v$.

\item{-} $\xx_v$, if $v$ is not an endpoint, is the family of all
space-time points associated with one of the endpoints following
$v$.

\ed

\subsection{Convergence of the RG expansion}\lb{ss2.2}

In order to control the RG expansion, it is sufficient to show
that $\bar\l_{h} \= \max_{h\le j \le N} |\l_j|$ stays small if
$\l=\l_N$ is small enough. This property is surely true if $|h-N|$
is at most of order $\l^{-1}$, but to prove that it is true for
any $h,N$ is quite nontrivial. In \S\ref{ss4.2}, by using WTi and
SDe, we shall prove the following Theorem, essentially taken from
\cite{BM4}.

\begin{theorem} \lb{th6} There exists a constant $\e_1$, independent
of $N$, such that, if $|\l|\le \e_1$, the constants $\l_j$, $Z_j$,
$Z^{(2)}_j$ and $\m_j$ are well defined for any $j\le N$; moreover
there exist suitable sequences $\hat\l_j$, $\hat Z_j$, $\hat
Z^{(2)}_j$ and $\hat\m_j$, defined for $j\le 0$ and independent of
$N$, such that $\l_j= \hat\l_{j-N}$, $Z_j= \hat Z_{j-N}$,
$Z^{(2)}_j= \hat Z^{(2)}_{j-N}$ and $\m_j= \hat\m_{j-N}$. The
sequence $\hat\l_j$ converges, as $j\to -\io$, to a function
$\l_{-\io}(\l)= \l +O(\l^2)$, such that
\be
|\hat\l_j- \l_{-\io}| \le C\l^2 \g^{j/4}\;.\lb{2.42a}
\ee
Finally, there exist $\h_\m= -a_\m \l +O(\l^2)$ and $\h_z= a_z
\l^2 +O(\l^3)$, with $a_\m$ and $a_z$ strictly positive, such
that, for any $j\le 0$, $|\log_\g( \hat Z_{j-1}/ \hat Z_j) - \h_z|
\le C\l^2 \g^{j/4}$, $|\log_\g( \hat Z^{(2)}_{j-1}/ \hat
Z^{(2)}_j) - \h_z| \le C\l^2 \g^{j/4}$ and $|\log_\g(\hat\m_{j-1}/
\hat\m_j) - \h_\m| \le C|\l| \g^{j/4}$.

\end{theorem}

\0 {\it Remark.} Note that the definitions of $\l_j$, $\m_j$,
$Z_j$ and $Z^{(2)}_j$ are independent of the $\m$ value; however,
in the theory with $\m\not= 0$, there appear only their values
with $j\ge \bar h^*$.

\*\* The above result implies that we can remove the cutoffs and
take the limit $N,-h\to\io$, by choosing the {\it normalization
conditions}
\be \quad Z_{0}=1, \quad \m_0=\m\;. \lb{2.43}\ee
In fact, by using \pref{2.43}, it is easy to prove that, if $Z_N=
Z^{(2)}_N = [\prod_{i=1}^N (Z_{j-1}/Z_j)]^{-1}$ and $\m_N=
[\prod_{i=1}^N (\m_{j-1}/\m_j)]^{-1}$, then
\be
\m_j = \m \g^{-\h_\m j} F_{1,j,N}(\l), \quad  Z_j = \g^{-\h_z j}
F_{2,j,N}(\l), \quad Z^{(2)}_j= \z(\l) \g^{-\h_z j}
F_{3,j,N}(\l)\;, \lb{2.45}\ee
where
\be
\z(\l) = \prod_{j=-\io}^0 {\hat Z^{(2)}_{j-1} \hat Z_j \over \hat
Z^{(2)}_j \hat Z_{j-1}}\lb{2.45a}
\ee
and $F_{i,j,N}(\l)$, $i=1,2,3$, satisfy the conditions
\be
F_{i,0,N}(\l)=1, \qquad |F_{i,j,N}(\l)-1| \le
C|\l|^2\g^{-[N-\max\{j,0\}]/4}\;.\lb{2.45b}\ee
Note also that the first of \pref{2.45} implies that, in the limit
$N,-h\to\io$, if $[x]$ denotes the largest integer $\le x$,
\be \bar h^* = \left[ {\log_\g |\m| \over 1-\h_\m} \right]
\;.\lb{2.45c}\ee
Moreover, the proof of Theorem \ref{th6} implies that the critical
indices $\h_z$ and $\h_\m$ are given by tree expansions, such that
everywhere the constants $\l_j$ and $Z_j$ are substituted with
$\l_{-\io}$ and $\g^{-\h_z j}$. In particular $\h_z$ is the
solution of an equation of the form
\be
\h_z = a_z\l_{-\io}^2 + \l_{-\io}^4 H(\l_{-\io},\h_z)\;,
\ee
which allows to explicitly calculate the perturbative expansion of
$\h_z$ through an iteratively procedure.

\*

\0{\it Remark.} The normalization conditions \pref{2.43} could
also include the value of $Z^{(2)}_j$ for $j=0$, but we have
chosen to fix the value of $Z^{(2)}_j$ for $j=N$, by putting it
equal to $Z_N$. A different choice would only change the value of
$\z(\l)$ by an arbitrary finite constant.

\*\*

\subsection{The Schwinger functions}\lb{ss2.3}

Theorem \ref{th6} allows us to control the expansion of the
Schwinger functions, by using the following bound for the kernels
appearing in the expansion \pref{2.26}:
\bea
&&\int d\xxx |S_{2m^\f,n^J,\t,\uo}(\xxx)| \le L^2 C^{2m_\f+n_J}
(C\bar\l_{j_0})^n \g^{-j_0(-2+m^\f+n^J)}
\cdot\nn\\
&& \cdot \prod_{i=1}^{2m^\f} {\g^{-h_i}\over (Z_{h_i})^{1/2}}
\prod_{r=1}^{n^J }{Z_{\bar h_r}^{(2)}\over Z_{\bar h_r}}
\prod_{\rm v\ not\ e.p.} \lft({Z_{h_v}\over
Z_{h_v-1}}\rgt)^{|P_v|/2} \g^{-d_v} \;,\lb{2.27}\eea
where $h_i$ is the scale of the propagator linking the $i$-th
endpoint of type $\f$ to the tree, $\bar h_r$ is the scale of the
$r$-th endpoint of type $J$ and
\be d_v = -2+|P_v|/2+n_v^J +\tilde z(P_v)\;,\lb{2.28}\ee
with
\be \tilde z(P_v)=
\cases{$3/4$ & if $|P_v|=4$, $n_v^\f=0,1$, $n^J_v=0$,\cr
       $3/2$ & if $|P_v|=2$, $n_v^\f=0,1$, $n^J_v=0$,\cr
       $3/4$ & if $|P_v|=2$, $n_v^\f=0$, $n^J_v=1$,\cr
       $0$ & otherwise.\cr} \lb{2.29}\ee

\*

The above bound has a simple dimensional interpretation; how to
prove it rigorously has been explained in detail in the very
similar model studied in \cite{BM1} (see also \S 3 of \cite{BM2}).
We simply remark here that, had we defined $\LL=0$, we would have
obtained a bound similar to \pref{2.27} with $\tilde z(P_v)=0$ in
\pref{2.29}. The regularization procedure has the effect that the
{\it vertex dimension} $d_v$ gets an extra $\tilde z(P_v)$, whose
value can be understood in the following way. If we apply the
regularizing operator $1-\LL_0$ to the kernel associated with the
vertex $v$, the bound improves by a dimensional factor
$\g^{h_{v'}-h_v}$, if $v'$ is the first non trivial vertex
preceding $v$; if we apply $1-\LL_0-\LL_1$, the bound improves by
a factor $\g^{ 2(h_{v'}-h_v)}$. Moreover, if to a kernel
associated with the vertex $v$ the operator $1-\PP_0$ is applied,
the bound improves by a factor
\bea
&&|\m_{h_v}| \g^{-h_v}\le |\m_{h^*}| |\m_{h_v}/ \m_{h^*}|
\g^{-h_v}\le \g^{h^*} \g^{c\bar \l_{j_0} (h_v-h^*)} \g^{-h_v} =\\
&&=\g^{(1-c \bar \l_{j_0})(h^*-h_v)} \le \g^{{3\over
4}(h_{v'}-h_v)}\;;\nn \eea
if $1-\PP_0-\PP_1$ is applied, the bound improves by a factor
$(|\m_{h_v}| \g^{-h_v})^2 \le\g^{{3\over 2}(h_{v'}-h_v)}$.

By suitably modifying the analysis leading to the bound
\pref{2.27}, we can derive a bound for all the Schwinger functions
and get a relatively simple tree expansion for their removed
cutoffs limit. We shall here consider in detail the Schwinger
functions with $n^J=0$, at fixed non coinciding points; we shall
get a bound sufficient to prove two of the OSa, the boundedness
and the cluster property. Since relativistic invariance is obvious
by construction, to complete the proof of OSa there will remain to
prove only positive definiteness.

Given a set $\xxx= \{\xx_1, \ldots,\xx_k\}$ of $k$ (an even
integer) space-time points, such that $\d \= \min_{\xx\not= \yy
\in \xxx} |\xx-\yy| >0$, and a set $\uo= \{\o_1, \ldots, \o_k \}$
of $\o$-indices, the $k$-points Schwinger function
$S_{k,\uo}(\xxx)$ is defined as the $k$-th order functional
derivative of the generating function \pref{2.26} with respect to
$\f^+_{\xx_1, \o_1}, \ldots, \f^+_{\xx_{k/2}, \o_{k/2}}$ and
$\f^-_{\xx_{k/2+1}, \o_{k/2+1}}, \ldots, \f^-_{\xx_k, \o_k}$ at
$J=\f=0$, see \pref{1.10} and item 2) in Theorem \ref{th1}. By
using \pref{2.26}, we can write
\be
S_{k,\uo}(\xxx) = \lim_{|h|,N\to \io} \sum_{\p(\xxx,\uo)}
 \sum_{n=0}^\io \sum_{j_0=h^*-1}^{N-1} \sum_{\uo}
\sum_{\t\in\TT_{j_0,n,k,0}} \sum_{\bP\in \PP\atop |P_{v_0}|=k}
S_{k,0,\t,\uo}(\xxx)\;, \lb{2.26a}\ee
where $\sum_{\p(\xxx,\uo)}$ denotes the sum over the permutations
of the $\xx$ and $\o$ labels associated with the $k/2$ endpoints
of type $\f^+$, as well as those associated with the $k/2$
endpoints of type $\f^-$.

We need some extra definitions. Given a tree $\t$ contributing to
the r.h.s. of \pref{2.26a}, we call $\t^*$ the tree which is
obtained from $\t$ by erasing all the vertices which are not
needed to connect the $k$ special endpoints (all of type $\f$).
The endpoints of $\t^*$ are the $k$ special endpoints of $\t$,
which we denote $v^*_i$, $i=1,\ldots,k$; with each of them a
space-time point $\xx_i$ is associated. Given a vertex $v\in\t^*$,
we shall call $\xx^*_v$ the subset of $\xxx$ made of all points
associated with the endpoints following $v$ in $\t^*$; we shall
use also the definition $D_v = \max_{\xx, \yy \in \xx^*_v}
|\xx-\yy|$. Moreover, we shall call $s^*_v$ the number of branches
following $v$ in $\t^*$, $s^{*,1}_v$ the number of branches
containing only one endpoint and $s^{*,2}_v = s^*_v- s^{*,1}_v$.
Note that $\xx_v^* \subset \xx_v$ and $s^*_v \le s_v$.

The bound of $S_{k,0,\t,\uo}(\xxx)$ can be obtained by slightly
modifying the procedure described in detail in \S 3 of \cite{BM1},
which allowed us to prove the integral estimate \pref{2.27}, in
order to take into account the fact that the points in $\xxx$ are
not integrated. First of all, we note that it is possible to
extract a factor $e^{-c'\sqrt{\g^{h_v} D_v}}$ for each non trivial
(that is with $s^*_v\ge 2$, n.t. in the following) vertex
$v\in\t^*$, by partially using the decaying factors
$e^{-c\sqrt{\g^j|\xx-\yy|}}$ appearing in the bounds \pref{2.21},
which are used for the propagators of the spanning tree $T_\t
=\bigcup_v T_v$ of $\t$ (see (3.81) of \cite{BM1}); we can indeed
use the bound
\be
e^{-c\sqrt{\g^h|\xx|}} \le e^{-\frac{c}{2}\sqrt{\g^h|\xx|}} \cdot
e^{-c'\sum_{j=-\io}^h \sqrt{\g^j|\xx|}} \virg c'={c\over 2
\sum_{j=0}^\io \g^{-j/2}} \lb{2.30}
\ee
and the remark that, given a n.t. $v\in\t^*$, there is a subtree
$T_v^*$ of $T_\t$, connecting the points in $\xx_v^*$ (together
with a subset of the internal points in $\xx_v$), made of
propagators of scale $j\ge h_v$. It follows that, given two points
$\xx,\yy\in \xx_v^*$, such that $D_v=|\xx-\yy|$, there is a path
connecting $\xx$ and $\yy$, made of propagators in $T_v^*$, whose
length is at least $D_v$; the decomposition of the decaying
factors in the r.h.s. of \pref{2.30} allows us to extract, for
each of these propagators, a factor $e^{-c' \sqrt{\g^{h_v}|\xx|}}$
and the product of these factors can be bounded by
$e^{-c'\sqrt{\g^{h_v} D_v}}$.

Note that, after this operation, there will remain a factor
$e^{-(c/2)\sqrt{\g^j|\xx-\yy|}}$ for each propagator of $T_\t$, to
be used for the integration over the internal vertices. Moreover,
there will be $1+ \sum_{v\in \t^*} (s^*_v-1)= k$ integrations less
to do; by suitably choosing them, the lacking integrations produce
in the bound an extra factor $\prod_{v\in \t^*} \g^{2 h_v
(s_v^*-1)}L^{-2}$ so that we get
\bea
&&|S_{k,0,\t,\uo}(\xxx)| \le  C^k(C\bar\l_{j_0})^n \g^{-j_0(-2+
k/2)} \left[ \prod_{n.t. v\in\t^*} \g^{2 h_v (s^*_v-1)}
e^{-c'\sqrt{\g^{h_v} D_v}} \right]\cdot\nn\\
&&\cdot \prod_{i=1}^k {\g^{-h_i}\over (Z_{h_i})^{1/2}} \prod_{\rm
v\ not\ e.p.} \lft({Z_{h_v}\over Z_{h_v-1}}\rgt)^{|P_v|/2} \g^{-d_v}
\;.\lb{2.31}\eea

Let $E_i$ be the family of trivial vertices belonging to the
branch of $\t^*$ which connects $v^*_i$ with the higher non
trivial vertex of $\t^*$ preceding it; the definition of
$s^{*,1}_v$ and the fact that, by assumption, $1/Z_{h_i}\le
\g^{h_i\h}$, with $\h\le c\bar\l^2_{j_0}$, imply that, if
$E=\cup_i E_i$,
\be
\prod_{i=1}^k {\g^{-h_i}\over (Z_{h_i})^{1/2}} \le \prod_{v\in E}
\g^{-(1-\h/2)} \prod_{n.t. v\in \t^*} \g^{-h_v(1-\h/2) s^{*,1}_v}
\;.\lb{2.32}\ee

Let $v^*_0$ the first vertex following $v_0$ (the vertex
immediately following the root of $\t$, of scale $j_0+1$) with
$s^*_v\ge 2$; then we have, if $k_v$ denotes the number of
elements in $\xx_v^*$ (hence $k_v=k$, if $v_0 \le v \le v^*_0$),
\be
\g^{-j_0(-2+ k/2}) \prod_{v_0 \le v <v^*_0} \g^{-d_v} =
\g^{-h_{v^*_0} (-2+ k_{v^*_0}/2)} \prod_{v_0 \le v <v^*_0}
\g^{-\tilde d_v}\;, \lb{2.33}\ee
where we used the definition,
\be
\tilde d_v = d_v - \left(-2+ {k_v\over 2} \right) = \frac{|P_v|
-k_v}{2}\lb{2.34}\;;
\ee
note that $\tilde d_v \ge 1/2$, for any $v\in \t^*$.

By inserting \pref{2.32} and \pref{2.33} in the r.h.s. of
\pref{2.31}, we get
\bea
&&|S_{k,0,\t,\uo}(\xxx)| \le  C^k(C\bar\l_{j_0})^n \left[
\prod_{n.t. v\in\t^*} e^{-c'\sqrt{\g^{h_v} D_v}} \right]
\left[ \prod_{v\ not\ e.p.} \lft({Z_{h_v}\over Z_{h_v-1}}\rgt)^{|P_v|/2}
\right]\;\cdot\nn\\
&&\cdot \left[ \prod_{v\notin \t^*\atop v \ not\ e.p.} \g^{-d_v}
\right]
\left[ \prod_{v\in E} \g^{-d_v-1+\h/2} \right] \left[ \prod_{v_0
\le v <v^*_0} \g^{-\tilde d_v}\right] \cdot F_\t \;,\lb{2.35}
\eea
where
\be
F_\t = \g^{-h_{v^*_0} (-2+ k_{v^*_0}/2)}
\prod_{n.t. v\in\t^*} \g^{h_v [2(s^*_v-1) - (1-\h/2) s^{*,1}_v]}
\left[ \prod_{v^*_0 \le v\in\t^*\atop v\notin E} \g^{-d_v} \right]
\;.\lb{2.36}
\ee

Given a n.t. vertex $v\in\t^*$, let $s=s^*_v$, $s_1=s^{*,1}_v$,
$\tilde s=s-s_1$ and $v_1, \ldots, v_{\tilde s}$ the n.t. vertices
immediately following $v$ in $\t^*$. Note that $k_v=s_1 +
\sum_{i=1}^{\tilde s} k_{v_i}$; hence, given $\e>0$, we can write
\bea
&& -(-2+\e + k_v/2) + [2(s-1) - (1-\h/2) s_1] =\nn\\
&& =2-\e -k_v/2 +\e(s-1) +(2-\e)(s_1+ \tilde s-1)  - (1-\h/2)
s_1=\nn\\
&& = - \frac12 \left(s_1 + \sum_{i=1}^{\tilde s} k_{v_i}\right) +
\e(s-1) +(2-\e)\tilde s +s_1(2-\e-1+\h/2) =\nn\\
&&= \e(s-1) + s_1(1/2-\e +\h/2) - \sum_{i=1}^{\tilde s} (-2 +\e
+k_{v_i}/2)\;.\lb{2.37}
\eea
This identity, applied to the vertex $v^*_0$, implies that, if
$v_1, \ldots, v_{\tilde s}$, $\tilde s=s^*_{v_0^*} -
s^{*,1}_{v_0^*}$, are the n.t. vertices immediately following
$v^*_0$ in $\t^*$, then
\bea
&& \g^{-h_{v^*_0} (-2+ k_{v^*_0}/2)} \g^{h_{v^*_0}
[2(s^*_{v^*_0}-1) - (1-\h/2) s^{*,1}_{v^*_0}]}  =\nn\\
&& = \g^{\e h_{v^*_0}} \g^{\a_{v^*_0} h_{v^*_0}}
\prod_{i=1}^{\tilde s} \left[ \g^{-h_{v_i} (-2+\e+ k_{v_i}/2)}
\cdot \prod_{v\in \CC_i} \g^{-2+\e+ k_v/2} \right] \;, \lb{2.38}
\eea
where $\CC_i$ is the path connecting $v^*_0$ with $v_i$ in $\t^*$
(not including $v_i$) and we used the definition
\be \a_v= \e(s^*_v-1) + s^{*,1}_v(1/2-\e+\h/2)\;. \lb{2.39}\ee
The presence of the factor $\g^{-h_{v_i} (-2+\e+ k_{v_i}/2)}$ for
each vertex $v_i$ in the r.h.s. of  \pref{2.38} implies that an
identity similar to \pref{2.38} can be used for each n.t. vertex
$v\in\t^*$. It is then easy to show that
\be
F_\t = \g^{\e h_{v^*_0}} \left[ \prod_{n.t. v\in\t^*} \g^{\a_v
h_v} \right] \left[ \prod_{v\in\t^*, v\notin E} \g^{-\tilde
d_v+\e} \right] \;.\lb{2.40}
\ee
By inserting this equation in \pref{2.35}, we get
\bea
&&|S_{k,0,\t,\uo}(\xxx)| \le  C^k (C\bar\l_{j_0})^n \g^{\e
h_{v^*_0}} \left[ \prod_{n.t. v\in\t^*} \g^{\a_v h_v}
e^{-c'\sqrt{\g^{h_v} D_v}} \right]\;\cdot\nn\\
&&\cdot \left[ \prod_{v\ not\ e.p.} ({Z_{h_v}\over
Z_{h_v-1}})^{|P_v|/2} \g^{-\bar d_v}\right] \;,\lb{2.35a}
\eea
where
\be
\bar d_v = \cases{
\tilde d_v & if $v_0\le v < v_0^*$\cr
\tilde d_v -\e & if $v\in\t^*$, $v_0^*\le v\notin E$\cr
d_v+1-\h/2 & if $v\in E$\cr
d_v & otherwise\cr }\lb{2.41a}\ee

Note that $\bar d_v>0$ for any $v\in\t$, if $\e<1/2$; moreover, if
this condition is satisfied, $\a_v \ge \e>0$, for any n.t. vertex
$v\in\t^*$, uniformly in $\bar\l_{j_0}$. Moreover, since by
hypothesis $D_v\ge\d>0$, there is $c_0$ such that
\be
\g^{\a_v h_v} e^{-c'\sqrt{\g^{h_v} D_v}} \le \sup_{x>0} x^{2\a_v}
e^{-c' x\sqrt{\d} } \le \left( {c_0\over \d} \right)^{\a_v}
\a_v^{2\a_v}\;. \lb{2.43a}\ee
Note that
\be \sum_{n.t. v\in \t^*} \a_v = \frac12 k(1+\h) -\e \le
\frac12 k(1+ \h)\;. \lb{2.42aa}\ee
Hence, by using \pref{2.35a} and \pref{2.42aa}, we get
\bea
&&|S_{k,0,\t,\uo}(\xxx)| \le  C^k  (k!)^{1+ \h} (C\bar\l_{j_0})^n
\d^{-[k(1+\h)/2 -\e] +\a_{v_0^*}}\;\cdot\nn\\
&&\cdot \g^{ (\a_{v^*_0} +\e) h_{v^*_0}}
e^{-c'\sqrt{\g^{h_{v^*_0}} D_\xxx}} \left[ \prod_{v\ not\ e.p.}
\lft({Z_{h_v}\over Z_{h_v-1}}\rgt)^{|P_v|/2} \g^{-\bar d_v}\right]
\;,\lb{2.44a}
\eea
where $D_\xxx$ denotes the diameter of the set $\xxx$.

Let us now observe that, since the vertex dimensions $\bar d_v$
are all strictly positive, if we insert the bound \pref{2.44a} in
the r.h.s of \pref{2.26a}, we can easily perform all the sums (by
using the arguments explained, for instance, in \cite{BM1}), once
we have fixed the scale of the vertex $v_0^*$ and the values of
$s^*_{v^*_0}$ and $s^{*,1}_{v^*_0}$ (so that the value of
$\a_{v^*_0}$ is fixed) and we can take the limit $-h,N\to\io$. By
using Theorem \ref{th6} and the remark that the bound \pref{2.44a}
implies that the trees giving the main contribution to
$S_{k,\uo}(\xxx)$ are those with $\g^{h_{v^*_0}} D_\xxx$ of order
$1$, it is easy to prove that the limit can be expressed as an
expansion similar to \pref{2.26a}, with the sum over $j_0$ going
from $-\io$ to $+\io$, the sum over $\t$ including trees with
endpoints of arbitrary scale (satisfying the usual constraints)
and the values of $S_{k,0,\t,\uo}(\xxx)$ modified in the following
way:

\bd

\item{1)} in every endpoint there is the same constant $\l_{-\io}$
in place of $\l_{h_v}$;

\item{2)} the constants $Z_j$, and $\m_j$ are
substituted everywhere by $\g^{-\h_z j}$ and $\m \g^{-\h_\m j}$,
respectively, see \pref{2.45};

\item{3)} in the expansion which defines the constants $z_j$ and $s_j$
needed, respectively, in the definition of $\tilde Z_{j-1}(\kk)$
and $\m_{j-1}(\kk)$, see \pref{2.18}, one has to make the same
substitutions of items 1) and 2).

\ed

The bound \pref{2.44a} also implies the following one (valid for
$C_\e|\l|\le 1$, with $C_\e\to\io$ as $\e\to 1/2$):
\bea
&& |S_{k,\uo}(\xxx)| \le C^k  (C_\e|\l|)^{k/2-1} (k!)^{2+\h}
\d^{-[k(1+\h)/2 -\e]} \sum_{s=2}^k \sum_{s_1=0}^s \d^{\e (s-1)
+ s_1(1-2\e +\h)/2} \cdot\nn\\
&& \cdot \sum_{h=-\io}^{+\io} \g^{[\e s + s_1(1-2\e +\h)/2]h}
e^{-c'\sqrt{\g^{h} D_\xxx}} \le\nn\\
&& \le C^k (C_\e|\l|)^{k/2-1} (k!)^{3+ 2\h}  \d^{-k(1+\h)/2}
\sum_{s=2}^k \sum_{s_1=0}^s \left(  {\d\over D_\xxx} \right)^{\e s
+ s_1(1-2\e+\h)/2} \;. \lb{2.45aa}\eea
Since $\d/ D_\xxx\le 1$, the sum over $s$ and $s_1$ is bounded by
$Ck^2(\d/ D_\xxx)^{2\e}$; hence we get the bound
\be
|S_{k,\uo}(\xxx)| \le C^k (C_\e|\l|)^{k/2-1} (k!)^{3+2\h}
\d^{-[k(1+\h)/2 -2\e]} {1\over 1+D_\xxx^{2\e}}\;, \lb{2.45b}\ee
which proves both the boundedness and the cluster property, see
Appendix \ref{ss6}.

\*

In conclusion,we have proved the following result.

\begin{theorem} \lb{th5}
If $\e_1$ is defined as in Theorem \ref{th6}, there exists
$\e_2\le \e_1$ such that,if the normalization conditions
\pref{2.43} are satisfied and $|\l|\le \e_2$, then the Schwinger
functions $S_{k,\uo}(\xxx)$ are well defined at non coinciding
points and verify all the OS axioms, possibly except the axiom of
positive definiteness.
\end{theorem}

 \*\*

The positivity property will be proved in \S\ref{ss4}, together
with the claim in item 4) of Theorem \ref{th1}. Moreover, it is
easy to derive from the previous bounds (see for instance
\cite{BM4} for the case $\m=0$) the bound for the two point
Schwinger functions \pref{1.15}. Finally, the previous arguments
can be extended to prove that also the Schwinger functions with
$n_J>0$ are well defined in the limit of removed cutoffs, so
completing the proof of Theorem \ref{th1}, except for eq.
\pref{1.16}, which will be proved in \S\ref{ss2.5} below.

\subsection{Bounds for the Fourier transform of the
Schwinger functions}\lb{ss2.4}

The main bound \pref{2.27} can be also used to get bounds on the
Fourier transform of the Schwinger functions at non zero external
momenta; these bounds are uniform in the cutoffs and allow, in
particular, to prove (by some obvious technicality, that we shall
ship) that the removed cutoffs limit is well defined. Here we
shall only consider, as an example, the function
$\hG^{2,1,N,h}_{\o,\o'}(\pp;\kk)$ in the massless case.

By using \pref{2.26}, we can write
\be
\hG^{2,1,N,h}_{\o,\o'}(\pp;\kk) = \sum_{n=0}^\io
\sum_{j_0=h-1}^{N-1} \sum_{\t\in\TT_{j_0,n,2,1}} \sum_{\bP\in \PP
\atop |P_{v_0}|=2} \hG^{2,1}_\t(\pp,\kk)\;, \lb{2.60}\ee
with an obvious definition of $\hG^{2,1}_\t(\pp,\kk)$. Let us
define, for any $\kk\not= 0$, $h_\kk= \min\{j: f_j(\kk)\not= 0\}$
and suppose that $\pp$, $\kk$, $\pp-\kk$ are all different from
$0$. It follows that, given $\t$, if $h_-$ and $h_+$ are the scale
indices of the $\psi$ fields belonging to the endpoints associated
with $\f^+$ and $\f^-$, while $h_J$ denotes the scale of the
endpoint of type $J$, $\hG^{2,1}_\t(\pp,\kk)$ can be different
from $0$ only if $h_-=h_\kk, h_\kk+1$, $h_+=h_{\kk-\pp},
h_{\kk-\pp} +1$ and $h_J\ge h_\pp- \log_\g 2$. Moreover, if
$\TT_{j_0,n,\pp,\kk}$ denotes the set of trees satisfying the
previous conditions and $\t\in \TT_{j_0,n,\pp,\kk}$,
$|\hG^{2,1}_\t(\pp,\kk)|$ can be bounded by $\int d\zz d\xx
|G^{2,1}_\t(\zz; \xx, \yy)|$. Hence, by using \pref{2.27} and
\pref{2.45}, we get
\bea
&&|\hG^{2,1,N,h}_{\o,\o'}(\pp;\kk)| \le C \g^{-h_\kk(1-\h_z/2)}
\g^{-h_{\kk-\pp}(1-\h_z/2)}\;\cdot\nn\\
&& \cdot \sum_{n=0}^\io \sum_{j_0=h-1}^{N-1}
\sum_{\t\in\TT_{j_0,n,\pp,\kk}} \sum_{\bP\in \PP \atop
|P_{v_0}|=2} (C|\l|)^n \prod_{\rm v\ not\ e.p.} \g^{-d_v} \;.
\lb{2.61}\eea

The bound of the r.h.s. of \pref{2.61} could be easily performed
by using the procedure described in \S3 of \cite{BM1}, if $d_v$
were greater than $0$ for any $v$; however, by looking at
\pref{2.28}, one sees that this is not true. Given $\t\in
\TT_{j_0,n,\pp,\kk}$, let $v^*_0$ the higher vertex preceding all
three special endpoints and $v^*_1\ge v^*_0$ the higher vertex
preceding either the two endpoints of type $\f$ (to be called
$v_{\f,+}$ and $v_{\f,-}$) or one endpoint of type $\f$ and the
endpoint of type $J$ (to be called $v_J$). It turns out that
$d_v>0$, except for the vertices belonging to the path $\CC^*$
connecting $v_1^*$ with $v^*_0$, where, if $|P_v|=4$ and $n_v^J=0$
or $|P_v|=2$ and $n_v^J=1$, $d_v=0$. Hence, we can perform as in
\S3 of \cite{BM1} the sums over the scale and $P_v$ labels of
$\t$, only if we fix the scale indices $h^*_0$ and $h^*_1$  of
$v^*_0$ and $v^*_1$, after multiplying by $\g^{-\d(h^*_1-h^*_0)}$
the r.h.s. of \pref{2.61}, $\d$ being any positive number. Of
course, we have also to perform the sum over $h^*_0$, $h^*_1$ of
$\g^{\d (h^*_1-h^*_0)}$, which is divergent, if we proceed exactly
in this way.

In order to solve this problem, we note that, if $v\notin \CC^*$,
$d_v-1/4>0$. Hence, before performing the sums over the scale and
$P_v$ labels, we can extract from each $\g^{-d_v}$ factor
associated with the vertices belonging to the paths connecting the
three special endpoints with $v^*_0$ or $v^*_1$, a $\g^{-1/4}$
piece, to be used to perform safely the sums over $h^*_0$, $h^*_1$
in the following way.

Let us consider first the family $\TT^{(1)}_{j_0,n,\pp,\kk}$ of
trees such that the two special endpoints following $v^*_1$ are
$v_{\f,+}$ and $v_{\f,-}$ and let us suppose that $|\kk| \ge
|\kk-\pp|$. In this case, before doing the sums over the the scale
and $P_v$ labels, we fix also the scale $h_J$ of $v_J$. We get, if
$h_J^*\= \max\{h_\pp+2, h^*_0+1\}$:
\bea
&& \sum_{n=0}^\io \sum_{j_0=h-1}^{N-1}
\sum_{\t\in\TT^{(1)}_{j_0,n,\pp,\kk}} \sum_{\bP\in \PP \atop
|P_{v_0}|=2} (C|\l|)^n \prod_{\rm v\ not\ e.p.} \g^{-d_v} \le
\lb{2.62}\\
&& \le C \sum_{h_1^*=-\io}^{h_{\kk-\pp}} \sum_{h_0^*=-\io}^{h^*_1}
\sum_{h_J=h_J^*}^{+\io} \g^{\d(h^*_1- h^*_0)} \g^{-\frac14
[(h_\kk-h^*_1) +(h_{\kk-\pp} -h^*_1) + (h_J-h^*_0)]}\;,\nn\eea
and it is easy to prove that the r.h.s. of \pref{2.62} is bounded
by $C\g^{\d(h_\kk -h_\pp)}$, if $\d\le 1/8$. If $|\kk-\pp| \ge
|\kk|$, we get a similar result, with $h_{\kk-\pp}$ in place of
$h_\kk$.

Let us consider now the family $\TT^{(2,+)}_{j_0,n,\pp,\kk}$ of
trees such that the two special endpoints following $v^*_1$ are
$v_J$ and $v_{\f,+}$. We get, if $h_J^*\= \max\{h_\pp+2,
h^*_1+1\}$ and $\bar h_0=\min\{h_{\kk-\pp}, h^*_1\}$:
\bea
&& \sum_{n=0}^\io \sum_{j_0=h-1}^{N-1}
\sum_{\t\in\TT^{(1,+)}_{j_0,n,\pp,\kk}} \sum_{\bP\in \PP \atop
|P_{v_0}|=2} (C|\l|)^n \prod_{\rm v\ not\ e.p.} \g^{-d_v} \le
\lb{2.63}\\
&& \le C \sum_{h_1^*=-\io}^{h_{\kk}} \sum_{h_0^*=-\io}^{\bar h_0}
\sum_{h_J=h_J^*}^{+\io} \g^{\d(h^*_1- h^*_0)} \g^{-\frac14
[(h_\kk-h^*_1) +(h_{\kk-\pp} -h^*_0) + (h_J-h^*_1)]}\;,\nn\eea
and it is easy to prove that, if $\d\le 1/8$, the r.h.s. of
\pref{2.63} is bounded by $C\g^{\d(h_\kk -h_{\kk-\pp})}$, if
$|\kk| \ge |\kk-\pp|$, by a constant, otherwise. The family
$\TT^{(2,-)}_{j_0,n,\pp,\kk}$ of trees such that the two special
endpoints following $v^*_1$ are $v_J$ and $v_{\f,-}$ can be
treated in a similar way and one obtains a bound $C\g^{\d(
h_{\kk-\pp}- h_\kk)}$, if $|\kk-\pp| \ge |\kk|$, or a constant,
otherwise.

By putting together all these bounds, we get, for any positive
$\d\le 1/8$:
\bea
&& |\hG^{2,1,N,h}_{\o,\o'}(\pp;\kk)| \le {C_\d\over |\kk|^{1-\h_z}
|\kk-\pp|^{1-\h_z}} \;\cdot\lb{2.64}\\
&&\left[ \left( {|\kk|\over |\pp|} \right)^\d + \left( {
|\kk-\pp|\over |\pp|} \right)^\d + \left( {|\kk| \over |\kk -\pp|}
\right)^\d + \left( {|\kk-\pp|\over |\kk|} \right)^\d
\right]\;,\nn\eea
with $C_\d\to \io$ as $\d\to 0$.

\subsection{Calculation of $\hG_\o^2{(\kk)}$ in the massless case}
\lb{ss2.5}

We want now to discuss the structure of the limit $-h,N \to\io$ of
the interacting propagator $\hG^{2,N,h}_\o(\kk)$ for $\m=0$.

By using \pref{2.26}, we can write
\be
\hG^{2,N,h}_\o(\kk) = \sum_{n=0}^\io \sum_{j_0=h-1}^{N-1}
\sum_{\t\in\TT_{j_0,n,2,0}} \sum_{\bP\in \PP \atop |P_{v_0}|=2}
\hG^{2}_\t(\kk)\;, \lb{2.65}\ee
with an obvious definition of $\hG^{2}_\t(\kk)$.

Let us define $h_\kk$ as in \S\ref{ss2.4} and suppose that
$\kk\not= 0$. It follows that, given $\t$, if $h_-$ and $h_+$ are
the scale indices of the $\psi$ fields belonging to the endpoints
associated with $\f^+$ and $\f^-$, $\hG^{2}_\t(\kk)$ can be
different from $0$ only if $h_\pm=h_\kk, h_\kk+1$. Moreover, if
$\TT_{j_0,n,\kk}$ denotes the set of trees satisfying the previous
conditions and $\t\in \TT_{j_0,n,\kk}$, $|\hG^{2}_\t(\kk)|$ can be
bounded by $\int d\xx |G^{2}_\t(\xx, \yy)|$. Hence, by using
\pref{2.27} and \pref{2.45}, we get
\bea
&&|\hG^{2,N,h}_{\o}(\kk)| \le C \g^{-(h_\kk-j_0)}
{\g^{-h_\kk}\over Z_{h_\kk}} \;\cdot\nn\\
&& \cdot \sum_{n=0}^\io \sum_{j_0=h-1}^{N-1}
\sum_{\t\in\TT_{j_0,n,\kk}} \sum_{\bP\in \PP \atop |P_{v_0}|=2}
(C|\l|)^n \prod_{\rm v\ not\ e.p.} \g^{-d_v} \;, \lb{2.66}\eea
where $d_v>0$, except for the vertices belonging to the path
connecting the root with $v^*$, the higher vertex preceding both
the two special endpoints, where $d_v$ can be equal to $0$. These
vertices can be regularized by using the factor
$\g^{-(h_\kk-j_0)}$ in the r.h.s. of \pref{2.66}; hence, by
proceeding as in \S\ref{ss2.4}, we can easily perform the sum over
the trees with a fixed value of the scale label $h^*$ of $v^*$ and
we get the bound
\be
|\hG^{2,N,h}_{\o}(\kk)| \le C {\g^{-h_\kk}\over Z_{h_\kk}}
\sum_{h^*=-\io}^{h_\kk} \g^{-(h_\kk - h^*)/2} \le C
{\g^{-h_\kk}\over Z_{h_\kk}}\;.\lb{2.67}\ee

By using Theorem \ref{th6}, it is not hard to argue, as in
\S\ref{ss2.3}, that the removed cutoffs limit $\hG^{2}_{\o}(\kk)$
is well defined and is given by an expansion similar to
\pref{2.65}, with the sum over $j_0$ going from $-\io$ to $+\io$
and the quantity $\hG^{2}_\t(\kk)$ modified by substituting, in
every endpoint, $\l_j$ with $\l_{-\io}$, and, in every propagator,
$Z_j$ with $\g^{-\h j}$, $\h\=\h_z$; this property easily implies
that $\hG^{2}_{\o}(\g\kk) = \g^{\h-1} \hG^{2}_{\o}(\kk)$. On the
other hand, the symmetries of the model imply that there is a
function $g(x,\l)$, defined for $x>0$ and $\l$ small enough, such
that $\hG^{2}_{\o}(\kk)= D_\o^{-1}(\kk) g(|\kk|,\l)$; by the
previous scaling property, $g(\g x,\l)= \g^{\h} g(x,\l)$. We want
to show that $g(x,\l)=x^{\h} f(\l)$, with $f(\l)$ independent of
$x$.

To prove this claim, first of all note that
$\hG^{2,N,h}_{\o}(\kk)$ is independent of $\g$, since the cutoff
function $C^{-1}_{h,N}(\kk)$ only depends on $\g_0$ and $\g\ge
\g_0$, see \S\ref{ss1.3}. This property is then valid also for
$\hG^{2}_{\o}(\kk)$, hence for $g(x,\l)$. However, since the
expansion heavily depends on $\g$, the value of $\h$ is apparently
a function of $\g$; we want to show that this is not true.

Note that, for any $\g$ and any integer $j$, $g(\g^j,\l)=
\g^{j\h(\g)} g(1,\l)$; it follows that, if there exist, given
$\g_1$ and $\g_2$, two integers $j_1, j_2$, such that $\g_1^{j_1}
=\g_2^{j_2}$, then $\h(\g_1)= \h(\g_2)$. Hence, given an interval
$I=[\g_0, \bar\g]$ and $\g\in I$, the set $\{\g'\in
I:\h(\g')=\h(\g)\}$ is dense in $I$, as the set of rational
numbers is dense in the interval $[\log_\g \g_0, \log_\g \bar\g]$.
Since $\h(\g)$ is obviously continuous in $\g$, it follows that it
is constant.

Let us now put $g(x,\l)= x^\h f(x,\l)$; we see immediately that
$f(\g x,\l)= f(x,\l)$. Hence, by varying $\g$ in the interval
$[2,4]$ and by choosing $x=1/\g$, we see that $f(1,\l)=f(x,\l)$,
if $x\in [1/4, 1/2]$. By using this equation, by varying $x$ in
the interval $[1/4,1/2]$ and by choosing $\g=2$, we get also
$f(1,\l)=f(x,\l)$, if $x\in [1/2, 1]$. By proceeding in this way,
it is easy to show that $f(1,\l)=f(x,\l)$, for any $x>0$.

The previous discussion and the fact that, in the expansion
\pref{2.66}, $d_v>1/4$ for any $v>v^*$, imply also that
\be
\hG^{2,N,h}_\o(\kk) = {|\kk|^\h \over D_\o(\kk)} [f(\l) + O(|\kk|
\g^{-N})^{1/4}]\;.\lb{2.68}\ee

\section{ Ward--Takahashi Identities}\lb{ss3}

\subsection{Proof of Theorem \ref{th2}}\lb{ss3.1}

In order to derive WTi in the massless case $\m=0$ from the
generating functional \pref{1.6} (in the continuum limit $a=0$),
it is convenient to introduce a cutoff function
$[C^\e_{h,N}(\kk)]^{-1}$ equivalent to $[C_{h,N}(\kk)]^{-1}$ as
far as the scaling features are concerned, but such that the
support of $[C^\e_{h,N}(\kk)]^{-1}$ is the whole set $\DD_0$ and
$\lim_{\e\to 0} [C^\e_{h,N}(\kk)]^{-1}= [C_{h,N}(\kk)]^{-1}$; we
refer to \cite{BM2} \S 2.2 for its exact definition. We then
substitute $[C_{h,N}(\kk)]^{-1}$ with $[C^\e_{h,N}(\kk)]^{-1}$ in
the r.h.s. of \pref{1.6} and perform the gauge transformation
$\psi^\pm_{\xx,\o}\to e^{\pm \a_{\xx,\o}} \psi^\pm_{\xx,\o}$
(equivalent to the usual phase and chiral transformations). The
change in the cutoff function has the effect that the Lebesgue
measure $d\hat\psi^{[h,N]}$ is invariant under this transformation
and we get the WTi \pref{1.17}, where, if $< .
>_{h,N}$ denotes the expectation with respect to measure $\NN^{-1}
P_{Z_N}(d\psi) e^{-\l_N Z_N^2 V(\psi)}$ (see \pref{1.7}-\pref{1.9}
for the definitions), $\hD^{2,1,N,h}_{\o,\o'} (\pp,\kk)$ is the
Fourier transform of
\be \D^{2,1,N,h}_{\o,\o'}(\xx;\yy,\zz)\defi
\la\psi^-_{\yy,\o'};\psi^+_{\zz,\o'};\d
T_{\xx,\o}\ra_{h,N}\;,\lb{3.1}\ee
where $\la -; -; -\ra_{h,n}$ denotes the truncated expectation
with respect to the measure \pref{1.7},
\be \d T_{\xx,\o}\defi
{Z_N \over L^4}\sum_{\kk^+,\kk^-\atop\kk^+\not=\kk^-}
 e^{i(\kk^+-\kk^-)\xx} C^\e_{h,N;\o}(\kk^+,\kk^-)
\hp^+_{\kk^+,\o}\hp^-_{\kk^-,\o}\;,\lb{3.2}\ee
and
\be C^\e_{h,N;\o}(\kk^+,\kk^-)=[C_{h,N}^\e(\kk^-)-1]D_\o(\kk^-)
-[C_{h,N}^\e(\kk^+)-1]D_\o(\kk^+)\;.
\lb{3.3}\ee
Let us now suppose that $\pp$ is fixed independently of $h$ and
$N$, as well as $\kk$, and that $\pp$, $\kk$ and $\kk-\pp$ are all
different from $0$. This implies, in particular, that the
condition $\wt\c(\pp)=1$ is satisfied if $|h|$ and $N$ are large
enough and $\tilde\c(\pp)$ is the function appearing in
\pref{1.22}. Hence we can prove Theorem \ref{th2} by substituting
in \pref{1.18} $\hR^{2,1,N,h}_{\o,\o'}(\pp,\kk)$ with $\wt\c(\pp)
\hR^{2,1,N,h}_{\o,\o'}(\pp,\kk)$, which is the Fourier transform
of
\be {\partial\over\partial J_{\xx,\o}}
{\partial^2\over\partial \f^+_{\yy,\o'}\partial \f^-_{\zz,\o'}}
\WW_\D|_{J=\f=0}\;,\lb{3.4}\ee
where
\bea
e^{\WW_{\D} (J,\f)} &\defi& \int\! \der P^{[h, N]}(\ps)\
 \exp\Bigg\{ -\l_{N} V\left(\sqrt{Z_N}\ps\right) +\nn\\
&+& \sum_\o\int\!d\xx\
\lft[\f^{+}_{\xx,\o}\ps^{-}_{\xx,\o}+\ps^{+}_{\xx,\o}\f^{-}_{\xx,\o}
\rgt]\Bigg\} \cdot \lb{3.5}\\
&\cdot& \exp\left\{ \sum_\o \left[\bar T_{0,\o} - \n_{+,N}\bar
T_{+,\o} - \n_{-,N}\bar T_{-,\o} \right] \left(J, \sqrt{Z_N}
\ps\right) \right\}\nn\;,\eea
with
\bea
\bar T_{0,\o} \left(J,\ps\right) &=& {1\over L^4}\sum_{\kk,\pp}
J_{\pp,\o} \wt\c(\pp) {C^\e_{h,N;\o}(\kk,\kk-\pp)\over
D_{\o}(\pp)} \hp^+_{\kk,\o}\hp^-_{\kk-\pp,\o}\equiv {1\over
L^2}\sum_{\pp\not=0} J_{\pp,\o} \d\r_{\pp,\o}\;,\nn\\
\bar T_{\pm,\o} \left(J,\ps\right) &=& {1\over L^4} \sum_{\kk,\pp}
J_{\pp,\o} \wt\c(\pp) {D_{\pm\o}(\pp)\over D_{\o}(\pp)}
\hp^+_{\kk,\pm\o}\hp^-_{\kk-\pp,\pm\o}\;.\lb{3.6}
\eea
The coefficients $\n_{\pm,N}$ will be fixed by the requirement
that \pref{1.20} holds. A crucial role in the analysis is played
by the function
\be \D_\o^{(i,j)}(\kk^+,\kk^-)
={C^\e_{h,N;\o}(\kk^+,\kk^-)\over D_\o(\kk^+-\kk^-)}\hat
g^{\;(i)}_\o(\kk^+) \hat g^{\;(j)}_\o(\kk^-)\;,\lb{3.7}\ee
where $\hat g^{(j)}_\o \= \hat g^{\;(j)}_{\o,\o}$. By proceeding
as in \S(4.2.) of \cite{BM2} (where only the case $N=0$ is
considered), one can show that, if $\pp=\kk^+-\kk^-\not=0$ and
$|\pp|\ge 2\g^{h+1}$ (which is true since $\wt\c(\pp)=1$):

\bd

\item[1.]
if $h<i,j<N$, since $[C^\e_{h,N}(\kk^\pm)]^{-1}=1$,
\be \D^{(i,j)}_\o(\kk^+,\kk^-)=0\;;\lb{3.8}\ee
\item[2.]
if $h< j\le N$,
\be \D^{(N,j)}_\o(\kk^+,\kk^-)=
{\pp\over D_\o(\pp)} {\bf S}_\o^{(j)}(\kk^+,\kk^-) \;;\lb{3.9}\ee
where ${\bf S}_\o^{(j)}(\kk^+,\kk^-)
\defi\big(S^{(j)}_{\o,0}(\kk^+,\kk^-),S^{(j)}_{\o,1}(\kk^+,\kk^-)\big)$
is a vector of smooth functions such that
\be |\dpr_{\kk^+}^{m_+} \dpr_{\kk^-}^{m_-} S^{(j)}_{\o,i}(\kk^+,\kk^-)|\le
C_{m_+ +m_-} {\g^{-N(1+m_+)}\g^{-j(1+m_-)}\over Z_N
Z_{j-1}}\;;\lb{3.10}\ee
\item[3.]
if $h\le i \le N$,
\be |\D^{(i,h)}_\o(\kk^+,\kk^-)|\le C \g^{-(i-h)} {\g^{-h-i}\over
Z_{i-1} Z_N} \;;\lb{3.11}\ee
\item[4.]
if $i=j=h$,
\be \D^{(h,h)}_\o(\kk^+,\kk^-)=0\;.\lb{3.8a}\ee

\ed

Note that, in the r.h.s. of \pref{3.11}, there is apparently a
$Z_N/ Z_{h-1}$ factor missing, but the bound can not be improved;
this is a consequence of the fact that $\tilde Z_{h-1}(\kk)=0$ for
$|\kk|\le \g^{h-1}$, see eq. (63) of \cite{BM2}.

The multiscale integration of $\WW_{\D}$ has been described in
detail in \S 4 of \cite{BM2} (of course the scale $0$ has to be
replaced with the scale $N$). After the integration of $\psi^{(N)}$ we
get an expression like \pref{2.1} and the terms linear in $J$ and
quadratic in $\psi$ in the exponent will be denoted by
$K_{J}^{(N-1)}(\sqrt{Z_{N-1}} \psi^{[h,N-1]})$; we write
$K_{J}^{(N-1)}=K_{J}^{(a,N-1)}+K_{J}^{(b,N-1)}$, where
$K_{J}^{(a,N-1)}$ was obtained by the integration of $\bar T_0$
and $K_{J}^{(b,N-1)}$ from the integration of $\bar T_\pm$. We can
write $K_{J}^{(a,N-1)}$ as
\bea
&& K_{J}^{(a,N-1)}(\sqrt{Z_{N-1}} \psi) =\sum_\o Z_{N}
\int d\xx J_{\xx,\o} \Bigg\{\bar T_{0,\o} (J,\psi)+ \lb{3.12}\\
&&+ \sum_{\tilde \o}\int d\yy d\zz \left[
F^{(N-1)}_{2,\o,\tilde\o}(\xx,\yy,\zz) +
F^{(N-1)}_{1,\o}(\xx,\yy,\zz) \d_{\o,\tilde\o} \right]
[\psi^+_{\yy,\tilde\o} \psi^-_{\zz,\tilde\o}] \Bigg\} \nn\eea
where $F^{(N-1)}_{2,\o,\tilde\o}$ and $F^{(N-1)}_{1,\o}$ are the
analogous of eq. (132) of \cite{BM2}; they represent the terms in
which both or only one of the fields in $\d \r_{\pp,+}$,
respectively, are contracted. Both contributions to the r.h.s. of
\pref{3.12} are dimensionally marginal; however, the
regularization of $F^{(N-1)}_{1,\o}$ is trivial, as it is of the
form
\be
F_{1,\o}^{(N-1)}(\kk^+,\kk^-)= {[C_{h,N}(\kk^-)-1] D_\o(\kk^-) Z_N
\hat g^{(N)}_\o(\kk^+)- u_N(\kk^+) \over D_\o(\kk^+-\kk^-)}
G_\o^{(2)}(\kk^+)\lb{3.13}\ee
or the similar one, obtained exchanging $\kk^+$ with $\kk^-$;
$u_N(\kk)=0$ if $|\kk|\le\g^N$ and $u_N(\kk)=1-f_N(\kk)$ for
$|\kk|\ge\g^N$. By the oddness of the propagator in the momentum,
$G_\o^{(2)}(0)=0$, hence we can regularize such term without
introducing any local term, by simply rewriting it as
\bea
&&\qquad F_{1,\o}^{(N-1)}(\kk^+,\kk^-)=
[G_\o^{(2)}(\kk^+)- G_\o^{(2)}(0)]\;\cdot\nn\\
&&{[C_{h,N}(\kk^-)-1] D_\o(\kk^-) Z_N \hat g^{(N)}_\o(\kk^+)-
u_N(\kk^+) \over D_\o(\kk^+-\kk^-)} \;.\lb{3.14}\eea

By using the symmetry property \pref{2.13a},
$F^{(N-1)}_{2,\o,\tilde\o}$ can be written as
\be F^{(N-1)}_{2,\o,\tilde\o}(\kk^+, \kk^-) = {1\over D_\o(\pp)}
\left[ p_0 A_{0,\o,\tilde\o}(\kk^+,\kk^-) + p_1
A_{1,\o,\tilde\o}(\kk^+,\kk^-) \right]\;,\lb{3.15}\ee
where $A_{i,\o,\tilde\o}(\kk^+,\kk^-)$ are functions such that, if
we define
\be \LL F^{(N-1)}_{2,\o,\tilde\o}={1\over D_\o(\pp)}
\left[ p_0 A_{0,\o,\tilde\o}(0,0) +p_1 A_{1,\o,\tilde\o}(0,0)
\right]\;,\lb{3.16}\ee
then
\be \LL F^{(N-1)}_{2,\o,\o}=Z_{N-1}^{3,+} \virg \LL F^{(N-1)}_{2,\o,-\o}=
{D_{-\o}(\pp)\over D_\o(\pp)} Z_{N-1}^{3,-}\;,\lb{3.17}\ee
where  $Z_{N-1}^{3,+}$ and $Z_{N-1}^{3,-}$ are suitable real
constants. Hence the local part of the marginal term in the second
line of \pref{3.12} is, by definition, equal to
\be \sum_\o [Z_N Z_{N-1}^{3,+} \bar T_{+,\o}(J,\psi^{[h,N-1]}) +
Z_N Z_{N-1}^{3,-} \bar T_{-,\o}(J,\psi^{[h,N-1]})]\;.\lb{3.18}\ee
The terms linear in $J$ and quadratic in $\psi$ obtained by the
integration of $\bar T_\pm$ have the form
\bea
&&K_{J}^{(b,N-1)}(\sqrt{Z_{N-1}}\psi)=Z_N {1\over
L^4}\sum_{\kk^+,\pp} \wt\c(\pp) J_{\pp,\o} \sum_{\o, \tilde\o}
\hat\psi_{\kk^+,\tilde\o}^+ \hat\psi_{\kk^+
-\pp,\tilde\o}^-\;\cdot\nn\\
&&\cdot\; \left[ -\n_{+,N} G^{(N)}_{\o,\tilde\o}(\kk^+, \kk^+-\pp)
- \n_{-,N} {D_{-\o}(\pp) \over D_\o(\pp)}
G^{(N)}_{-\o,\tilde\o}(\kk^+, \kk^+-\pp) \right] \;.\lb{3.19}\eea
By using the symmetry property of the propagators, it is easy to
show that $G^{(N)}_{\o,-\o}(0,0)=0$. Hence, if we regularize
\pref{3.19} by subtracting $G^{(N)}_{\o,\tilde\o}(0,0)$ to
$G^{(N)}_{\o,\tilde\o}(\kk^+, \kk^+-\pp)$, we still get a local
term of the form \pref{3.18}. Finally by collecting all the local
term linear in $J$ we can write
\bea
&&\qquad \LL K_J^{N-1}(\sqrt{Z_{N-1}}\psi^{[h,N-1]})= \sum_\o
\Big[ Z_N\bar T_{0,\o}\left(J,\ps^{[h,N-1]}\right)-\lb{3.20}\\
&&- \n_{+,N-1}\bar T_{+,\o}\left(J,\sqrt{Z_{N-2}}
\ps^{[h,N-1]}\right)- \n_{-,N-1}\bar
T_{-,\o}\left(J,\sqrt{Z_{N-2}} \ps^{[h,N-1]}\right) \Big]\;,
\nn\eea
where $Z_{N-2} \n_{\pm,N-1} =Z_{N-1}[\n_{\pm,N} - Z_{N-1}^{3,\pm}
+\n_{\pm,N} G^{(N)}_{\pm \o,\pm \o}(0,0)]$ (our definitions imply
that $Z_{N-1}=Z_N$). The above integration procedure can be
iterated with no important differences up to scale $h+1$. In
particular, for all the marginal terms such that one of the fields
in $\bar T_{0,\o}$ in \pref{3.12} is contracted at scale $j$, we
put $\RR=1$; in fact the second field has to be contracted at
scale $h$ and, by \pref{3.11}, the extra factor $\g^{h-j}$ has the
effect of automatically regularizing such contributions.

The above analysis implies that $\n_{+,j}$ gets no contributions
from trees with an endpoint of type $\n_{-,k}$, $k>j$, and
viceversa; moreover, if a tree has an endpoint corresponding to
$\bar T_{0,\o}$, this endpoint has scale index $N+1$. Hence we can
write, for $h+1\le j\le N-1$,
\be \n_{\pm,j-1}=\n_{\pm, j}+
\b_{\pm,\n}^{j}(\l_j,\n_j..,\l_N,\n_N)\;,\lb{3.21}\ee
with

\be \b_{\pm,\n}^{j}(\l_j,\n_j..,\l_N,\n_N)=
\b_{\pm,\n}^{j}(\l_j,..,\l_N)+\sum_{j'=j}^N \n_{\pm,j'}
\b^{j,j'}_{\pm,\n}(\l_j,..,\l_N)\lb{3.22}\ee
and, given a positive $\th<1/4$,
\be |\b_{\pm,\n}^{j}(\l_j,..,\l_N)|\le
C\bar\l_j\g^{-2\th(N-j)}\quad,
|\b^{j,j'}_{\pm,\n}(\l_j,..,\l_N)|\le
C\bar\l_j^2\g^{-2\th|j-j'|}\;.\lb{3.23}\ee
We fix $\n_{\pm,N}$ so that
\be \n_{\pm,N}=-\sum_{j=h+1}^N
\b_{\pm,\n}^{j}(\l_j,\n_j..,\l_N,\n_N)\;.\lb{3.21a}\ee
By a fixed point argument (see \S 4.6 of \cite{BM4}), one can show
that, if $\bar\l_h$ is small enough, it is possible to choose
$\n_{\pm,N}$ so that
\be |\n_{\o,j}| \le c_0 \bar\l_h \g^{-\th (N-j)}\;,\lb{3.25}\ee
for any $h+1\le j\le N$.

The convergence of $\n_{\pm,N}$ as $|h|,N\to \io$ is an easy
consequence of the previous considerations. Moreover, from an
explicit computation of \pref{3.21}, we get
$\n_-={\l\over4\pi}+O(\l^2)$ and $\n_+=c_+\l^2+O(\l^3)$ with
$c_+<0$.

\insertplot{250}{60}%
{\footnotesize
\ins{40pt}{59pt} {$\o$}
\ins{45pt}{17pt} {$\o$}

\ins{75pt}{47pt} {$-\o$}
\ins{75pt}{15pt} {$-\o$}

\ins{190pt}{57pt} {$\o$}
\ins{195pt}{17pt} {$\o$}

\ins{235pt}{53pt} {$\o$}
\ins{235pt}{11pt} {$\o$}

\ins{195pt}{35pt} {$-\o$}
\ins{217pt}{35pt} {$-\o$}
}%
{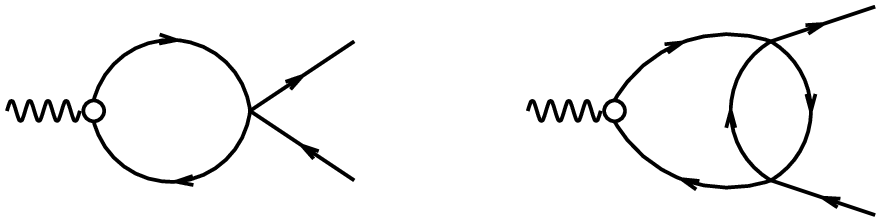}{\lb{f6}: Graphical representation of the lowest order
contribution to $\n^-$ and to $\n^+$; the small circle represent the operator $C$
\pref{3.3}}{0}

The convergence of $\hG^{2,1,N,h}_{\o,\o'}(\pp;\kk)$ was discussed
in \S \ref{ss2.4}. Hence, to complete the proof of Theorem
\ref{th2}, we have to prove that $\wt\c(\pp)
\hR^{2,1,N,h}_{\o,\o'}(\pp,\kk)\to 0$ , if $\pp$, $\kk$ and
$\kk-\pp$ are all different from $0$. In fact, since
$\wt\c(\pp)=1$ for $\pp\not=0$ and $|h|,N$ large enough, this
implies \pref{1.20}.

This result can be obtained by a simple extension of the arguments
given in \S\ref{ss2.4} to prove that
$\hG^{2,1,N,h}_{\o,\o'}(\pp;\kk)$ is bounded uniformly in $h$ and
$N$. In fact, $\wt\c(\pp) \hR^{2,1,N,h}_{\o,\o'}(\pp,\kk)$ can be
written by a sum of trees essentially identical to the ones for
$\hG^{2,1,N,h}_{\o,\o'}$, with the only important difference that
there are three different special endpoints associated to the
field $J$, corresponding to the three different terms in
\pref{3.20}; we call these endpoints of type $T_0,T_+,T_-$
respectively.

The sum over the trees such that the endpoint is of type $T_\pm$
can be bounded as in \pref{2.61}, the only difference being that,
thanks to the bound \pref{3.25}, one has to multiply the r.h.s. by
a factor $|\l|\g^{-\th(N-h_J)}$, which has to be inserted also in
the r.h.s. of the bounds \pref{2.62} and \pref{2.63}. Hence, it is
easy to see that the contributions of these trees vanishes as
$N\to\io$.

Let us now consider the trees with an endpoint of type $T_0$. In
this case there are two possibilities. The first is that the
fields of the $T_0$ endpoint are contracted at scale $j,N$; this
implies that the sum over $h_J$ is missing in the r.h.s. of the
bounds \pref{2.62} and \pref{2.63} and $h_J=N$. Hence it is easy
to see that the sum over such trees goes to $0$ as $N\to\io$. The
second possibility is that the fields of the $T_0$ endpoint are
contracted at scale $j,h$; this implies that the sum over $j_0$ is
missing in the r.h.s. of \pref{2.61} and $j_0=h$. Since
$d_v-1/4>0$ for all vertices belonging to the path connecting the
root to the vertex $v^*_0$, we can add a factor $\g^{-(j_0-h))/4}$
to the r.h.s. of the bounds \pref{2.62} and \pref{2.63}, which
then go to $0$ as $h\to-\io$.

\section{ Schwinger-Dyson equations and new anomalies}\lb{ss4}

\subsection{Proof of Theorem \ref{th3}}\lb{ss4.1}

In this section we study the last addend of the Schwinger-Dyson
equation \pref{1.22}, so proving Theorem \ref{th3}; the analysis
rests heavily on \S 4 of \cite{BM4}.

Let us consider a fixed finite $\kk$ and let us define its scale
$h_\kk$ as in \S\ref{ss2.4}; then, if $-h$ and $N$ are large
enough, $\hg^{N,h}_\o(\kk)= \hg_\o(\kk)$. We start by putting (see
\S 4.1 of \cite{BM4}):
\be
\wt G^{2,N,h}_{\e,\o}(\kk) \defi \hg_\o(\kk) \int {d\pp\over
(2\p)^2} \wt\c(\pp) {{D_{\e\o}(\pp)} \over D_{-\o}(\pp)}
\hR^{2,1,N,h}_{\e\o;\o}(\pp,\kk) = {\dpr\WW_{T,\e} \over \dpr
\hf^+_{\kk,\o}\partial \hj_{\kk,\o}} \;, \lb{4.1}\ee
where $\e=\pm$ and $\WW_{T,\e}$ is defined (in the infinite volume
limit) by the equation:
\bea
&& e^{\WW_{T,\e} (J,\f)} \defi\int\! \der P_{Z_N}(\ps)\
\exp\Bigg\{ -\l_{N}Z_N^2 V(\ps) \int\!d\xx\
\Big[\f^{+}_{\xx,\o}\ps^{[h,N]-}_{\xx,\o}
+\nn\\
&& + \ps^{[h,N]+}_{\xx,\o}\f^{-}_{\xx,\o}\Big]+ Z_N
\left[T^{(\e)}_1 -\n_{+,N} T_{-\e}- \n_{-,N} T_{\e} \right]
\left(\ps, J\right)\Bigg\}\;,\lb{4.1a}\eea
with
\bea
T^{(\e)}_1(\ps,J) &\defi& \int {d\pp d\kk'\over (2\p)^4}
\wt\c(\pp) \hj_{\kk,\o} \hg_\o(\kk) {C_{\e\o}(\kk',
\kk'-\pp)\over D_{-\o}(\pp)} \cdot\nn\\
&& \cdot \hp^+_{\kk-\pp,\o} \hp^+_{\kk',\e\o}
\hp^-_{\kk'-\pp,\e\o}\;,\lb{4.1b}\\
T_{+}(\ps,J) &\defi& \int {d\pp d\kk'\over (2\p)^4} \wt\c(\pp)
\hj_{\kk,\o} \hg_\o(\kk) \hp^+_{\kk-\pp,\o}
\hp^+_{\kk',-\o} \hp^-_{\kk'-\pp,-\o}\;,\nn\\
T_{-}(\ps,J) &\defi& \int {d\pp d\kk'\over (2\p)^4} \wt\c(\pp)
\hj_{\kk,\o} \hg_\o(\kk) {D_{\o}(\pp)\over D_{-\o}(\pp)}
\hp^+_{\kk-\pp,\o} \hp^+_{\kk',\o} \hp^-_{\kk'-\pp,\o}\;,\nn\eea
and $\n_{\pm,N}$ are defined as in \pref{3.21a}.

The calculation of $\wt G^{2,N,h}_{\e,\o}(\kk)$ is done again via
a multiscale expansion, very similar to the one described in \S 4
of \cite{BM4}. The main differences are that here we are
considering a quantity with two external lines, instead of four,
and that the external momenta are on the scale $h_\kk$, instead of
the infrared cutoff scale $h$. However, the last remark implies
that the integration of the fields of scale $j>h_\kk +1$ differs
from that discussed in \cite{BM4} only for trivial scaling
factors; in particular, there is no contribution to $\wt
G^{2,N,h}_{\e,\o}(\kk)$ associated with a tree, whose root has
scale higher than $h_\kk$.

Let us call $\bar\VV^{(N-1)}(\psi^{[h,N-1]})$ the sum over the
terms linear in $J$, obtained after the integration of the field
$\psi^{(N)}$; we put:
\bea
&&\bar\VV^{(N-1)}(\psi^{[h,N-1]}) =
\bar\VV^{(N-1)}_{a,1}(\psi^{[h,N-1]}) +
\bar\VV^{(N-1)}_{a,2}(\psi^{[h,N-1]}) +\nn\\
&&\bar\VV^{(N-1)}_{b,1}(\psi^{[h,N-1]}) +
\bar\VV^{(N-1)}_{b,2}(\psi^{[h,N-1]}) \;,\lb{4.2}\eea
where $\bar\VV^{(N-1)}_{a,1} + \bar\VV^{(N-1)}_{a,2}$ is the sum
of the terms in which the field $\hat\psi^+_{\kk-\pp,\o}$
appearing in the definition of $T^{(\e)}_1(\psi)$ or $T_\pm(\psi)$
is contracted, $\bar\VV^{(N-1)}_{a,1}$ and $\bar\VV^{(N-1)}_{a,2}$
denoting the sum over the terms of this type containing a $T_1$ or
a $T_\pm$ vertex, respectively; $\bar\VV^{(N-1)}_{b,1} +
\bar\VV^{(N-1)}_{b,2}$ is the sum of the other terms, that is
those where the field $\hat\psi^+_{\kk-\pp,\o}$ is an external
field, the index $i=1,2$ having the same meaning as before.

Let us consider first $\bar\VV^{(N-1)}_{a,1}$; we shall still
distinguish different group of terms, those where both fields
$\hat\psi_{\kk',\e\o}^+$ and $\hat\psi_{\kk'-\pp,\e\o}^-$ are
contracted, those where only one among them is contracted and
those where no one is contracted. If no one of the fields
$\hat\psi_{\kk',\e\o}^+$ and $\hat\psi_{\kk'-\pp,\e\o}^-$ is
contracted, we can only have terms with at least four external
lines; for the properties of $\D_{\e\o}^{(i,j)}$ (see \pref{3.7}), at
least one of the fields $\hat\psi_{\kk',\e\o}^+$ and
$\hat\psi_{\kk'-\pp,\e\o}^-$ must be contracted at scale $h$. If
one of these terms has four external lines, hence it is marginal,
it has the following form
\bea
&&Z_N \int {d\pp d\kk'\over (2\p)^4} \hat\psi^+_{\o,\kk-\pp}
G^{(N)}_2(\kk-\pp) \hat
g^{N,h}_\o(\kk-\pp) \;\cdot\nn\\
&& \cdot \hj_{\kk,\o} \hg_\o(\kk) \wt\c(\pp)
{C_{\e\o}(\kk',\kk'-\pp) \over D_{-\o}(\pp)}
\hat\psi^+_{\kk',\e\o}\hat\psi^-_{\kk'-\pp,\e\o}\;, \lb{4.3}\eea
where $G^{(N)}_2(\kk)$ is a suitable function which can be
expressed as a sum of graphs with an odd number of propagators,
hence it vanishes at $\kk=0$. This implies that $G^{(N)}_2(0)=0$,
so that we can regularize it without introducing any running
coupling.

\insertplot{100}{80}%
{\footnotesize
\ins{10pt}{49pt} {$\kk,\o$}
\ins{15pt}{65pt} {$\kk-\pp,\o$}
\ins{75pt}{47pt} {$\kk',\e\o$}
\ins{45pt}{10pt} {$\kk'-\pp,\e\o$}
}%
{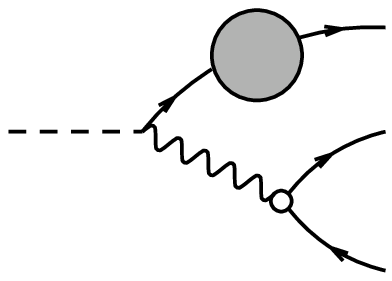}{\lb{f8}: Graphical representation of \pref{4.3}}{0}

If both $\hat\psi_{\kk',\e\o}^+$ and $\hat\psi_{\kk'-\pp,\e\o}^-$
in $T^{(\e)}_1(\psi)$ are contracted, we get terms of the form (up
to an integral over the external momenta)
\be
\hj_{\kk,\o} g_\o(\kk) \tilde W_{n+1}^{(N-1)} (\kk,\kk_1,..,\kk_n)
(\sqrt{Z_N})^{n-1} \prod_{i=1}^n \hat\psi^{\e_i}_{\kk_i}\;,
\lb{4.4}\ee
where $n$ is an odd integer. We want to define an $\RR$ operation
for such terms. There is apparently a problem, as the $\RR$
operation involves derivatives and any term contributing to
$\tilde W_{n+1}^{(N-1)}$ contains the $\D_{\e\o}^{(N,N)}$ and the
cutoff function $\wt\c(\pp)$. Hence one can worry about the
derivatives of the factor $\wt\c(\pp) \pp D_{-\o}(\pp)^{-1}$.
However, as $\kk$ is fixed independently from $N$ (and far enough
from $\g^N$) and $\kk-\pp$ is fixed at scale $N$, then $|\pp|\ge
\g^{N-1}/2$, so that we can freely multiply by a smooth cutoff
function $\bar\chi(\pp)$ restricting $\pp$ to the allowed region;
this allows us to pass to coordinate space and shows that the
$\RR$ operation can be defined in the usual way. We define
\be \LL\tilde W_4^{(N-1)} (\kk,\kk_1,\kk_2, \kk_3) =
\tilde W_4^{(N-1)}(0,..,0)\;,\lb{4.5}\ee
\be \LL\tilde W_2^{(N-1)} (\kk)=\tilde W_2^{(N-1)}(0)+
\kk\partial_\kk \tilde W_2^{(N-1)}(0)\;. \lb{4.6}\ee
Note that by parity the first term in \pref{4.6} is vanishing;
this means that there are only marginal terms.

\insertplot{310}{80}%
{\footnotesize
\ins{10pt}{50pt}{$\kk,\o$}
\ins{30pt}{65pt}{$\kk-\pp,\o$}
\ins{110pt}{67pt}{$\kk_1,\o$}
\ins{115pt}{52pt}{$\kk_2,-\o$}
\ins{120pt}{33pt}{$\kk_3,-\o$}

\ins{180pt}{50pt}{$\kk,\o$}
\ins{200pt}{65pt}{$\kk-\pp,\o$}
\ins{285pt}{50pt}{$\kk,\o$}

}%
{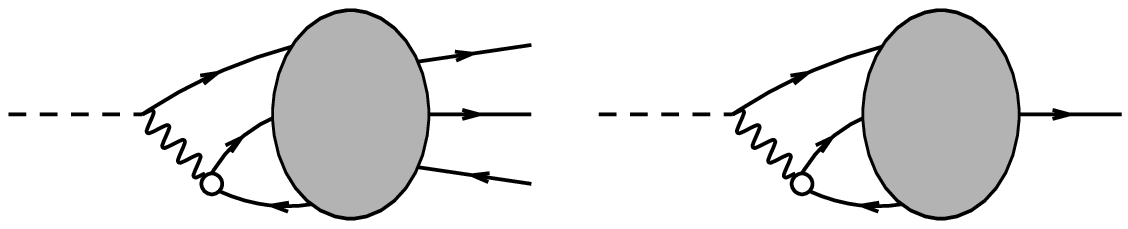}{\lb{f9}: Graphical representation of $\tilde W_4^{(-1)}$
and $\tilde W_2^{(-1)}$}{0}

If only one among the fields $\hat\psi_{\kk',\e\o}^+$ and
$\hat\psi_{\kk'-\pp,\e\o}^-$ in $T_1(\psi)$ is contracted, we get
terms with four external lines of the form (up to an integral over
the external momenta):
\bea
&& Z_N \hg_\o(\kk) \hj_{\kk,\o} \hat\psi^+_{\kk_1,\o_1}
\hat\psi^-_{\kk^-,\e\o} \hat\psi^+_{\kk^- +\kk- \kk_1,\o_2} \int
d\kk^+ \tilde\chi(\kk^+ -\kk^-) \;\cdot\nn\\
&& \cdot \hat g^{N,h}_{\o}(\kk- \kk^+ +\kk^-) G_4^{(N)}(\kk^+,
\kk_1, \kk+\kk_- -\kk_1) \;\cdot\nn\\
&& \cdot \left\{ {[C_{h,N}(\kk^-)-1] D_{\e\o}(\kk^-) \hat
g^{N,h}_{\e\o}(\kk^+)\over D_{-\o}(\kk^+-\kk^-)} - {u_N(\kk^+)
\over D_{-\o}(\kk^+-\kk^-)} \right\}\;, \lb{4.7}\eea
or the similar one with the roles of $\kk^+$ and $\kk^-$
exchanged. Note that the indices $\o_1$ and $\o_2$ must satisfy
the constraint $\o_1 \o_2=\e$.

\insertplot{120}{60}%
{\footnotesize
\ins{10pt}{40pt}{$\kk,\o$}
\ins{20pt}{51pt}{$\kk-\pp,\o$}
\ins{60pt}{22pt}{$\kk^+,\e\o$}
\ins{55pt}{2pt}{$\kk^-,\e\o$}
\ins{101pt}{55pt}{$\kk_1,\o$}
\ins{101pt}{28pt}{$\kk_2,-\o$}

}%
{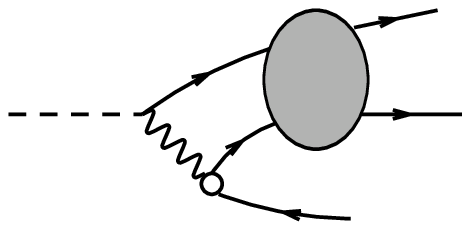}{\lb{f10}: Graphical representation of a single addend in
\pref{4.7}}{0}

The two terms in \pref{4.7} must be treated differently, as
concerns the regularization procedure. The first term is such that
one of the external lines is associated with the operator
$[C_{h,N}(\kk^-)-1] D_{\e\o}(\kk^-) D_{-\o}(\pp)^{-1}$. We define
$\RR=1$ for such terms; in fact, when such external line is
contracted (and this can happen only at scale $h$), the factor
$D_{\e\o}(\kk^-) D_{-\o}(\pp)^{-1}$ produces an extra factor
$\g^{h-N}$ in the bound, with respect to the dimensional one. The
second term in \pref{4.7} can be regularized as above, by
subtracting the value of the kernel computed at zero external
momenta, \ie for $\kk^-=\kk=\kk_1=0$. Note that such quantity
vanishes, if the four $\o$-indices are all equal, otherwise it is
given by the product of the field variables times
\be -Z_N \hg_\o(\kk) \hj_{\kk,\o} \int d\kk^+ \tilde\chi(\kk^+)
\hat g^{N,h}_{\o}(\kk^+) G_4^{(N)}(\kk^+, 0, 0) {u_N(\kk^+) \over
D_{-\o}(\kk^+)} \;,\lb{4.8}\ee
and there is no singularity associated with the factor
$D_{-\o}(\kk^+)^{-1}$, thanks to the support on scale $N$ of the
propagator $\hat g^{N,h}_\o(\kk^+)$. The terms with two external
lines can be produced only if $\e=+1$ and can be treated in a
similar way; they have the form
\bea
&& \hat\psi^-_{\kk,\o} \hg_\o(\kk) \hj_{\kk,\o} \int d\kk^+
\tilde\chi(\kk^+-\kk) G_1^{(N)}(\kk_+)\;\cdot\nn\\
&& \cdot\; \left\{ {[C_{h,N}^\e(\kk)-1] D_{\e\o}(\kk) \hat
g^{(N)}_{\o}(\kk^+)\over D_{-\o}(\kk^+ -\kk)} - {u_N(\kk^+) \over
D_{-\o}(\kk^+ -\kk)} \right\}\;, \lb{4.9}\eea
where $G_1^{(N)}(\kk_+)$ is a smooth function of order $0$ in
$\l$. However, the first term in the braces is equal to $0$, since
we keep $\kk$ fixed and far from the cutoffs, hence
$C_{h,N}^\e(\kk)-1=0$, and the second term can be regularized as
above.

A similar (but simpler) analysis holds for the terms contributing
to $\bar\VV^{(N-1)}_{a,2}$, which contain a vertex of type $T_+$
or $T_-$ and are of order $\l\n_\pm$. Now, the only thing to
analyze carefully is the possible singularities associated with
the factors $\tilde\chi(\pp)$ and  $\pp D_{-\o}(\pp)^{-1}$.
However, since in these terms the field $\hat\psi^+_{\kk-\pp,\o}$
is contracted, $|\pp|\ge \g^{N-1}/2$; hence the regularization
procedure can not produce bad dimensional bounds.

We will define $\tilde z^{(\e)}_{N-1}$ and
$\tilde\l^{(\e)}_{N-1}$, so that (recall that $Z_{N-1}=Z_N$)
\bea
&& \LL[\bar\VV^{(N-1)}_{a,1} +
\bar\VV^{(N-1)}_{a,2}](\psi^{[h,N-1]}) = -\tilde\l^{(\e)}_{N-1}
{Z_{N-2}^2\over Z_N} \bar F_\l^{[h,N-1]}(\psi^{[h,N-1]}, J) -\nn\\
&&- \tilde z^{(\e)}_{N-1} {Z_{N-1}\over Z_N}
\hat\psi^{[h,N-1]+}_{\kk,\o} D_\o(\kk) \hg_\o(\kk) \hj_{\kk,\o}
\;,\lb{4.10}\eea
where we used the definition
\be \bar F_\l(\psi^{[h,N-1]},J)= \int {d\kk_1 d\kk_2\over
(2\p)^4} \hj_{\kk,\o} \hg_\o(\kk)
 \hp^+_{\kk_2,\o} \hp^+_{\kk+\kk_1- \kk_2,-\o}
\hp^-_{\kk_1,-\o}. \lb{4.11}\ee

Let us consider now the terms contributing to
$\bar\VV^{(N-1)}_{b,i}$, that is those where
$\hat\psi^+_{\bar\kk-\pp,\o}$ is not contracted. Such terms can be
analyzed exactly as in \S 4.3 of \cite{BM4}; it turns out that
\bea
&& \LL[\bar\VV^{(N-1)}_{b,1} +
\bar\VV^{(N-1)}_{b,2}](\psi^{[h,N-1]}) = -\n_{+,N-1}
Z_{N-2} T_{-\e}(\psi^{[h,N-1]},J) -\nn\\
&&-\n_{-,N-1} Z_{N-2} T_\e(\psi^{[h,N-1]},J) \;,\lb{4.12}\eea
$\n_{\pm,N-1}$ being exactly the same constants appearing in
\pref{3.20}.

The integration over subsequent scales is performed in a similar
way; as described in more details in \S 4.4 of \cite{BM4}, it
turns out that, if $j\ge h_\kk$, the local part of the terms
linear in $J$ has the form (coinciding with eq. (131) of
\cite{BM4} for $Z_N=1$ and $\o=\e=-1$):
\bea
&& \LL\bar\VV^{(j)}( \ps^{[h,j]})=Z_N T^{(\e)}_1(\psi^{[h,j]},J) -
\n_{+,j} Z_{j-1} T_{-\e} (\ps^{[h,j]}, J)-\nn\\
&&- \n_{-,j} Z_{j-1} T_{\e}((\psi^{[h,j]},J) -\tilde\l^{(\e)}_j
{Z_{j-1}^2\over Z_N} \bar F_\l^{[h,j]}(\psi^{[h,j]}, J) -\nn\\
&&- \sum_{i=j}^{N-1} \tilde z^{(\e)}_i {Z_i\over Z_N}
\hat\psi^{[h,j]+}_{\kk,\o} D_\o(\kk) \hg_\o(\kk) \hj_{\kk,\o}\;.
\lb{4.13}\eea
If $j< h_\kk$, $\LL\bar\VV^{(j)}$ has the same structure, but
there is indeed no term with two external legs, since
$\hat\psi^{[h,j]+}_{\kk,\o}=0$; for a similar reason the term with
four external legs is different from $0$ only if $3\g^{j-1}
>\g^{h_\kk}$. However, the constants $\tilde\l^{(\e)}_j$ and
$\tilde z^{(\e)}_i$ are defined for any $j>h$ and their value is
independent of $\kk$. Note also that, as in the expansion of a
normal Schwinger function, we do not localize the terms with four
external legs, containing both a $J$ vertex and a $\f$ vertex.

It follows that we can write $\wt G^{2,N,h}_{\e,\o}(\kk)$ as a sum
of trees with two special endpoints, similar to those described in
detail in \S 4.5 of \cite{BM4}; they differ from those present in
the expansion of the function $\hG^{2,N,h}_{\o}(\kk)$, see
\S\ref{ss2.5}, since one of the special endpoints corresponds to
one of the addenda in \pref{4.13}, to be called of type $T$,
$T_+$, $T_-$, $\wt\l^{(\e)}$, $\wt z^{(\e)}$. By construction the
constants $\n_{\pm,j}$ coincide with those introduced in
\S\ref{ss3.1}, hence they verify \pref{3.25}. Moreover, it was
shown in \S 4.6) of \cite{BM4}, by a fixed point argument, that,
if $\bar\l_h$ is small enough, it is possible to choose
$\a_{+,h,N}=c_1\l+O(\l^2)$ and $\a_{-,h,N}=c_3+O(\l)$ so that
there exist two positive constant, $C$ and $\th$, independent of
$h$ and $N$, such that, if $h+1\le j\le N-1$,
\be |z_i \; \a_{\e,h,N} -\wt z^{(\e)}_i|\le C\bar \l_h\g^{-\th(N-i)}\;,
\quad |\l_i \; \a_{\e,h,N}-\wt\l^{(\e)}_i|\le
C\bar\l_h\g^{-\th(N-i)}\;.\lb{4.15}\ee


\insertplot{290}{90}%
{\footnotesize
\ins{98pt}{77pt}
{$ \o $}
\ins{118pt}{92pt}
{$ \o $}
\ins{1400pt}{65pt}
{$ \o $}
\ins{152pt}{45pt}
{$ \o $}
\ins{125pt}{47pt}
{$ \n^- $}
\ins{141pt}{65pt}
{$ \o$}
\ins{157pt}{72pt}
{$ -\o $}
\ins{159pt}{87pt}
{$ -\o $}

\ins{15pt}{35pt}
{$\o$}
\ins{45pt}{55pt}
{$\o$}
\ins{45pt}{27pt}
{$-\o$}
\ins{45pt}{9pt}
{$-\o$}
\ins{20pt}{8pt}
{$ \n^- $}
\ins{83pt}{58pt}
{$-\o$}
\ins{83pt}{27pt}
{$\o$}
\ins{81pt}{9pt}
{$-\o$}

\ins{205pt}{55pt}
{$\o$}
\ins{237pt}{67pt}
{$\o$}
\ins{230pt}{37pt}
{$-\o$}
\ins{235pt}{19pt}
{$-\o$}
\ins{215pt}{26pt}
{$ \n^- $}
\ins{270pt}{48pt}
{$-\o$}
\ins{271pt}{30pt}
{$-\o$}
\ins{261pt}{9pt}
{$\o$}
}
{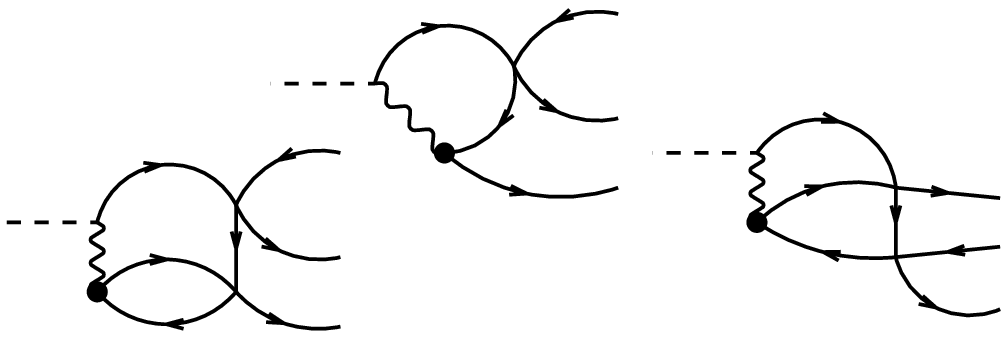}{\lb{f12}: Graphical representation of the leading terms contributing
to $c_1\l$; other four graphs contributing to $c_1\l$, as well as the
graph contributing to $c_3$, are vanishing
in the limit  of short tail ($\g_0\to 1$)
of the cutoff function (see definition \pref{1.5bis}).
The two graphs giving the
0--th order expansion in $\l$ of $\a^+_{h,N}$ cancell
each other by symmetry.}{0}

Then we can write
\be \wt G^{2,N,h}_{\e,\o}(\kk)
\defi A^{\wt\l,N,h}_{\e,\o}(\kk)+A^{\wt z,N,h}_{\e,\o}(\kk)
+A^{T_\pm,N,h}_{\e,\o}(\kk)+A^{T,N,h}_{\e,\o}(\kk)\;,\lb{4.16}\ee
where $ A^{\wt\l,N,h}_{\e,\o}$, $A^{\wt z,N,h}_{\e,\o}$,
$A^{T_\pm,N,h}_{\e,\o}$ and $A^{T,N,h}_{\e,\o}$ contain
respectively one endpoint of type $\wt\l^{(\e)}$, $\wt z^{(\e)}$,
$T_\pm$, $T$.

In order to bound $A^{T,N,h}_{\e,\o}$, we repeat the analysis in
\S 4.8 in \cite{BM4}. It follows that it is bounded by an
expression similar to the r.h.s. of \pref{2.66}, with the
following differences. Given a tree $\t$ contributing to
$A^{T,N,h}_{\e,\o}$, the dimensional bound differs from that of a
tree contributing to $\hG^{2,N,h}_\o(\kk)$ for the following
reasons:

\bd

\item[(1)] there is an extra factor
$Z_{h_\kk}/Z_N$, because one external propagator is substituted by
the free one, $Z_N^{-1}  \hg_\o(\kk)$ (see the definition of
$T_1^{(\e)}$);
\item[(2)] since there is no external field
renormalization for $T_1^{(\e)}$ (which is dimensionally
equivalent to a term with four external fields), there is an extra
factor $(Z_N/Z_{j_T})^2$, if $j_T$ is the scale of the endpoint of
type $T$;
\item[(3)] if at least one of fields in $T_1^{(\e)}$) is
contracted on scale $h$, there is an extra factor $Z_{h}/Z_N$,
because of the bound \pref{3.11};
\item[(4)] because of \pref{3.8}, either $j_T=N+1$ or the root
of $\t$ has scale $h-1$.

\ed

Hence, $A^{T,N,h}_{\e,\o}$ can be bounded by an expression equal
to the r.h.s. of \pref{2.66}, multiplied by a factor $Z_h
Z_{h_\kk}/ Z_{j_T}^2 \le \g^{C\l^2 [(j_T-h) + |j_T-h_\kk|]}$,
which takes into account the items (1)-(3) above. This factor can
be absorbed in the sum over the scale labels, since all vertices
have an "effective" positive dimension (see remark before
\pref{2.67}). Then, by taking into account the remark in item (4)
above, it is easy to show that
\be |A^{T,N,h}_{\e,\o}(\kk)|\le C {\g^{-h_\kk}\over Z_{h_\kk}}
\lft(\g^{-(N-h_\kk)/4} + \g^{-(h_\kk-h)/4}\rgt)\;.\lb{4.17}\ee

Let us now consider $A^{T_\pm,N,h}_{\e,\o}$. We still have some
extra factors with respect to the bound \pref{2.66}, the same
factor of item (1) above and a factor $(Z_N/Z_{j_T})$, due to the
partial field renormalization of $T_\pm$; the product of these
factors can be treated as before. We do not have anymore a
condition like item (4) above, but we have to take into account
that the running constant associated with the special vertex of
type $T_\pm$ satisfies the bound \pref{3.25}. It follows that
\be |A^{T_\pm,N,h}_{\e,\o}(\kk)|\le C{\g^{-h_\kk}\over
Z_{h_\kk}}\g^{-(N-h_\kk)/4}\;.\lb{4.18}\ee

Let us now consider $A^{\wt z,N,h}_{\e,\o}$ and let us suppose
that $|\kk|=\g^{h_\kk}$, so that (see \pref{2.18} and \pref{2.21})
$\hg^{(j)}_\o(\kk)= [D_\o(\kk) Z_{h_\kk-1}]^{-1}$, if $j=h_\kk$,
while $\hg^{(j)}_\o(\kk)=0$, if $j\not= h_\kk$. This condition,
which greatly simplifies the following discussion, is not really
restrictive. In fact, since the external momentum $\kk$ is fixed
in this discussion, one could modify the definition \pref{1.15} of
the cutoff functions $f_j(\pp)$, by substituting it with
$f_j(\pp/p_0)$, $p_0$ being a fixed positive number $\le \g$, to
be chosen so that $p_0^{-1} |\kk|=\g^{h_\kk}$, for some integer
$h_\kk$. Since our bounds would be clearly uniform in this new
parameter and the removed cutoffs limit is independent of $\g$
(see \S\ref{ss2.5}), this procedure can not produce any trouble.

By using \pref{4.13}, we can write
\be A^{\wt z,N,h}_{\e,\o}(\kk)= A^{1,\wt z,N,h}_{\e,\o}(\kk)
-{\hg_\o(\kk)\over Z_N} \lft[{1\over Z_{h_\kk-1}} \sum_{i=
h_\kk}^{N-1}\wt z^{(\e)}_i Z_i \rgt] \;,\lb{4.19}\ee
where $A^{1,\wt z,N,h}_{\e,\o}$ contains the contributions to
$A^{\wt z,N,h}_{\e,\o}$ coming from trees with at least one $\l$
endpoint. Since $Z_{j-1}=Z_j(1+z_j)$ and $Z_{N-1}=Z_N$,
\be Z_{h_\kk-1}-\sum_{j=h_\kk}^{N-1}Z_j z_j=Z_N\;, \lb{4.20}\ee
hence we can write
\be
\sum_{i=h_\kk}^{N-1} {\wt z^{(\e)}_i Z_i \over Z_N} =\sum_{i=
h_\kk}^{N-1} \lft(\wt z^{(\e)}_i- z_i \a_{\e,h,N} \rgt) {Z_i\over
Z_N} +\a_{\e,h,N} \lft({Z_{h_\kk-1}\over Z_N} -1 \rgt)
\;.\lb{4.21}\ee
The first term in the r.h.s. of \pref{4.21} can be written as
\bea
&& \sum_{i=h_\kk}^{N-1} (\wt z^{(\e)}_i-\a_{\e,h,N} z_i) {Z_i\over
Z_N}=\sum_{j=h}^{N-1} (\wt z^{(\e)}_j-\a_{\e,h,N} z_j) {Z_j\over
Z_N}- \sum_{j=h}^{h_\kk-1} (\wt z^{(\e)}_j-\a_{\e,h,N} z_j)
{Z_j\over Z_N}\nn\\
&&\defi -\r_{\e,h,N} +R^{2,N,h}_{\e}(\kk)\;,\lb{4.22}\eea
where $\r_{\e,h,N}$ is independent of $\kk$ and satisfies, by
\pref{4.15}, the bound
\be
|\r_{\e,h,N}| \le C |\l|\sum_{j=h}^{N} \g^{-(\th-c\bar
\l_h^2)(N-j)} \le C|\l|\;,\lb{4.23}\ee
implying that there exists the limit $\r_\e= \lim_{-h,N\to\io}
\r_{\e,h,N}$. By an explicit computation one can show that $\r_{+}
= c_2\l+O(\l^2)$ and $\r_{-}=c_4+O(\l)$, with $c_2$ and $c_4$
strictly positive constants.
On the contrary, $R^{2,N,h}_{\e}(\kk)$ is vanishing for
$-h,N\to\io$; in fact
\be
|R^{2,N,h}_{\e}(\kk)| \le  C |\l|\sum_{j=h}^{h_\kk}
\g^{-(\th-c\bar\l_h^2)(N-j)} \le C |\l| \g^{-(\th/2)(N-h_\kk)}\;.
\lb{4.24}\ee

\insertplot{300}{30}%
{\footnotesize
\ins{15pt}{27pt}
{$ \o $}
\ins{40pt}{42pt}
{$ \o $}
\ins{50pt}{20pt}
{$-\o $}
\ins{47pt}{3pt}
{$-\o $}
\ins{85pt}{27pt}
{$ \o $}

\ins{105pt}{17pt}
{$ \o $}
\ins{135pt}{33pt}
{$ \o $}
\ins{175pt}{17pt}
{$ \o $}

\ins{215pt}{17pt}
{$ \o $}
\ins{245pt}{33pt}
{$ \o $}
\ins{285pt}{17pt}
{$ \o $}
}%
{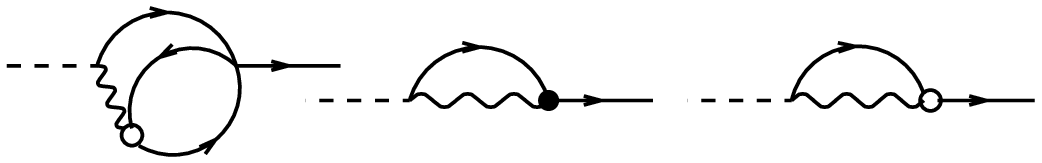}{\lb{f13}: The first two graphs are the
graphical representation of $c_2\l$; the last is the
graph for $c_4$.}{0}

By collecting all terms we get
\bea
A^{\wt z, N,h}_{\e,\o}(\kk) = A^{1,\wt z,N,h}_{\e,\o}(\kk)
-\a_{\e,N,h}{\hg_\o(\kk)\over Z_N} +\lb{4.19a}\\
+(\r_{\e,N,h}+\a_{\e,N,h}) \hg^{(h_\kk)}_\o(\kk) - R^{2,
N,h}_{\e}(\kk) \hg^{(h_\kk)}_\o(\kk)\;, \nn\eea
with $\lft|R^{2,N,h}_{\e}(\kk)\hg^{(h_\kk)}_\o(\kk)\rgt| \le C
|\l| \g^{-h_\kk} Z_{h_\kk}^{-1} \g^{-\th(N-h_\kk)}$.

We now consider $A^{1,\wt z,N,h}$ together to $A^{\wt \l,N,h}$. We
proceed as in \cite{BM3}, formulas (161)--(165); to summarize,
given a tree $\t\in\TT_{\l,n}$, $n\ge 1$, we can associate to it a
tree $\t'\in\TT_{z,n+1}$, substituting the endpoint $v^*$, on
scale $j^*$, of type $\wt\l$ with an endpoint of type $\l$, and
linking the endpoint $v^*$ to an endpoint of type $\wt z$. If we
define
\be \hG^{2,N,h}_\o(\kk)\defi
\hg^{(h_\kk)}_\o(\kk)+ \sum_{j^*=h_\kk}^N \l_{j^*}
B^{2,N,h}_{\o,j^*}(\kk)\;,\ee
then it is easy to check that
\bea
A^{\wt \l,N,h}_{\e,\o}(\kk) &=& {Z_{h_\kk-1}\over Z_N}
\sum_{j^*=h_\kk}^{N-1} \wt\l^{(\e)}_{j^*}
B^{2,N,h}_{\o,j^*}(\kk)\;,
\nn\\
A^{1,\wt z,N,h}_{\e,\o}(\kk) &=& - {Z_{h_\kk-1}\over
Z_N}\sum_{j^*=h_\kk}^{N-1} \l_{j^*}\lft({\sum_{i=h_\kk}^{N-1}
 \wt z^{(\e)}_i Z_i\over Z_{h_\kk-1}}\rgt) B^{2,N,h}_{\o,j^*}(\kk)\;.
\lb{4.26}\eea
Using \pref{4.20} and the definitions of $\r_{\e,h,N}$ and
$R^{2,N,h}_{\e}(\kk)$, we get:
\bea
&&\wt\l^{(\e)}_{j^*}-\l_{j^*} {\sum_{j=h_\kk}^{N-1}\wt z^{(\e)}_j
Z_j\over Z_{h_\kk-1}}= (\wt\l_{j^*}^{(\e)}-\a_\e\l_{j^*})+\nn\\
&&+\l_{j^*}{Z_N\over Z_{h_\kk-1}}
\lft[\a_{\e,h,N}+\r_{\e,h,N}-R^{2,N,h}_\e(\kk)\rgt]
\;.\lb{4.27}\eea
By the usual arguments, one can see that
$|B^{2,N,h}_{\o,j^*}(\kk)| \le C \g^{-h_\kk} Z_{h_\kk}^{-1}
(\g^{-\th|j^*-h_\kk|})$; hence, by summing the two equations in
\pref{4.26}, we get:
\be A^{\wt \l,N,h}_{\e,\o}(\kk)+
A^{1,\wt z,N,h}_{\e,\o}(\kk)= (\a_{\e,h,N}+\r_{\e,h,N})
\lft[\hG^{2,N,h}_\o(\kk) -\hg^{(h_\kk)}_\o(\kk)\rgt] +R^{3,
N,h}_{\e}(\kk)\;, \lb{4.28}\ee
where
\be R^{3,N,h}_\e(\kk)\defi {Z_{h_\kk-1}\over Z_N}
 \sum_{j^*=h_\kk}^N B^{2,N,h}_{\o,j^*}(\kk)
\lft[\lft(\wt\l_{j^*}^{(\e)}-\a_{\e,h,N}\l_{j^*}\rgt) - {Z_N\over
Z_{h_\kk-1}} \l_{j^*} R^{2,N,h}_\e(\kk)\rgt]\ee
is bounded by $C |\l| \g^{-h_\kk} Z_{h_\kk}^{-1} \g^{-(\th/2)
(N-h_\kk)}$.

Finally, the summation of all terms in the r.h.s. of \pref{4.16}
gives
\be \wt G^{2,N,h}_{\e,\o}(\kk)=
-\a_{\e,h,N} {g_\o(\kk)\over Z_N}+ (\a_{\e,h,N}+\r_{\e,h,N})
G^{2,N,h}_\o(\kk)+R^{4,N,h}_{\e}(\kk)\;,\lb{1.24a}\ee
with
\be |R^{4,N,h}_{\e}(\kk)|\le C|\l|{\g^{-h_\kk}\over Z_{h_\kk}}
\lft(\g^{-(\th/2)(h_\kk-h)}+
\g^{-(\th/2)(N-h_\kk)}\rgt)\;.\lb{1.25a}\ee
This ends the proof of Theorem \ref{th3}.

\subsection{Proof of Corollary \ref{cor1}}\lb{ss4.3}

If we insert the identity \pref{1.24a} in the r.h.s. of
\pref{1.22} and we take the limit $h\to -\io$, we get
\be
\hG^{2,N}_{\o}(\kk) D_\o(\kk)  = {B_N \over Z_N} - b_N \int
{d\pp\over (2\p)^2} \bar\chi_N(\pp) {\hG^{2,N}_\o(\kk-\pp)\over
D_{-\o}(\pp)} + H_{N,\o}(\kk)\;, \lb{4.31}\ee
where $\bar\chi_N(\pp)$ is the function appearing in \pref{1.21},
$B_N\= (1 -\l_N \sum_{\e} A_{\e,N} a_{\e,N}) [1-\l_N \sum_{\e}
A_{\e,N} (a_{\e,N}+\r_{\e,N})]^{-1}$, $b_N\= \l_N A_{+,N} [1-\l_N
\sum_{\e} A_{\e,N} (a_{\e,N}+\r_{\e,N})]^{-1}$ and $H_{N,\o}(\kk)$
is a function satisfying the bound
\be |H_{N,\o}(\kk)| \le C|\l| Z_{h_\kk}^{-1}
\g^{-(\th/2)(N-h_\kk)}\;.\lb{4.32}\ee
On the other hand, by \pref{2.68}, there is a function $f(\l)$,
independent of $\kk$, such that
\be \hG^{2,N}_{\o}(\kk)={|\kk|^{\h_z} \over D_\o(\kk)} F_{N}(\kk)
\virg F_{N}(\kk)= f(\l) +O(\g^{-N}|\kk|)^\th\;;\lb{4.46}\ee
hence, we can rewrite \pref{4.31} as
\be
|\kk|^{\h_z}F_{N}(\kk) = {B_N\over Z_N} +b_N \int{d\pp\over
(2\p)^2} \bar\chi_N(\pp+\kk) |\pp|^{\h_z} {F_{N}(\pp)\over
D_{-\o}(\pp+\kk)D_\o(\pp)} +{H_{N,\o}(\kk)}\lb{4.46a}\ee
and, subtracting the equation with $\kk=0$, we obtain
\bea
&&|\kk|^{\h_z} F_{N}(\kk) = b_N \int{d\pp\over (2\p)^2}
{|\pp|^{\h_z}\over D_\o(\pp)} F_{N}(\pp)
\left[{\bar\chi_N(\kk+\pp)\over D_{-\o}(\kk+\pp)}-
{\bar\chi_N(\pp)\over D_{-\o}(\pp)} \right]+ \nn\\
&&+H_{N;\o}(\kk)-H_{N,\o}(0)\;.\lb{4.46b}\eea
The integral can be written as the sum of two terms
\bea
&&\int {d\pp\over (2\p)^2} {|\pp|^{\h_z}\over |\pp|^2}
F_{N}(\pp)\bar\chi_N(\kk+\pp){D_{-\o}(\kk)\over
D_{-\o}(\kk+\pp)}-\nn\\
&& -\int {d\pp\over (2\p)^2} {|\pp|^{\h_z}\over |\pp|^2}
 F_{N}(\pp)[\bar\chi_N(\kk+\pp)-\bar\chi_N(\pp)]\;,\lb{4.47}\eea
and the second addend is vanishing in the $N\to\io$ limit, as it
can be written as
\be \g^{\h_z N}\int {d\pp\over (2\p)^2} {|\pp|^{\h_z}\over |\pp|^2}
 F_{N}(\pp)[\bar\chi_0(\g^{-N}\kk+\pp)- \bar\chi_0(\pp)]\lb{4.48}\ee
and $\bar\chi_0(\g^{-N}\kk+\pp)- \bar\chi_0(\pp)$ is
$O(\g^{-N}|\kk|)$ and with compact support. On the other hand, by
\pref{4.46}, the integral we obtain, if we substitute $F_{N}(\pp)$
with $f(\l)$, is vanishing as $N\to\io$. Hence, in the limit
$N\to\io$ we get the identity:
\be |\kk|^{\h_z}= b_\io \int {d\pp\over (2\p)^2}
{|\pp|^{\h_z}\over |\pp|^2} {D_{-\o}(\kk)\over D_{-\o}(\kk+\pp)}=
{b_\io\over 2\p} \int_0^\io {d\r\over\r^{1-\h_z}} \int_0^{2\pi}
{d\th\over 2\p} {|\kk|\over |\kk|+\r e^{i\th}}\;, \lb{4.49}\ee
that is
\be 1={b_\io\over 2\p} \int_0^\io {d\r\over\r^{1-\h_z}}
\int_0^{2\pi} {d\th\over 2\p} {1\over 1+\r e^{i\th}}= {b_\io\over
2\p} \int_0^1 {d\r\over\r^{1-\h_z}} ={b_\io\over 2\p \h_z}\;,
\lb{4.50}\ee
which proves \pref{1.27}.

\subsection{Proof of Theorem \ref{th6}}\lb{ss4.2}
The {\it Beta function equations} for the running coupling or
renormalization constants are
\bea
\l_{j-1}&
=&\l_j+\b^j_\l(\l_j,...,\l_N)\;,\nn\\
{Z_{j-1}\over Z_j} &=& 1+\b_z^{(j)}(\l_{j},..,\l_N)\;,\nn\\
{Z^{(2)}_{j-1}\over Z^{(2)}_j} &=& 1+ \b_{z_2}^{(j)}(\l_{j},
\ldots,\l_N)\;,\lb{2.44}\\
{\m_{j-1}\over \m_j} &=& 1+ \sum_{k\ge j} {\m_k\over \m_j}
\b_{\m}^{(j,k)}(\l_{j},..,\l_N)\;,\nn \eea
with $\b_z^{(j)},\b_{z_2}^{(j)}, \b_{\m}^{(j,k)}$ {\it independent
from $\m$} and, if $a_\m$, $a_z$, $a_{z_2}$ are suitable positive
constants,
\bea
\b_\m^{(j,k)}(\l_j,..\l_j) &=& a_\m\l_j\d_{j,k} + O(\bar\l_j^2)\;,\nn\\
\b_z^{(j)}(\l_{j},.., \l_j) &=& a_z\l_j^2+O(\bar\l_j^4)\;,\\
\b_{z_2}^{(j)}(\l_{j},..,\l_j) &=&
a_{z_2}\l_j^2+O(\bar\l_j^4)\;.\lb{2.44f}\eea
Moreover, these functions do not depend directly of $Z_N$, but
only depend on the ratios $Z_{j-1}/Z_j$, $j\le N$; hence the value
of $\l_j$ is a function of $\l_N=\l$ and the number of RG steps
needed to reach scale $j$ starting from scale $N$. It follows
that, if we call $\hat\l_j$, $j\le 0$, the constants we get for
$N=0$, then, for any $N>0$ and $j\le N$, $\l_j=\hat\l_{j-N}$. The
problem with $N=0$ was studied in detail in \cite{BM4}, where it
has been proved (see Theorem 2 of that paper) that there exist
constants $c_1,\e_1$ (independent of $N,h$), such that, if
$|\l|\le \e_1$, then $|\l_j|\le c_1\e_1$ for any $j$. The proof of
this statement is based on the analogue of SDe equation
\pref{1.22} for the four point function; if the momenta are
calculated at the infrared cut-off scale $\g^j$, a relation is
obtained between $\l_j$ and $\l$ implying that $\l_j=\l+O(\l^2)$.
This properties implies, see (3.48) of \cite{BM3}, that
\be |\b^j_\l(\l_j,...,\l_j)|\le C |\l_j|^2\g^{-(N-j)/4}\lb{5.13}\ee
From \pref{2.44} and \pref{5.13} one gets immediately, see \S 4.10
of \cite{BM1}, the bound \pref{2.42a} with
$\l_{-\io}(\l)=\l+O(\l^2)$ together with $|\log_\g(Z_{j-1}/ Z_j) -
\h_z| \le C\l^2 \g^{-(N-j)/4}$, $|\log_\g(\m_{j-1}/ \m_j) - \h_\m|
\le C|\l| \g^{-(N-j)/4}$; finally by the WTi \pref{1.17} with
momenta calculated at the infrared cut-off scale $\g^j$ one gets,
see \cite{BM2}, $|Z^{(2)}_j/ Z_j-1|\le C|\l|$.

\section{ Lattice Wilson fermions}\lb{ss4b}

\subsection{Integration of the doubled fermions}\lb{ss5.1}

In order to prove Theorem \ref{th4}, we have to compare the
Schwinger functions of the continuum model with ultraviolet cutoff
scale $N$ with those of the lattice model \pref{1.28} with $a=
\pi/ (4 \g^{N+1})$. In this model the momentum $\kk$ belongs to
the two-dimensional torus $\DD_a$ of size $2\p/a$ and we shall
denote by $|\kk-\kk'|$ the corresponding distance.

To begin with, we define $\bar f(\kk)$ so that
\be C_{N}^{-1}(\kk)+ \bar f(\kk)=1\;,\lb{5.1}\ee
where $C_{N}^{-1}(\kk)=\sum_{j=-\io}^N f_j(\kk)$, with $f_j(\kk)$
as in \pref{1.5}; since $C_{N}^{-1}(\kk)=0$ for $|\kk|\ge
\g^{N+1}=\pi/ (4 a)$, the support of the function $\bar f(\kk)$ is
given by the set $\{\kk:|\kk-\p/a|\le 3\pi/ 4 a\}$. Therefore, it
is possible to decompose the propagator $\wh r_{\o,\o'}(\kk)$,
defined in \pref{1.28a}, as the sum of $\wh r^{(\leq
N)}_{\o,\o'}(\kk) = C_{N}^{-1}(\kk) \wh r_{\o,\o'}(\kk)$ and $\wh
r^{( N+1)}_{\o,\o'}(\kk) = \bar f(\kk) \wh r_{\o,\o'}(\kk)$. With
this decomposition we associate the following decomposition of the
measure \pref{1.28}
\be P_{Z_a}(d\psi)=P_{Z_a}(d\psi^{(\le N)})P_{Z_a}
(d\psi^{(N+1)})\;.\lb{5.4}\ee
Note that the second integration has a ``very massive''
propagator; in fact, since the function $\bar f(\kk)$ is a Gevrais
function of class $2$, with a compact support of size $a^{-2}$,
and $[1-\cos(k_0a)+1-\cos(ka)]/a\geq \tilde C a^{-1} =C\g^N$ on
its support, it is easy to show that
\be |r^{(N+1)}_{\o,\o'}(\xx)|\le C\g^N e^{-c\sqrt{\g^N|\xx|}}\;.
\lb{5.5}\ee

The integration of $\WW(\f,J)$ is performed in a way very similar
to the one presented in \S\ref{ss2}, except for the first step,
made with $Z_{N+1}=Z_a$, $\l_{N+1}=\l_a$, $\n_{N+1}=\n_a$,
$\m_{N+1}(\kk)=\m_a(\kk)$. We define all localization operators as
in \S\ref{ss2}, except $\LL_1$, which is defined as
\be \LL_1\hW_{2,\o,\o'}^{(h)}(\kk)=\fra14
\sum_{\h,\h'=\pm 1}\hW_{2,\o,\o'}^{(h)}(\bk\h{\h'}) \big[\h {\sin
k_0a\over \sin{\p a\over L}} + \h'{\sin k a\over \sin{\p a\over
L}}\big]\;,\lb{5.6}\ee
in order to take into account the lattice structure of the space
coordinates; hence the localization procedure is essentially
unchanged. However, the presence in the interaction of the term
proportional to $\n_{N+1}$ has the effect (see below) that in the
effective potential a new type of vertex will appear (which we
shall call $\n$ vertex); this new vertex changes the symmetry
properties of the functions $\hW^{(j)}_{2,\o,\o'}$, so that, in
particular, $\PP_0\hW^{(j)}_{2,\o,-\o}\not=0$.

To be more precise, we note that $\hW^{(j)}_{2,\o,\o}$ is given by
the sum of graphs with
\bd
\item{1.}
either an even number of $\n$ vertices, an even number of non
diagonal propagators and an odd number of diagonal propagators;
\item{2.} or an odd number of $\n$ vertices, an odd number of non
diagonal propagators and an odd number of diagonal propagators.
\ed
Moreover $\hW^{(j)}_{2,\o,-\o}$ is given by the sum of graphs with
\bd
\item{3.} either an even number of $\n$ vertices, an odd number
of non diagonal propagators and an even number of diagonal
propagators;
\item{4.}
or an odd number of $\n$ vertices, an even number of non diagonal
propagators and an even number of diagonal propagators. \ed
As the diagonal propagators are odd in the exchange $\kk\to-\kk$
while the non diagonal ones are even, we get
$\LL_0\PP_0\hW^{(j)}_{2,\o,\o}=\LL_0\PP_1\hW^{(j)}_{2,\o,\o}=0$
and $\LL_1\PP_0\hW^{(j)}_{2,\o,-\o}=0$. Then
\be \LL \hW^{(j)}_{2,\o,\o}=\LL_1\PP_0
\widehat W^{(j)}_{2,\o,\o},\quad \LL \widehat W^{(j)}_{2,\o,-\o}
=\LL_0\PP_0 \hW^{(j)}_{2,\o,-\o}+\LL_0
\PP_1\hW^{(j)}_{2,\o,-\o}\;.\lb{5.7}\ee
This implies that we can write
\be \LL\VV^{(j)}(\psi^{[h,j]})=z_j F_\z^{[h,j]}+(s_j+\g^j n_j)
F_\s^{[h,j]} +l_j F_\l^{[h,j]} \;,\lb{5.8} \ee
where $\g^j n_j=\LL_0\PP_0 \hW^{(j)}_{2,\o,-\o}$, while $s_j=
\LL_0\PP_1 \hW^{(j)}_{2,\o,-\o}$, as in \pref{2.9}-\pref{2.10}.

The renormalization of the free measure is done exactly as in
\S\ref{ss2}, see \pref{2.18}, that is we do not put the term
proportional to $n_j$ in the free measure, but we define a new
running coupling constant $\n_j= n_j (Z_j/Z_{j-1})$. It follows
that the rescaled potential $\hat \VV^{(j)}(\psi^{[h,j]})$ differs
from that of \pref{2.24} because its local part contains the term
$\g^j \n_j F_\s^{[h,j]}$.

For $j\le N$, the renormalized measure takes the form:
\be \wh r^{(j)}_{\o,\o'}(\kk)\defi
{\tilde f_j(\kk)\over e_+(\kk)e_-(\kk) -\m^2_j(\kk)}
  \pmatrix
  {e_-(\kk) & -\tilde\m_j(\kk) \cr
  -\tilde\m_j(\kk)  & e_+(\kk)\cr}_{\o,\o'}\;,\lb{58bis}\ee
with $\tilde\m_j(\kk)= \hat\m_j(\kk) +[Z_N/Z_j(\kk)] [1-\cos(k_0a)
+1-\cos(ka)]/a$, $\hat\m_j(\kk)$ being a function equal to $\m$
for $j=N$, which satisfies the same recursion relation as
$\tilde\m_j(\kk)$ in \pref{2.18}.

It is convenient to split the propagator \pref{58bis} as
\be r^{(j)}_{\o,\o'}(\xx,\yy)= g^{(j)}_{\o,\o'}(\xx,\yy)+
g^{R,(j)}_{\o,\o'}(\xx,\yy)\lb{5.9}\ee
where $g^{(j)}_{\o,\o'}$ is obtained from $r^{(j)}_{\o,\o'}$ by
substituting $\tilde\m_j(\kk)$ with $\hat\m_j(\kk)$ (hence it has
the same form as the propagator of \pref{2.2}). We shall prove
below that the flow of the running couplings and the free measure
can be controlled as in \S\ref{ss2}, if the value of $\n_a$ is
suitable chosen. This implies that there is $h^*$, satisfying a
bound like \pref{2.45a}, such that, as far as $h>h^*$,
$|\tilde\m_j(\kk)|\le \g^j$, so that
\be |g^{R,(j)}_{\o,\o'}(\xx,\yy)|
\le C\g^{-(N-j)} \g^je^{-c\sqrt{\g^j|\xx-\yy|}}\;.\lb{5.10}\ee

The flow equation for $\l_j$ can be written, for $j\le N+1$, as
\bea \l_{j-1}
=&&\l_j+\b^j_\l(\l_N,...,\l_j)+r^j_\l(\l_a,\l_N,...,\l_j)\\
&& + \sum_{k\ge j} \n_k
\tilde\b^{j,k}_\l(\l_a,\n_a,\l_N,\n_N,...,\l_j,\n_j)\;,\lb{5.11}\eea
where the functions in the r.h.s. can be represented as sums over
trees similar to those of \pref{2.26}; in particular, we have
included the sum over all trees with at least one $\n$-endpoint in
the last term in the r.h.s. of \pref{5.11} and we have split the
sum of all trees with no $\n$-endpoints as $\b^j_\l+r^j_\l$, where
$\b^j_\l$ contains the trees with propagator $g^{(j)}_{\o,\o'}$
(the decomposition \pref{5.9} is used), while all other terms are
included in $r^j_\l$. The fact that the contribution of a single
tree satisfies a bound similar to that of \pref{2.27}, with
$d_v>0$ for any $v$, easily implies that, if $|\n_j| \le C|\l_a|$
for any $j$,
\be |\tilde\b^{j,k}_\l|\le C \bar\l_j \g^{-(k-j)/4}\virg
|r^j_\l|\le C \bar\l_j^2 \g^{-(N-j)/4}\;.\lb{5.12}\ee
Note also that \pref{5.13} still holds, as the only difference
comes from the fact that in the continuum model the delta function
of conservation of momenta is $L^2 \d_{k,0}\d_{k_0,0}$, while in
the lattice model is $L^2 \sum_{n,m\in Z^2} \d_{k,2\pi
n/a}\d_{k_0,2\pi m/a}$. However, the difference between the two
delta functions has no effect on the local part $\LL\VV^N$,
because of the compact support of $\psi^{\le N}$ and only slightly
affects the non local terms. To see that, let us consider a
particular tree $\t$ and a vertex $v\in\t$ of scale $h_v$ with
$2n$ external fields of space momenta $\kk_i$; the conservation of
momentum implies that $\sum_i \e_i \kk_i={\bf m}{2\pi\over a}$,
with $m$ an arbitrary integer. On the other hand, $\kk_i$ is of
order $\g^{h_v}$ for any $i$, hence $m$ can be different from $0$
only if $n$ is of order $\g^{N-h_v}$. Since the number of
endpoints following a vertex with $2n$ external fields is greater
or equal to $n-1$ and there is a small factor (of order
$\bar\l_j$) associated with each endpoint, we get an improvement,
in the bound of the terms with $|m|>0$, with respect to the
others, of a factor $\exp(-C\g^{N-h_v})$. Hence, by using the
remark preceding \pref{5.12}, it is easy to show that the
difference between the two beta functions is of order
$\bar\l_j^2\g^{-(N-j)/4}$.

\begin{lemma}
For any given $\l_{N+1}$ small enough, it is always possible to
fix $\n_{N+1}$ so that, for any $j\le N+1$,
\be
|\n_j|\le C |\l_a|\g^{-(N-j)/8}\;, \qquad |\l_j-\l_a|\le
C\l^2_a\;.\lb{5.13}
\ee
\end{lemma}

{\it Proof.} We consider the Banach space $\MM_\xi$ of sequences
$\underline\n= \{\n_j\}_{j\le N+1}$ such that
\be
||{\underline\n}||_\xi=\sup_{j\le N+1} \g^{(N-j)/8}|\n_j|\le \xi
|\l_a|\;,\ee
with $\xi$ to be fixed later. From \pref{5.11}, \pref{5.12} and
\pref{5.13} it follows, see \S 4 of \cite{BM1} or Appendix 5 of
\cite{GiM} for details, that there exists $\e_0$ such that, if
both $|\l_a|$ and $\xi|\l_a|$ are smaller than $\e_0$, then, for
any $\underline\n$, $\underline\n'\in\MM_\xi$,
\be |\l_j(\underline\n)-\l_{a}|\le C\l_{a}^2
\virg |\l_j(\underline\n)-\l_j(\underline\n')|\le C |\l_a|
||\underline\n -\underline{\n'}||_\xi\;. \lb{5.14}\ee
We want to show that it is possible to choose $\n_{N+1}$ so that
$\underline\n\in\MM_\xi$. Note that $\underline\n$ verifies by
construction the equation
\be \n_{j-1}=\g\n_j+\b_\n^{(j)}\big(\l_a,\n_a;\l_N,\n_N;...;\l_j,\n_j\big)
\lb{5.15}\ee
and that, if $\underline \n \in \MM_\xi$, $\lim_{j\to
-\io}\n_j=0$; by some simple algebra, this implies that
\be \n_{j}=-\sum_{k\le j}\g^{k-j-1}\b_\n^{(k)}
\big(\l_a,\n_a;\l_N,\n_N;...;\l_j,\n_j\big)\;.\lb{5.15a}\ee
Hence, we look for a fixed point of the operator ${\bf T}:
\MM_\xi\to\MM_\xi$ defined as
\be {\bf T}(\underline\n)_{j}\defi-\sum_{k\le j}\g^{k-j-1}
\b_\n^{(k)}\Big(\l_a,\n_a, \l_N(\underline\n),
\n_N,..,\l_j(\underline\n),\n_j\Big)\;.\lb{5.17}\ee
Note that
\bea
&& \b_\n^{(j)}(\l_a,\n_a,\l_N,\n_N,...,\l_j,\n_j) =
\b_\n^{(1,j)}(\l_N,...,\l_j)+\nn\\
&&+\sum_{k\ge j} \n_k\tilde\b_\n^{(j,k)}
\big(\l_a,\n_a,\l_N,\n_N,...,\l_j,\n_j\big)\;,\lb{5.1a}\eea
where $\b_\n^{(j,1)}$ is a sum over trees with no endpoints of
type $\n$ and no endpoints of scale $N+1$. By using the
decomposition \pref{5.9}, the parity properties of
$g^{(j)}_{\o,\o'}(\xx,\yy)$ and the remark preceding \pref{5.12},
we get the bounds
\be |\b_\n^{(1,j)}|\le C |\l_a| \g^{-(N-j)/4} \virg
|\tilde\b_\n^{(j,k)}|\le C |\l_a| \g^{-(k-j)/4}\;, \lb{5.19}\ee
which implies that
\be |{\bf T}(\underline\n)_j|\le \sum_{k\le j} C |\l_a|
\g^{-(j-k)} \g^{-(N-k)/8} \le c_0|\l_a|
\g^{-(N-j)/8}\;.\lb{5.20}\ee
Hence the operator ${\bf T}:\MM_\xi\to\MM_\xi$ leaves $\MM_\xi$
invariant, if $\xi\ge c_0$ and $\l_a$ is sufficiently small, and
it is also a contraction since $|{\bf T}(\underline\n)_j-{\bf
T}(\underline\n')_j|\le C|\l_a| ||\n-\n'||_\xi$.
It follows that there is a unique fixed point in $\MM_\xi$,
satisfying the flow equation \pref{5.15}.\qed

\*\*

An important consequence of the bound \pref{5.13} is that, if we
construct as in \S\ref{ss2} the Schwinger functions, by imposing
the normalization conditions \pref{2.43}, we get, as $N\to \io$,
exactly the same expansion in terms of trees, containing only $\l$
endpoints with a fixed coupling constant $\tilde\l_{-\io}(\l_a) =
\lim_{j\to -\io} \l_j$; in fact, the trees containing at least one
$\n$ vertex vanish in this limit.

By a fixed point argument, one can show that we can fix $\l_a$ so
that $\tilde\l_{-\io}(\l_a)$ has the same value as $\l_{-\io}(\l)$
in the continuum model; this remark completes the proof of Theorem
\ref{th4}.

\appendix

\section{Osterwalder-Schrader axioms}\lb{ss6}

Osterwalder-Schrader axioms were partially stated in \cite{OS1}
and completed in \cite{OS2} by the ``linear growth property''. We
show here that they are satisfied by the Schwinger functions of
our model.

\subsection{Linear growth condition and Clustering}\lb{ss6.1}
In order to verify the linear growth property, see the bound (4.1)
of \cite{OS2}, for $s=3$, let us consider the space
$\SS_0(\RRR^{2k})$ of the test functions such that, for any $m\in
\NNN$,
\be
\|f\|_m\defi \sup_{\xxx\in\rrr^{2k}\atop |\underline\a|\le m}
\left|(1+|\xxx|^2)^{m/2}\big(D^{\underline\a}f\big)(\xxx) \right|
< \io \lb{A.1}\ee
and which vanish, together with all their partial derivatives, if
at least two among the points in the set
$\xxx=\{\xx_1,\dots,\xx_k\}$ are coinciding. By \pref{2.45b}
\be
\lft|\big(S_{k,\underline\o},f\big)\right| \le
C^k(k!)^{3+2\h}\sum_{i<j} \int\!d\xx_1\cdots d\xx_k\
{\big|f(\xxx)\big| \over  |\xx_i-\xx_j|^{k(1+\h)/2
-2\e}}\;.\lb{A.2}\ee
On the other hand, by \pref{A.1}, $|f(\xxx)|\le \|f\|_{4k+1}
(1+|x|^{4k+1})^{-1}$ and, for any $i\not= j$, $|f(\xxx)| \le 2^{k}
[(2k)!]^{-1} |\xx_i -\xx_j|^{2k} \|f\|_{2k}$; hence, since
$\|f\|_{2k} \le \|f\|_{4k+1}$,
\be
|f(\xxx)|\le \|f\|_{4k+1} \sqrt{(1+|\xxx|^{4k+1})^{-1}}
\sqrt{2^{k} [(2k)!]^{-1} |\xx_i -\xx_j|^{2k}}\;.\lb{A.2bis} \ee
It follows that
\bea
\lft|\big(S_{k,\underline\o},f\big)\right| &&\le C^k
(k!)^{2+2\h}\|f\|_{4k+1}\;.\lb{A.2ter}\eea

In order to prove the ``cluster property'', fixed any integer
$p\in [1,k-1]$, $\yy\in\RRR^2$ and $f\in \SS_0(\RRR^{2k})$, we
first prove that $\big(S_{k,\underline\o},f_{p,\yy} \big)$ goes to
$0$ as $|\yy|\to \io$, if $f_{p,\yy}(\xxx) \=
f(\xx_1,\dots\xx_p,\xx_{p+1}-\yy,\dots,\xx_k-\yy)$. Let us
consider the characteristic functions $\c_{\yy}(\xxx)$ and
$\c'_{\yy}(\xxx)$ of the set
\be
M\defi \left\{\xxx\in \RRR^{2k}:\max_{1\le j\le p}|\xx_j|\le
|\yy|/4\;,\ \max_{p+1\le j\le k}|\xx_j-\yy|\le
|\yy|/4\right\}\lb{A.3}\ee
and of its complementary, respectively. Since $D_\xxx\geq |\yy|/2$
in $M$, by using \pref{2.45b} and \pref{A.2bis}, we see that
$|\big(S_{k,\underline\o},f_{p,\yy} \c_{\yy}\big)| \le [1+
(|\yy|/2)^{2\e}]^{-1} C^k (k!)^{2+2\h}\|f\|_{4k+1}$, so that
$\big(S_{k,\underline\o},f_{p,\yy} \c_{\yy}\big)$ is uniformly
bounded and vanishes as $|\yy|\to\io$. On the other hand, by
\pref{A.2bis}, $|\big(S_{k,\underline\o},f_{p,\yy} \c'_{\yy}
\big)| \le C^k (k!)^{2+2\h}\|f\|_{4k+1} \int d\xxx
\sqrt{(1+|\xxx|^{4k+1})^{-1}} \c'_0(\xxx)$, so that even
$\big(S_{k,\underline\o},f_{p,\yy} \c'_{\yy}\big)$ is uniformly
bounded and vanishes in the limit $|\yy|\to\io$, as well as
$\big(S_{k,\underline\o},f_{p,\yy}\big)$.

The cluster property E0, defined in \S3 of \cite{OS1}, now simply
follows, by decomposing the connected Schwinger functions as
finite linear combinations of the truncated Schwinger functions, .

\subsection{Symmetry, Euclidean invariance and Reflection positivity}
\lb{ss6.2} From the explicit definition of  the generating
functional, \pref{1.6}, two properties immediately follow. First,
since the fields anticommute, the Schwinger functions are
antisymmetric in the exchange of their arguments. Moreover, the
generating functional \pref{1.6} is invariant under the Lorentz
transformation of the fields by construction.

Finally the ``reflection positivity'' E2, defined in \S6 of
\cite{OS1}, is verified in the lattice regularization \pref{1.30},
as proved in \cite{OSe}, hence it holds even in the removed
cutoffs limit of the regularized model \pref{1.6}, which we have
shown to be equivalent to the $a=0$ limit of the lattice model,
see Theorem 1.4.

\section{Lowest order computation of $\n_-$ and $\n_+$}\lb{ss7}

\subsection{Lowest order computation of $\n_-$}\lb{ss7.1}
Calling  $\hat g_{\o,\o}(\kk)\defi\hat g_{\o}(\kk)$ and
$u_0(t)\defi 1-\c_0(t)$, the lowest order contibution to the
$\n_{-,N}$, appearing in \pref{3.5}, is obtained, from \pref{3.16}
and \pref{3.17}, by taking the $\pp\to 0$ limit of the following
expression (see the first graph in Fig. \ref{f6}), whose value is
independent of the infrared cutoff for any fixed $\pp$ and $|h|$
large enough:
\bea
&&\l\int\!{d\kk\over (2\pi)^2}\
 {C_{h,N;\o}(\kk,\kk-\pp)\over D_{-\o}(\pp)}
\hat g^{(\le N)}_\o(\kk)\hat g^{(\le N)}_\o(\kk-\pp) =\nn\\
=&&-\l{D_\o(\pp)\over D_{-\o}(\pp)}\int {d\kk\over (2\pi)^2}
{u_0\big(\g^{-N}|\kk- \pp|\big)\chi_0\big(\g^{-N}|\kk|\big)\over
D_\o(\kk-\pp) D_{\o}(\kk)} +\nn\\
&&+\l\int {d\kk\over (2\pi)^2}
{\chi_0\big(\g^{-N}|\kk|\big)-\chi_0\big(\g^{-N}|\kk-\pp|\big)\over
D_\o(\kk-\pp) D_{-\o}(\pp)}\lb{B.1}\eea
where we have used \pref{3.3} and rearranged the terms. In the
limit $|\pp|\to 0$, the first contribution in the r.h.s. of
\pref{B.1} vanishes by the symmetry  $\hat g_\o(\kk)=-i\o \hat
g_\o(\kk^*)$, $\kk^*=(-k_0,k)$. As regards the second term, if we
write the first order Taylor expansion in $\pp$ of the numerator
as a linear combination of $D_{-\o}(\pp)$ and $D_\o(\pp)$, the
term proportional to $D_\o(\pp)$ also vanishes, again for the
symmetry $\kk\to\kk^*$, so that
\be \n_-= -{\l\over 2}\int\!{d\kk\over (2\pi)^2}\
{\chi'_0(|\kk|) \over |\kk|} = -{\l\over 4\pi}\int_1^\io\!d\r\
\chi'_0(\r)={\l\over 4\pi}\;. \lb{B.2}\ee

\subsection{Lowest order computation of $\n_+$}\lb{ss6.2}

If we define
\be
I_{\o}\lft(\g^{-N}\kk\rgt)=\int\!{d\kk'\over (2\pi)^2}\ \hat
g_{\o}^{(\le N)}(\kk') \hat g_{\o}^{(\le N)}(\kk'+\kk)\;,
\ee
then the lowest order contribution to the anomaly coefficient
$\n_{+,N}$, appearing in \pref{3.5}, is is obtained, from
\pref{3.16} and \pref{3.17}, by taking the $\pp\to 0$ limit and,
after that, the $h\to -\io$ limit of the following expression (see
the second graph in Fig. \ref{f6}):
\bea
&&-\l^2 \int {d\kk\over (2\pi)^2} {C_{h,N:\o}(\kk,\kk-\pp)\over
D_{\o}(\pp)} g^{(\le N)}_\o(\kk)g^{(\le N)}_\o(\kk-\pp)
I_{-\o}\lft(\g^{-N}\kk\rgt)
=\nn\\
&& =\l^2\int {d\kk\over (2\pi)^2}
{u_0\big(\g^{-N}|\kk-\pp|\big)\chi_0\big(\g^{-N}|\kk|\big)\over
D_\o(\kk-\pp)
D_{\o}(\kk)}I_{-\o}\lft(\g^{-N}\kk\rgt)-\nn\\
&&-\l^2\int {d\kk\over (2\pi)^2}
{\chi_0\big(\g^{-N}|\kk|\big)-\chi_0\big(\g^{-N}|\kk-\pp|\big)\over
D_\o(\kk-\pp) D_{\o}(\pp)}I_{-\o}\lft(\g^{-N}\kk\rgt)\;.\lb{B.4}
\eea
In the limit $|\pp|\to 0$ and $h\to -\io$, we get
\bea
\n_+ &&=\l^2\int\!{d\kk\over (2\pi)^2}\
\left[{u_0(|\kk|)\chi_0(|\kk|)\over |\kk|^4}- {\chi'_0(|\kk|)\over
2 |\kk|^3}\right] I_{-\o}(\kk)D^2_{-\o}(\kk)\;,\lb{B.5}
\eea
where we are using the symbol $I_{-\o}(\kk)$ to denote even its
$h=-\io$ limit, which is finite. Note that the term in square
brackets is nonnegative; moreover, it is different from $0$ only
for $1\le|\kk|\le \g_0$ (defined in \pref{1.5bis}). We now fix
$\o=+$ for definiteness (the result is $\o$-independent); then if
$i k_0+k=y e^{i\phi}$ and $i k'_0+k'=x e^{i\th}$ we get:
\be
I_{-}(\kk)=e^{-2 i\phi}\int {dx d\th\over (2\pi)^2}
\chi_0(x){\chi_0(|x e^{-i\th}+y|)\over |x
e^{-i\th}+y|^2}e^{-i\th}(x e^{-i\th}+y)\;,
\ee
so that
\be
D^2_{-}(\kk)I_{-}(\kk)=y^2 \int {dx d\th\over (2\pi)^2}
\chi_0(x){\chi_0(|x e^{-i\th}+y|)\over |x e^{-i\th}+y|^2}(x\cos
2\th+y\cos\th)\;. \lb{B.7}\ee
The integral \pref{B.5} is easily shown to be strictly negative in
the limit $\g_0\to 1$; hence by continuity in $\g_0$, $\n_+<0$ for
$\g_0-1$ small enough. Indeed in the limit $\g_0\to 1$ \pref{B.5}
becomes
\be
{\pi\l^2\over (2\p)^4} \int_0^1 dx \int_0^{2\pi} d\th\ {\chi_0(|x
e^{-i\th}+1|)\over |x e^{-i\th}+1|^2}(x\cos 2\th+\cos\th)\;;
\lb{B.8}\ee
on the other hand, since $|x e^{-i\th}+1|\le 1$, $\cos\th< 0$ if
$x>0$ and $x\cos 2\th+ \cos\th =\cos\th (1+x\cos\th)-x\sin^2\th<
0$ if $0<x<1$; it follows that the integrand of \pref{B.8} is $<0$
for $x\not=0,1$.

A numerical calculation also shows that $|\n_+|$ is not constant
as a function of $\g_0$, but is a strictly decreasing function
near $\g_0=1$.
\vskip1cm
{\bf Acknowledgments} We are indebted with K Gawedzki
for enlightening discussions on the Thirring model which we have 
summarized
in the considerations after (1.4) in the introduction.
P.F. gratefully acknowledges
the hospitality and the financial 
support of the Erwin Schr\"odinger 
Institute for Mathematical Physics (Vienna)
during the 
preparation of this work.


\vskip.5cm
\begin{thebibliography}{999999}

\bibitem[A]{A} Adler S. L.:
Axial--Vector Vertex in Spinor Electrodynamics. {\it Phys. Rev.}
{\bf 177}, { 2426--2438}, 1969.

\bibitem[A1]{A1} Adler S. L.: Anomalies. {\it hep-th/0411038}
To appear in the Encyclopedia of Mathematical Physics, Elsevier,
2006.

\bibitem[AB]{AB} Adler S. L., Bardeen W.A.: Absence of higher order
corrections in the anomalous axial vector divergence equation.
{\it Phys. Rev.} {\bf 182}, { 1517-1536}, 1969.

\bibitem[AF]{AF} Akiyama A., Futami Y.: Two-fermion-loop contribution
to the axial anomaly in the massive Thirring model. {\it Phys.
Rev. D} {\bf 46}, { 798-805}, 1992.


\bibitem[BM1]{BM1} Benfatto G., Mastropietro V.:
Renormalization group, hidden symmetries and approximate Ward
identities in the $XYZ$ model. {\it Rev. Math. Phys.} {\bf 13}, {
1323--1435}, 2001.

\bibitem[BM2]{BM2} Benfatto G., Mastropietro V.:
On the density--density critical indices in interacting Fermi
systems. {\it Comm. Math. Phys.} {\bf 231}, { 97--134}, 2002.

\bibitem[BM3]{BM3} Benfatto G., Mastropietro V.:
Ward identities and vanishing of the Beta function for $d=1$
interacting Fermi systems. {\it J. Stat. Phys.} {\bf 115}, {
143--184}, 2004.

\bibitem[BM4]{BM4} Benfatto G., Mastropietro V.:
Ward identities and chiral anomaly in the Luttinger liquid. {\it
Comm. Math. Phys.} {\bf 258}, { 609--655}, 2005.


\bibitem[C]{C} Coleman S.: Quantum sine-Gordon equation as the massive
Thirring model. {\it Phys. Rev. D} {\bf 11}, { 2088-2097}, 1975.

\bibitem[CRW]{CRW} Carey A.L., Ruijsenaars S.N.M., Wrigth J.D.:
The massless Thirring model: Positivity of Klaiber's n-point
functions. {\it Comm. Math. Phys.} {\bf 99}, { 347--364}, 1985.

\bibitem[DFZ]{DFZ} Dell'Antonio G., Frishman Y., Zwanziger, D.:
Thirring Model in Terms of Currents: Solution and Ligth--Cone
Expansions. {\it Phys. Rev. D} {\bf 6}, { 988--1007} 1972.

\bibitem[DR]{DR} Disertori M., Rivasseau, V.:
Interacting Fermi Liquid in Two Dimensions at Finite Temperature.
Part I: Convergent Attributions. {\it Comm. Math. Phys.} {\bf
215}, { 251--290}, 2000.

\bibitem[FGS]{FGS} Furuya K., Gamboa Saravi S, Schaposnik F. A. :
Path integral formulation of chiral invariant fermion models in
two dimensions. {\it Nucl. Phys. B} {\bf 208}, { 159--181}, 1982.


\bibitem[FMRS]{FMRS} Feldman J., Magnen J., Rivasseau V, S\'en\'eor R.:
Massive Gross--Neveu Model: A Rigorous Perturbative Construction.
{\it Phys. Rev. Lett} {\bf 54}, { 1479--1481}, 1985.


\bibitem[G]{G} Gallavotti G.:
Renormalization theory and ultraviolet stability for scalar fields
via renormalization group methods. {\it Rev.Mod.Phys.} {\bf 57},
{471--562}, 1985.

\bibitem[GK]{GK} Gawedzki K., Kupiainen A.:
Gross--Neveu model through convergent perturbation expansions.
{\it Comm.Math.Phys.} {\bf 102}, { 1--30}, 1985.

\bibitem[GL]{GL} Gomes M., Lowenstein J.H.:
Asymptotic scale invariance in a massive Thirring model. {\it
Nucl. Phys. B} {\bf 45}, { 252--266}, 1972.

\bibitem[GR]{GR} Georgi H., Rawls J.M.:
Anomalies of the Axial--Vector Current in Two dimensions. {\it
Phys.Rev.D} {\bf 3}, { 874--879}, 1971.

\bibitem[GiM]{GiM} Giuliani A., Mastropietro V.:
Anomalous Universality in the Ashkin-Teller model. {\it Comm.
Math. Pys} 2003.

\bibitem[J]{J} Johnson K.:
Solution of the Equations for the Green's Functions of a two
Dimensional Relativistic Field Theory. {\it Nuovo Cimento} {\bf
20}, {773--790}, 1961.


\bibitem[K]{K} Klaiber B.: The Thirring model. In: Quantum theory and
statistical physics. Vol X A editors Barut A.O. and Brittin. W.F.
{\it Gordon and Breach.}, 1968.

\bibitem[Le]{Le} Lesniewski A.:
Effective action for the Yukawa$_2$ quantum field theory. {\it
Comm. Math. Phys.} {\bf 108}, { 437--467}, 1987.


\bibitem[M]{M} Mastropietro V.: preprint 2006


\bibitem[MM]{MM} Montvay I., M\"unster G.: Quantum Fields on a
Lattice. {Cambridge University Press}, 1994.

\bibitem[OS1]{OS1} Osterwalder K., Schrader R.: Axioms for Euclidean
Green's Functions. {\it Comm. Math. Phys.} {\bf 31}, { 83--112},
1973.


\bibitem[OS2]{OS2} Osterwalder K., Schrader R.: Axioms for Euclidean
Green's Functions II. {\it Comm. Math. Phys.} {\bf 42},
{281--305}, 1975.

\bibitem[OSe]{OSe} Osterwalder K., Seiler E.: Gauge Field Theories on
a Lattice. {\it Ann.Phys.} {\bf 110}, {440--471}, 1978.

\bibitem[S]{S} Seiler E:  {\it Phys. Rev. D}
{\bf 22}, {2412--2418}, 1980.



\bibitem[T]{T} Thirring W.: A soluble relativistic field theory.
{\it Ann.Phys.} {\bf 3}, {91--112}, 1958.

\bibitem[Wi]{Wi} Wilson K.G.:
Non--Lagrangian Models of Current Algebra. {\it Phys. Rev.} {\bf
179}, { 1499--1512}, 1969.

\bibitem[W]{W} Wightman W.: { Cargese lectures}, 1976.



\end{thebibliography}
\end{document}